\documentclass[11pt,a4paper]{article}
\pdfoutput=1

\usepackage{jheppub}
\usepackage{latexsym}
\usepackage{color}
\usepackage[usenames,dvipsnames,svgnames,table]{xcolor}
\usepackage{graphicx}
\usepackage{epsfig}  
\usepackage{epsf}   
\usepackage{dcolumn}
\usepackage{bm}
\usepackage[toc,page]{appendix}
\usepackage{dcolumn}
\usepackage{textcomp}
\usepackage{float}
\usepackage{hypcap}
\usepackage[]{hyperref}
\usepackage{booktabs}
\usepackage{multirow}
\usepackage{diagbox}
\usepackage{alphalph}
\usepackage{textcomp}
\usepackage{epstopdf}

\usepackage{lineno}
\usepackage{soul} 
\usepackage[normalem]{ulem}

\newcommand{\beq}{\begin{equation}}
\newcommand{\eeq}{\end{equation}}
\newcommand{\beqa}{\begin{eqnarray}}
\newcommand{\eeqa}{\end{eqnarray}}
\definecolor{orcidlogocol}{HTML}{A6CE39}

\preprint{IP/BBSR/2021-11}

\title{A close look on 2-3 mixing angle with DUNE in light of current neutrino oscillation data}

\author[a,d,e]{Sanjib Kumar Agarwalla,}
\author[a,d]{Ritam Kundu,}
\author[b]{Suprabh Prakash,}
\author[a,c]{Masoom Singh}

\affiliation[a]{Institute of Physics, Sachivalaya Marg, Sainik School Post, Bhubaneswar 751005, India}
\affiliation[b]{The Institute of Mathematical Sciences, C.I.T. Campus, Taramani, Chennai 600113, India}
\affiliation[c]{Utkal University, Vani Vihar, Bhubaneswar 751004, India}
\affiliation[d]{Homi Bhabha National Institute, Training School Complex, Anushakti Nagar, Mumbai 400094}
\affiliation[e]{International Centre for Theoretical Physics, Strada Costiera 11, Trieste 34151, Italy}

\emailAdd{sanjib@iopb.res.in (ORCID: 0000-0002-8367-8401)}
\emailAdd{ritam.k@iopb.res.in (ORCID: 0000-0003-3258-4357)}
\emailAdd{suprabh@imsc.res.in (ORCID: 0000-0002-1529-4588)}
\emailAdd{masoom@iopb.res.in (ORCID: 0000-0002-8363-7693)}

\abstract{Recent global fit analyses of world oscillation data under 3$\nu$ hypothesis show a preference for normal mass ordering (NMO) at 2.5$\sigma$ and provide 1.6$\sigma$ indications for lower $\theta_{23}$ octant ($\sin^2\theta_{23} < 0.5$) and leptonic CP violation ($\sin\delta_{\mathrm{CP}} < 0$). A high-precision measurement of 2-3 mixing angle is pivotal to convert these hints into discoveries and to address the long-standing flavor problem. In this work, we study in detail the capabilities of the long-baseline experiment DUNE to establish the deviation from maximal $\theta_{23}$ and to resolve its octant at high confidence levels in light of the current neutrino oscillation data. We exhibit the possible correlations and degeneracies among $\sin^2\theta_{23}$, $\Delta m^2_{31}$, and $\delta_{\mathrm{CP}}$ in $\nu_{\mu} \to \nu_{\mu}$ and $\nu_{\mu} \to \nu_e$ oscillation channels at the probability and event levels. Introducing for the first time, a bi-events plot in the plane of total $\nu$ and $\bar\nu$ disappearance events, we discuss the impact of $\sin^2\theta_{23}$ - $\Delta m^2_{31}$ degeneracy in establishing non-maximal $\theta_{23}$ and show how this degeneracy can be resolved by exploiting the spectral shape information in $\nu$ and $\bar\nu$ disappearance events. A 3$\sigma$ (5$\sigma$) determination of non-maximal $\theta_{23}$ is possible in DUNE with an exposure of 336 kt$\cdot$MW$\cdot$years if the true value of $\sin^2\theta_{23} \lesssim 0.465~(0.450)$ or $\sin^2\theta_{23} \gtrsim 0.554~(0.572)$ for any value of true $\delta_{\mathrm{CP}}$ and true NMO. We study the individual contributions from appearance and disappearance channels, impact of systematic uncertainties and marginalization over oscillation parameters, importance of spectral analysis and data from both $\nu$ and $\bar\nu$ runs, while analyzing DUNE's sensitivity for the discovery of a non-maximal $\theta_{23}$. We observe that both $\nu$ and $\bar\nu$ data are essential to settle the $\theta_{23}$ octant at a high confidence level. DUNE can resolve the octant of $\theta_{23}$ at 4.2$\sigma$ (5$\sigma$) using 336 (480) kt$\cdot$MW$\cdot$years of exposure assuming the present best-fit values of $\sin^2\theta_{23}$ (0.455) and $\delta_{\mathrm{CP}}$ ($223^\circ$) as their true choices and with true NMO. DUNE can improve the current relative 1$\sigma$ precision on $\sin^2\theta_{23}$ ($\Delta m^2_{31}$) by a factor of  4.4 (2.8) using 336 kt$\cdot$MW$\cdot$years of exposure.}


\arxivnumber{2111.aaaaa}

\begin{document}
\maketitle

\section{Introduction and motivation}
\label{introduction}

A deeply-relevant and much-awaited result concerning neutrinos in recent times is the {\it hint} for violation of the CP symmetry in the leptonic sector. The T2K collaboration~\cite{T2K:2011qtm} in their 2019 results~\cite{T2K:2019bcf} have shown that their neutrino and antineutrino appearance data point towards CP being near-maximally violated $i.e.$ $\vert\sin\delta_{\rm CP}\vert$ is close to 1. They obtain a best-fit value of $\delta_{\rm CP}$ at $252^\circ$ while the CP-conserving values of $\delta_{\rm CP} = 0^\circ\,, \,\pm\, 180^\circ$ are ruled out at $95\%$ confidence level (C.L.). They also report a preference for the normal mass ordering (NMO) over the inverted mass ordering (IMO) at nearly $68\%$ confidence level. NMO and $\delta_{\rm CP} = 270^\circ$ is in fact one of the most favorable parameter combinations for which early hints regarding mass ordering and CP violation can be expected from the currently running long-baseline accelerator experiments~\cite{Prakash:2012az, Agarwalla:2012bv}. The same set of measurements are also being carried out by the NO$\nu$A experiment~\cite{NOvA:2007rmc,Ayres:2002ws,NOvA:2004blv} which operates at a longer baseline with more energetic neutrinos. The recent results from NO$\nu$A~\cite{NOvA:2021nfi} also show a preference for NMO, but their  best-fit to $\delta_{\rm CP}$ is not in conjunction with T2K. NO$\nu$A's $\delta_{\rm CP}$ best-fit value of $148^\circ$ is $2.5\sigma$ away from T2K's best-fit. However, the two experiments agree on $\delta_{\rm CP}$ measurements when they assume IMO to be true -- each reporting a best-fit value around $270^\circ$. The tension between these two data sets is not yet at a statistically significant level and we need to wait for further data from T2K and NO$\nu$A to see if this tension persists. In any case, a $5\sigma$ {\it discovery} of any of the current unknowns in neutrino oscillation physics does not seem to be within the reach of either of these experiments~\cite{Agarwalla:2012bv}. Nonetheless, these results are quite important and play an important role in the global fit studies.

\begin{figure}[htb!]
	\centering
	\includegraphics[width=\linewidth]{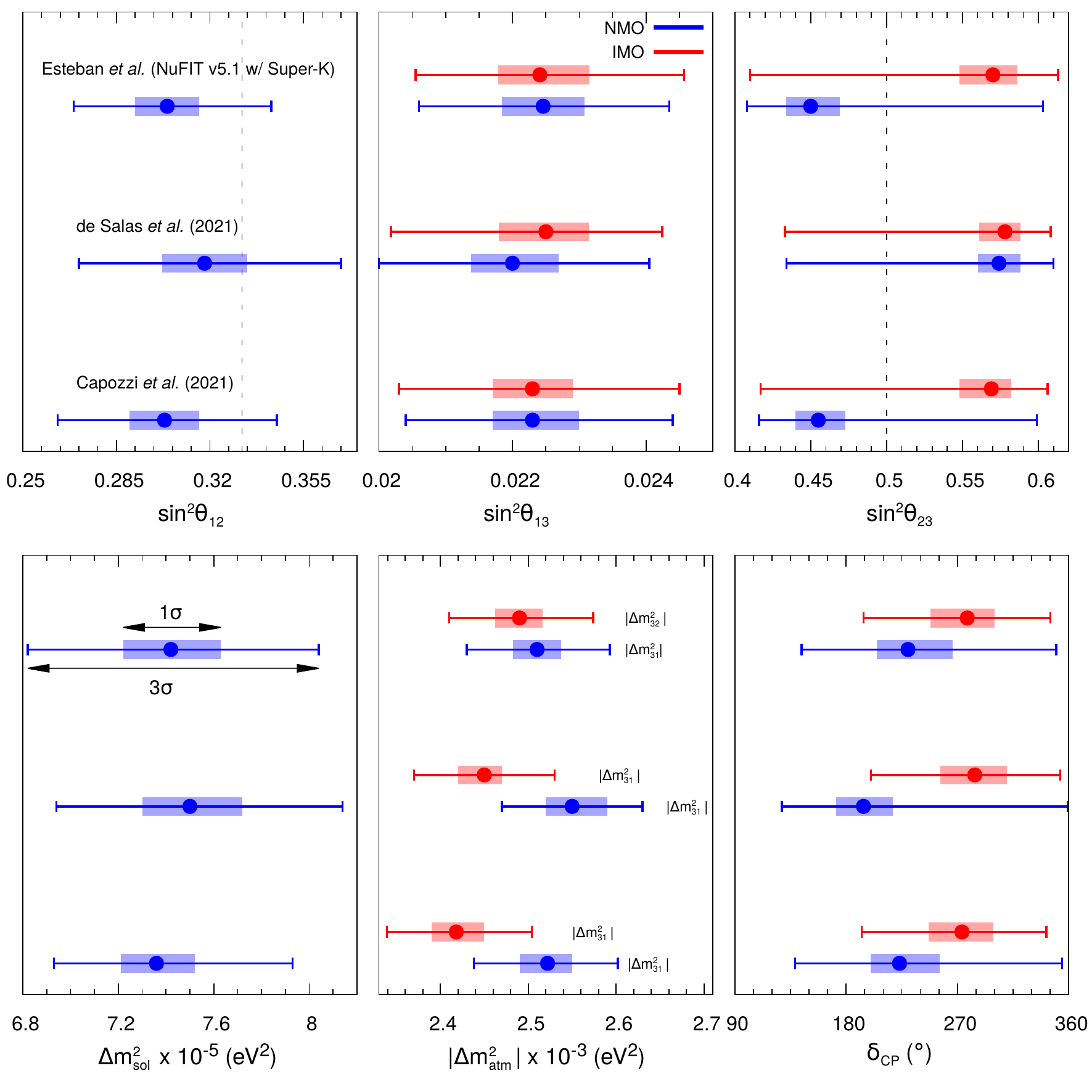}
	\caption{\footnotesize{Current $1\sigma$ (see rectangular boxes)  and $3\sigma$ (see horizontal lines) allowed ranges of the neutrino oscillation parameters obtained from the global fit studies performed by Esteban $et~ al.$~\cite{NuFIT}, de Salas $et~ al.$~\cite{deSalas:2020pgw}, and Capozzi $et~ al.$~\cite{Capozzi:2021fjo}. The blue (red) lines and boxes represent the values for NMO (IMO). In each panel, the best-fit value of respective oscillation parameter is shown by blue (red) dots for NMO (IMO).
Vertical black dashed lines in the panels related to $\sin^2\theta_{12}$ and $\sin^2\theta_{23}$ show their corresponding values in the tri-bimaximal mixing scheme. Note that the measurements of $\sin^{2}\theta_{12}$ and $\Delta m^{2}_{\mathrm{sol}}\, (\equiv \Delta m^{2}_{21})$ are not sensitive to the choice of mass ordering. }}
	\label{fig:1}
\end{figure}

Fig.~\ref{fig:1} summarizes our current understanding of the six neutrino oscillation parameters in the standard three-neutrino framework. It confirms that we have already attained a remarkable precision on solar oscillation parameters ($\Delta m^2_{21}$ and $\sin^2\theta_{12}$), atmospheric mass-splitting ($\Delta m^2_{31}$)~\cite{Nunokawa:2005nx}, and reactor mixing angle ($\theta_{13}$)~\cite{Blennow:2013oma,Minakata:2002jv}. In this figure, we compare the $1\sigma$ (shown with colored rectangular blocks) and $3\sigma$ allowed (shown with horizontal lines) regions of the oscillation parameters that have been calculated by doing a combined analyses~\cite{NuFIT,deSalas:2020pgw,Capozzi:2021fjo,Esteban:2020cvm} of the existing global neutrino data. These works take into account the data from the Solar (Gallex and GNO~\cite{Cleveland_1998}, SAGE~\cite{SAGE:2009eeu}, the four phases of Super-K (SK I-IV)~\cite{Super-Kamiokande:2005wtt,Super-Kamiokande:2008ecj, Super-Kamiokande:2010tar}, SNO~\cite{SNO:2011hxd}, and Borexino I-III~\cite{Bellini:2011rx,Borexino:2008fkj,BOREXINO:2014pcl}), atmospheric (IceCube/DeepCore~\cite{IceCube:2014flw,IceCube} and the four phases of Super-K (SK I-IV)~\cite{Super-Kamiokande:2017yvm,data-superk}), reactor (KamLAND~\cite{KamLAND:2013rgu}, Daya Bay~\cite{DayaBay:2018yms}, and RENO~\cite{RENO:2018dro,reno}), and the accelerator experiments (MINOS~\cite{MINOS:2013utc,MINOS:2013xrl}, T2K~\cite{t2k}, and NO$\nu$A~\cite{nova}). All the three studies find that the earlier tension between Solar and KamLAND data has been reduced considerably after incorporating the recent results from Super-K Phase IV 2970 days of solar data (energy spectra and day-night asymmetry)~\cite{sk}. Additionally, both Esteban $et~ al.$~\cite{Esteban:2020cvm, NuFIT} and Capozzi $et ~al.$~\cite{Capozzi:2021fjo} also consider the recent Super-K Phase IV atmospheric data~\cite{sk}.

In Fig.~\ref{fig:1}, the blue (red) regions are obtained assuming NMO (IMO). Note that in the case of the solar oscillation parameters $i.e.$ $\sin^2\theta_{12}$ and $\Delta m^2_{21}$, the IMO and NMO regions are identical. Vertical black dashed lines in the panels related to $\sin^2\theta_{12}$ and $\sin^2\theta_{23}$ depict their corresponding values in the tri-bimaximal mixing scheme~\cite{Harrison:2002er,Harrison:2002kp,King:2014nza}. 
Note that while Esteban $et~ al.$ and de Salas $et~ al.$ quote the values of atmospheric mass-splitting in terms of $\Delta m^{2}_{31}$, Capozzi $et~ al.$ express it in terms of $\Delta m^{2} = m^{2}_{3}- (m^{2}_{1} + m_{2}^{2})/2$, where $\Delta m^{2}_{31}$ = $\Delta m^{2} + \Delta m^{2}_{21}/2$ for both NMO and IMO.

For completeness, in Table~\ref{table:five} of Appendix~\ref{appendix_a}, we also give the numbers from the three global fit studies that we use to generate Fig.~\ref{fig:1}. A novel aspect of Fig.~\ref{fig:1} is that all three global fits now rule out $\delta_{\rm CP} \in \left[ 0, \sim135^\circ \right]$ at $3\sigma$ confidence level and $\delta_{\rm CP} \in \left[ 0, \sim180^\circ \right]$ at $1\sigma$ confidence level, while predicting the best-fit value to lie somewhere in the range $\left[ 200^\circ,~230^\circ \right]$. The constraint in $\delta_{\rm CP}$ is  essentially due to the data from T2K and NO$\nu$A as discussed earlier. As far as the neutrino mass ordering is concerned, all three global fits show preference for NMO, ruling out IMO at close to $2.5\sigma$~\cite{Esteban:2020cvm,NuFIT,Capozzi:2021fjo,deSalas:2020pgw}. Therefore, for the sake of simplicity, in this work, we show our results assuming NMO both in data and fit. We observe that the results do not change much for IMO.

Another important feature that emerges from Fig.~\ref{fig:1} is that the current $3\sigma$ allowed range in $\sin^2\theta_{23}$ is $\sim 0.4~ \rm{to}~ 0.6$. This range is still relatively large as compared to the current uncertainties on $\theta_{12}$ and $\theta_{13}$ and it spans on either sides of $\sin^2\theta_{23}  = 0.5$. The value $\sin^2\theta_{23}  = 0.5$ (or equivalently $\sin^22\theta_{23}  = 1$) corresponds to the case of maximal mixing (henceforth, referred to as {\it maximality}) between the $\nu_{2}, \, \nu_{3}$ and $\nu_{\mu}, \, \nu_{\tau}$ eigenstates, which can in principle allows for a complete flavor transition between $\nu_{\mu}$ and $\nu_{\tau}$. However, the recent global fit studies suggest that $\sin^2\theta_{23}$ $\neq$ 0.5 (see Fig.~\ref{fig:1}) or $\sin^22\theta_{23}  \neq 1$. This leads to the so-called octant degeneracy of $\theta_{23}$ $i.e.$ a lack of knowledge regarding whether $\theta_{23}$ is less than $45^\circ$ (denoted as lower octant, LO) or greater than $45^\circ$ (labelled as higher octant, HO)~\cite{Fogli:1996pv, Barger:2001yr, Minakata:2002qi, Minakata:2004pg,Hiraide:2006vh}. But before the question of octant of $\theta_{23}$ arises, it is vital to establish the exclusion of maximality at a high significance, which is the main thrust of this work. In Ref.~\cite{Capozzi:2021fjo}, the authors find a preference at 1.6$\sigma$ for $\theta_{23}$ in the LO with respect to the secondary best-fit in HO. They obtain a best-fit value of $\sin^{2}\theta_{23}$ = 0.455 in the LO assuming NMO and disfavor maximal $\theta_{23}$ mixing at $\sim$1.8$\sigma$. However, there is a slight disagreement between the three global fit studies as far as the measurement of $\theta_{23}$ is concerned (see top right panel in Fig.~\ref{fig:1}). In Ref. \cite{deSalas:2020pgw}, de Salas $et~ al.$ find a best-fit in the HO around $\sin^2\theta_{23} \sim 0.57$ assuming NMO, while Capozzi $et~ al.$~\cite{Capozzi:2021fjo} and Esteban $et~ al.$~\cite{Esteban:2020cvm} obtain the best-fit around $\sin^{2}\theta_{23} \sim 0.45$ in the LO. This difference in the best-fit value of $\sin^2\theta_{23}$ is probably due to the recent Super-K Phase I-IV 364.8 kt$\cdot$yrs of atmospheric data~\cite{sk} that only Capozzi $et~ al.$ and Esteban $et~ al.$ consider in their latest analyses.  

The issue of non-maximal $\theta_{23}$ and the resolution of its octant (if $\sin^{2}2\theta_{23} \neq 1$) have far reaching consequences as far as the models explaining neutrino masses and mixings are concerned~\cite{Mohapatra:2006gs,Albright:2006cw,Albright:2010ap,King:2013eh,King:2014nza}. Some examples of such models are quark-lepton complementarity~\cite{Raidal:2004iw,Minakata:2004xt,Ferrandis:2004vp,Antusch:2005ca}, $A_{4}$ flavor symmetry~\cite{Ma:2002ge,Ma:2001dn,Babu:2002dz,Grimus:2005mu,Ma:2005mw}, and $\mu$-$\tau$ permutation symmetry~\cite{Fukuyama:1997ky,Mohapatra:1998ka,Lam:2001fb,Harrison:2002et,Kitabayashi:2002jd,Grimus:2003kq,Ghosal:2003mq,Koide:2003rx,Mohapatra:2005yu}. The $\mu$-$\tau$ permutation symmetry is of particular interest since the current oscillation data strongly indicates that this symmetry is not exact in Nature. A high-precision measurement of 2-3 mixing angle and the measurement of its octant are inevitable to disclose the pattern of deviations from the above mentioned symmetries, which in turn will help us to explain tiny neutrino masses and one small and two large mixing angles in the lepton sector~\cite{Xing:2014zka,Xing:2015fdg}. It has also been shown that without an accurate measurement of $\theta_{23}$, a precise measurement of $\delta_{\rm CP}$ will not be possible~\cite{Minakata:2013eoa}.

There are several studies in the literature addressing the issues related to the 2-3 mixing angle in the context of various neutrino oscillation experiments. For example, see Refs.~\cite{Antusch:2004yx,Minakata:2004pg,GonzalezGarcia:2004cu,Choudhury:2004sv,Choubey:2005zy,Indumathi:2006gr,Kajita:2006bt,Hagiwara:2006nn,Samanta:2010xm,Agarwalla:2013hma,Minakata:2013eoa,Ge:2013zua,Ge:2013ffa,Chatterjee:2013qus,Choubey:2013xqa,Bass:2013vcg,Coloma:2014kca,Bora:2014zwa,Das:2014fja,Nath:2015kjg,Ghosh:2015ena,Agarwalla:2016xlg,Ballett:2016daj}. In this work, we analyze in detail the sensitivities of the next generation, high-precision long-baseline neutrino oscillation experiment DUNE (Deep Underground Neutrino Experiment)~\cite{DUNE:2015lol,DUNE:2020lwj,DUNE:2020ypp,DUNE:2020jqi,DUNE:2021cuw,DUNE:2021mtg} to establish the deviation from maximal $\theta_{23}$ and to resolve its octant at high confidence level in light of the current neutrino oscillation data. While estimating DUNE's capability for the discovery of non-maximal $\theta_{23}$, we shed light on some relevant issues such as: (i) the individual contributions from appearance and disappearance channels, (ii) impact of systematic uncertainties and marginalization over oscillation parameters, and (iii) importance of spectral analysis and data from both neutrino and antineutrino runs. We also study how much improvement DUNE can offer in the precision measurements of $\sin^{2}\theta_{23}$ and $\Delta m^2_{31}$ as compared to their current precision. While estimating the achievable precision on these parameters in DUNE, we also quantify the contribution from individual appearance and disappearance channels and demonstrate the importance of having both neutrino and antineutrino data.

The layout of this paper is as follows. In Sec.~\ref{probability}, we discuss the potential of DUNE's baseline and energy in establishing deviation from maximal $\theta_{23}$ at the level of probabilities. Next, in Sec.~\ref{events}, we describe the key features of DUNE which are relevant for our numerical simulation and disuss the impact of possible correlations and degeneracies among $\sin^{2}\theta_{23}$, $\Delta m^{2}_{31}$, and $\delta_{\mathrm{CP}}$ at the level of total event rates, bi-events, and event spectra. In Sec.~\ref{results}, we quantify the performance of DUNE to establish non-maximal $\theta_{23}$, to settle the correct octant of $\theta_{23}$, and to precisely measure the values of atmospheric oscillation parameters $-$ $\sin^{2}\theta_{23}$ and $\Delta m^{2}_{31}$. We also address several issues which are relevant to achieve the above mentioned goals. In Sec.~\ref{conclusion}, we summarize our findings and make concluding remarks. In Appendix~\ref{appendix_a}, we provide the best-fit values of the oscillation parameters along with their currently allowed 1$\sigma$ and 3$\sigma$ ranges obtained by the three global fit studies~\cite{Esteban:2020cvm,NuFIT,Capozzi:2021fjo,deSalas:2020pgw}.

\section{Discussion at the level of probabilities}
\label{probability}

\begin{table}[htb!]
		\label{tableprecision}

		\centering
		\resizebox{\columnwidth}{!}{%
			\begin{tabular}{|c|c|c|c|c|c|}
				\hline \hline
				\multirow{2}{*}{\textbf{Parameter}} & \multirow{2}{*}{\textbf{Best-fit}} & \multirow{2}{*}{\textbf{1$\sigma$ range}} & \multirow{2}{*}{\textbf{2$\sigma$ range}} & \multirow{2}{*}{\textbf{3$\sigma$ range}} & \textbf{Relative 1$\sigma$}\\
				& & & & &\textbf{Precision (\%)}\\
				\hline \hline
				$\Delta m^2_{21}/10^{-5}$ $\mathrm{eV^{2}}$ & 7.36 & 7.21 - 7.52 & 7.06 - 7.71 & 6.93 - 7.93 & 2.3\\
				\hline
				$\sin^{2}\theta_{12}/10^{-1}$ & 3.03 & 2.90 - 3.16 & 2.77 - 3.30 & 2.63 - 3.45 & 4.5\\
				\hline
				$\sin^{2}\theta_{13}/10^{-2}$ & 2.23 & 2.17 - 2.30 & 2.11 - 2.37 & 2.04 - 2.44 &  3.0\\
				\hline
				$\sin^2\theta_{23}/10^{-1}$ & 4.55 & 4.40 - 4.73 & 4.27 - 5.81 & 4.16 - 5.99 & 6.7 \\
				\hline
				$\Delta m^2_{31}/10^{-3}$ $\mathrm{eV^2}$ & 2.522 & 2.490 - 2.545 & 2.462 - 2.575 & 2.436 - 2.605 & 1.1\\
				\hline
				$\delta_{\text{CP}}$/$^\circ$ & 223  & 200 - 256 & 169 - 313 & 139 - 355 & 16\\
				\hline \hline
			\end{tabular}
			}
			\caption{\footnotesize{The benchmark values of the oscilation parameters and their corresponding ranges that we consider in our study assuming NMO. In second column, we mention the best-fit values as given in Ref.~\cite{Capozzi:2021fjo}. The third, fourth, and fifth columns depict the current 1$\sigma$, 2$\sigma$, and 3$\sigma$ allowed ranges, respectively under NMO scheme. The sixth column depicts the present relative 1$\sigma$ precision on various oscillation parameters as given in Ref.~\cite{Capozzi:2021fjo}.}}
			\label{table:one}
	\end{table}

In the three-neutrino framework, the flavor eigenstates $\vert \nu_{\alpha}\rangle~\left(\alpha = e, \mu, \tau \right)$ and the mass eigenstates $\vert \nu_{i} \rangle ~\left(i=1,2,3\right)$ are connected by the $3\times3$ unitary Pontecorvo-Maki-Nakagawa-Sakata (PMNS) matrix $U$:
\begin{equation}
\vert \nu_{\alpha}\rangle  = \sum_{i} U^\ast_{\alpha i}\vert\nu_{i} \rangle ~~~ {\rm and} ~~~
\vert \bar{\nu}_{\alpha} \rangle = \sum_{i} U_{\alpha i}\vert \bar{\nu}_{i} \rangle\, .
\end{equation}
Following the standard Particle Data Group convention~\cite{Zyla:2020zbs}, the vacuum PMNS matrix $U$ is parametrized in terms of the three mixing angles ($\theta_{23}$, $\theta_{13}$, $\theta_{12}$) and one Dirac-type CP phase ($\delta_{\mathrm{CP}}$). The probability that a neutrino, with flavor $\alpha$ and energy $E$, after traveling a distance $L$, can be detected as a neutrino with flavor $\beta$ is given by
\begin{equation}
P_{\alpha\beta} = \delta_{\alpha\beta} - 4 \sum_{j>i} \mathcal{R} \left( U^\ast_{\alpha j}U_{\beta j} U_{\alpha i} U^\ast_{\beta i}\right)\sin^2\frac{\Delta m^2_{ji} L}{4E} + 2 \sum_{j>i} \mathcal{I} \left( U^\ast_{\alpha j}U_{\beta j} U_{\alpha i} U^\ast_{\beta i}\right)\sin\frac{\Delta m^2_{ji} L}{2E}\, ,
\end{equation}
where, $\Delta m^2_{ji} = m^2_{j} - m^2_{i}$.
Approximate analytical expressions for oscillation probabilities including matter effect have been derived in Ref.~\cite{Akhmedov:2004ny}, retaining terms only up to second order in the small parameters $\sin^2\theta_{13}$ and $\alpha \left(\equiv \Delta m^2_{21}/\Delta m^2_{31}\right)$. The analytical expression for muon neutrino survival probability ($P_{\mu \mu}$) under the constant matter density approximation is given in Eq. 33 of Ref.~\cite{Akhmedov:2004ny}. Considering the current best-fit values of oscillation parameters (see second column in Table~\ref{table:one}), we have $\sin^2\theta_{13}\approx 0.02$, $\alpha \approx 0.03$, $\alpha\sin\theta_{13}\approx 0.004$, and $\alpha^2\approx 0.0008$. Therefore, ignoring the sub-leading terms which are of the order $\alpha^2$ and approximating $\cos \theta_{13}$ equal to 1, Eq. 33 of Ref. \cite{Akhmedov:2004ny} simplifies to
\begin{eqnarray}
\label{pmumushort} 
  P_{\mu\mu} &\approx & 1 - M \sin^22\theta_{23} - N \sin^2\theta_{23} - R \sin 2\theta_{23} + T \sin 4\theta_{23} \, ,
\end{eqnarray}
where,
\beqa \label{eqM}
\nonumber M &=& \sin^2\Delta \\ 
\nonumber   &-& \alpha \cos^2\theta_{12}\Delta\sin2\Delta \\
            &+& \frac{2}{\hat A -1}\sin^2\theta_{13}\bigg(\sin\Delta \cos\hat A \Delta\frac{\sin(\hat A - 1)\Delta}{\hat A - 1} - \frac{\hat A}{2} \Delta \sin 2 \Delta\bigg)\, ,
\eeqa
%
\beqa \label{eqN}
N = 4\sin^2\theta_{13}\frac{\sin^2(\hat A - 1) \Delta}{(\hat A-1)^2}\, ,
\eeqa
\beqa \label{eqR}
R = 2\alpha\sin\theta_{13}\sin2\theta_{12}\cos\delta_{\rm CP} \cos\Delta\frac{\sin\hat A \Delta}{\hat A}\frac{\sin(\hat A - 1)\Delta}{\hat A - 1}\, ,
\eeqa

and
\beqa \label{eqT}
T = \frac{1}{\hat{A}-1}\alpha\sin\theta_{13}\sin2\theta_{12}\cos\delta_{\rm CP}\sin\Delta\bigg(\hat A \sin\Delta - \frac{\sin \hat A \Delta}{\hat A}\cos(\hat A - 1)\Delta \bigg)\, .
\eeqa
In the above equations, $\Delta \equiv \Delta m^{2}_{31}L/4E$ and $\hat{A} \equiv A/\Delta m^{2}_{31}$. The Wolfenstein matter term, $A = 2\sqrt{2}G_{F}N_{e}E = 7.6\times10^{-5} \times \rho~(\mathrm{g/cm^{3}}) \times E$ (GeV), where $G_{F}$ is the Fermi coupling constant, $N_{e}$ is the ambient electron density, $E$ is the energy of neutrino, and $\rho$ is the constant matter density through which neutrino propagates. In Eq.~\ref{pmumushort}, all the terms containing $\theta_{23}$ (see Eqs.~\ref{eqM} to~\ref{eqT}) provide crucial information to establish non-maximal $\theta_{23}$ and contribute towards the precision measurement of $\theta_{23}$, which are the focus of this work. The first term in Eq.~\ref{eqM}, which is proportional to $\sin^{2} {\Delta}$, is the leading term in muon neutrino survival channel and contributes the most to address the above mentioned physics issues. The term in Eq.~\ref{eqN}, which is the leading term in $\nu_{\mu} \rightarrow \nu_{e}$ appearance channel, is suppressed by the small quantity $\sin^{2}\theta_{13}$, and the terms in Eqs.~\ref{eqR} and~\ref{eqT} are proportional to the quantitity $\alpha \sin \theta_{13}$ which is around $\sim 0.004$. Therefore, these terms provide sub-leading contributions towards establishing deviation from maximal mixing and to precisely measure the value of $\theta_{23}$. Note that the terms in Eqs.~\ref{eqR} and~\ref{eqT} are proportional to $\cos \delta_{\mathrm{CP}}$. These terms may help to measure the value of $\delta_{\mathrm{CP}}$, but they are blind to CP asymmetry. On the other hand, the terms in Eq.~\ref{eqN} and~\ref{eqT}, which are proportional to $\sin^{2}\theta_{23}$ and $\sin 4\theta_{23}$, respectively provide information on the octant of $\theta_{23}$. 

The main sensitivity to settle the octant of $\theta_{23}$ stems from $\nu_{\mu} \rightarrow \nu_{e}$ appearance channel ($P_{\mu e}$), which when expressed up to first order in $\alpha\sin\theta_{13}$ is given by (ignoring the term $\propto$ $\alpha^{2}$ and $\cos\theta_{13}\approx 1$)
\beqa \label{eqpmue}
P_{\mu e} \approx N \sin^2\theta_{23} + O \sin2\theta_{23}\cos\left(\Delta + \delta_{\mathrm {CP}}\right)\, ,
\eeqa
where, 
\beqa \label{eqO}
O = 2\alpha\sin\theta_{13}\sin2\theta_{12}\frac{\sin\hat A \Delta }{\hat A}\frac{\sin(\hat A -1)\Delta }{\hat A-1}\, .
\eeqa
%
Note that the first term in Eq.~\ref{eqpmue} is sensitive to octant of $\theta_{23}$, while the second term is sensitive to CP phase $\delta_{\mathrm {CP}}$. This leads to an octant\,-\,$\delta_{\mathrm {CP}}$ degeneracy in the measurements made via appearance channel. However, this degeneracy can be resolved with the help of balanced neutrino and antineutrino data in appearance mode as discussed for the the first time in Ref.~\cite{Agarwalla:2013ju}. Since, both the terms in $P_{\mu e}$ contain information on $\theta_{23}$ (see Eq.~\ref{eqpmue}), they contribute towards establishing deviation from maximal $\theta_{23}$ (see discussion in Sec.~\ref{appdisapp} and Fig.~\ref{fig:7}) and to precisely measure the value of $\sin^{2}\theta_{23}$ (see discussion in Sec.~\ref{precisionapp} and Fig.~\ref{fig:13}). 

	\begin{figure}[htb!]
	\centering
	\includegraphics[width=0.49\linewidth]{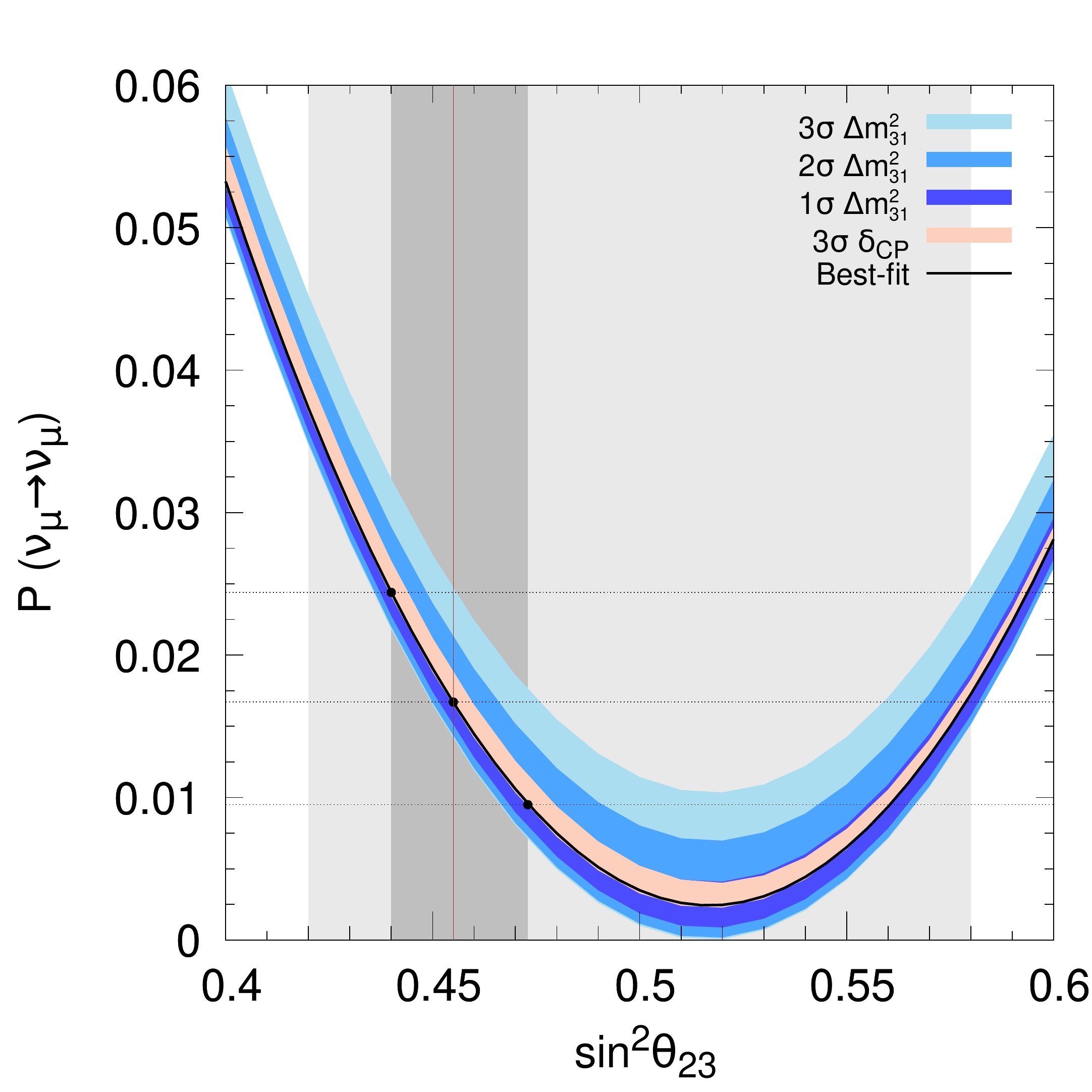}
	\includegraphics[width=0.49\linewidth]{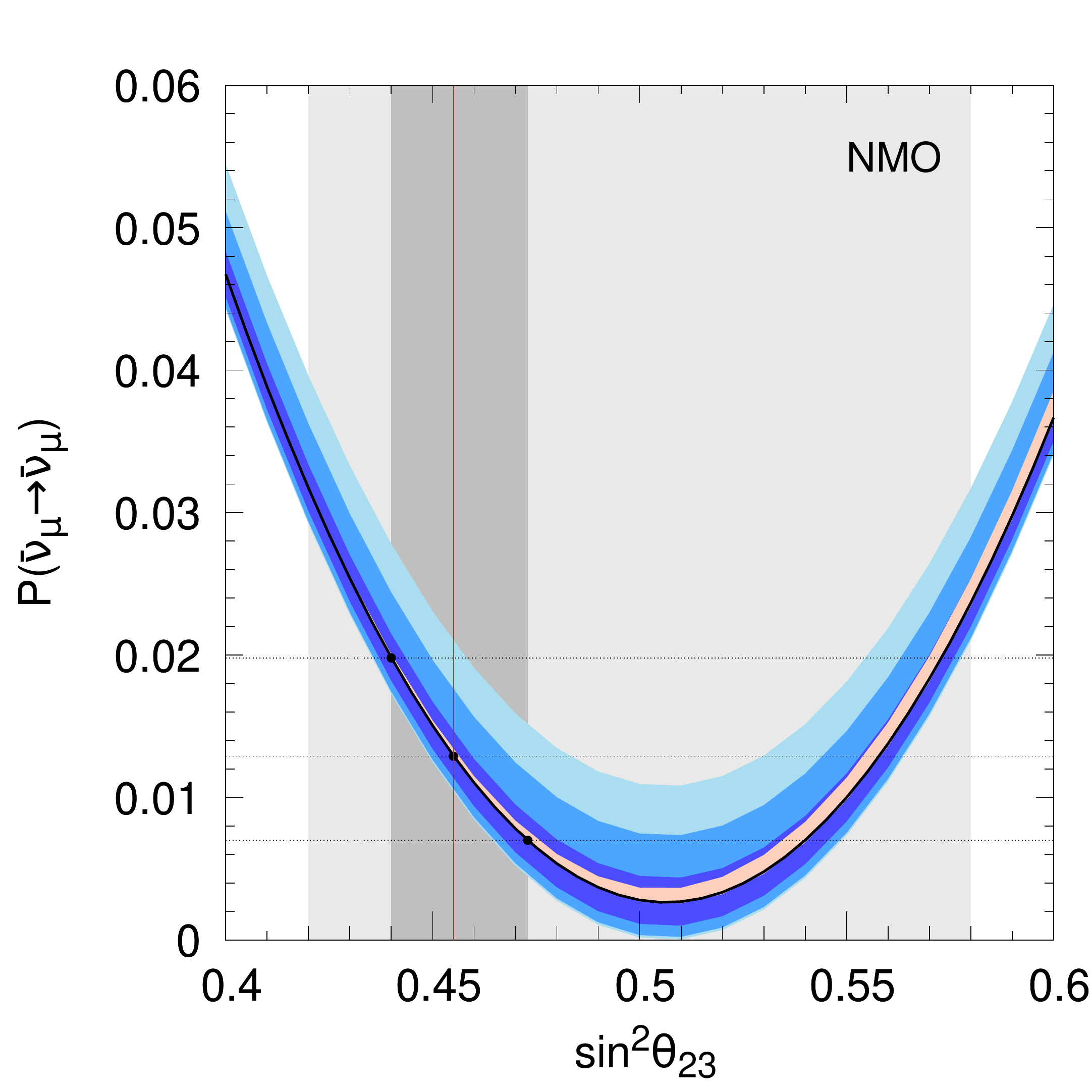}
	\includegraphics[width=0.49\linewidth]{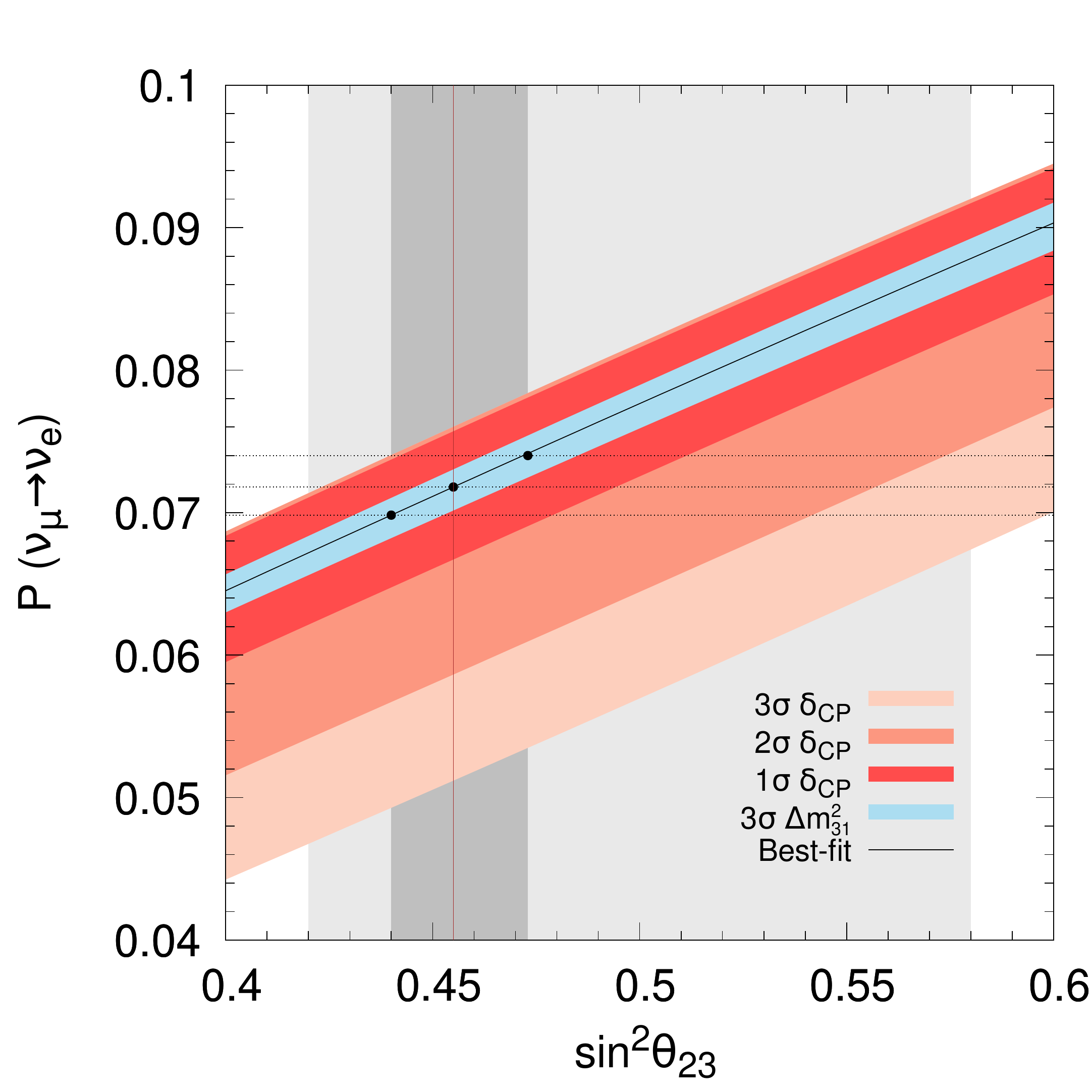}
	\includegraphics[width=0.49\linewidth]{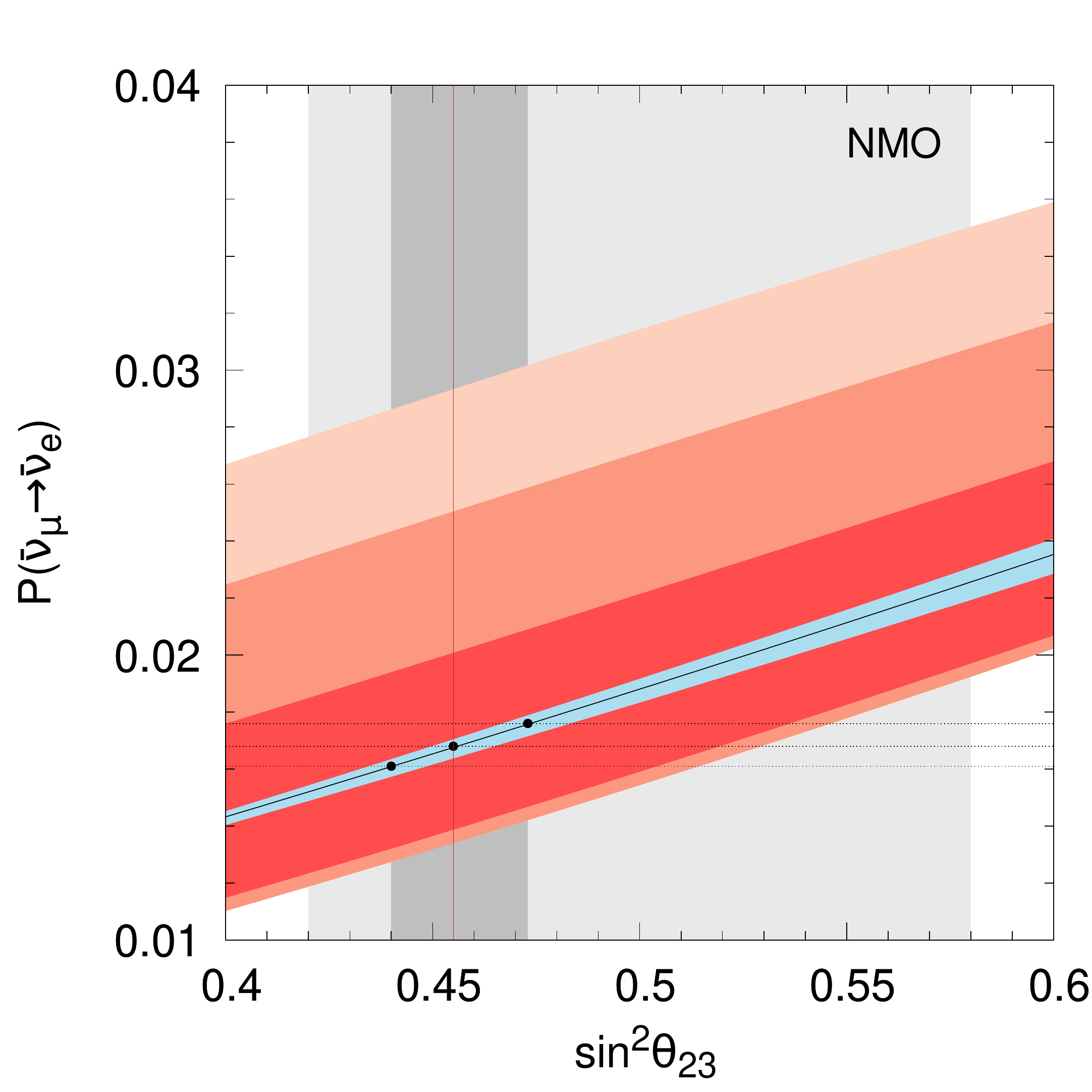}
	\caption{\footnotesize{Probability as a function of $\sin^2\theta_{23}$ for $E = 2.5$ GeV, $ L=1300$ km, and $\rho=2.87~ \rm{g/cm}^{3}$ assuming NMO. The top (bottom) panels are for disappearance (appearance) channel. The left (right) panels are for neutrino (antineutrino). The black solid curves show the probability considering the best-fit values of oscillation parameters as given in Table~\ref{table:one}. The three shaded blue (red) regions depict the variations in probability due to the present $1\sigma,~2\sigma,~\rm{and}~3\sigma$ allowed ranges in $\Delta m^2_{31}$ ($\delta_{\rm CP}$). The dark (light)-shaded grey area shows the currently allowed $1\sigma~ (2\sigma)$ region in $\sin^{2}\theta_{23}$ with the best-fit value of $\sin^{2}\theta_{23} = 0.455$ as shown by the vertical brown line. See Table~\ref{table:one} for details. Note that y-ranges are different in the bottom two panels.}}
	\label{fig:2}
\end{figure} 

To understand the role of oscillation channels $P_{\mu \mu}$ and $P_{\mu e}$ in distinguishing a non-maximal $\sin^{2}\theta_{23}$ from maximal mixing $i.e.$ $\sin^{2}\theta_{23} = 0.5$, we draw Fig.~\ref{fig:2}. In this figure, we show oscillation probabilities as a function of $\sin^2\theta_{23}$. To generate this figure, we use our benchmark values of the oscillation parameters and their corresponding ranges from Table~\ref{table:one}. The top panels are for $P_{\mu\mu}$ while the bottom panels are for $P_{\mu e}$. In the left (right) panels, we show the probabilities for neutrino (antineutrino).
The solid black curves in each of these figures show the probability corresponding to the best-fit values of $\Delta m^2_{31}$ and $\delta_{\rm CP}$ for each $\sin^2\theta_{23}$ whereas the different shades of blue and red bands correspond to the variation in probability due to $1,~2,$ and $3\sigma$ variations in $\Delta m^2_{31}$ and $\delta_{\rm CP}$, respectively. To generate these probabilities, we choose $ E = 2.5~\rm GeV$ and $ L = 1300~\rm km$ which correspond to the DUNE's baseline and peak-energy fluxes. $E \sim 2.5~\rm GeV$ also corresponds to the first oscillation maximum (minimum) in $P_{\mu e}$ ($P_{\mu\mu}$). From Fig.~\ref{fig:2}, we make the following observations.
\begin{itemize}
\item 
$P_{\mu\mu}$ varies a lot as $\Delta m^2_{31}$ is varied while it changes only marginally with respect to $\delta_{\rm CP}$\footnote{$P_{\mu\mu}$ depends only on $\cos\delta_{\rm CP}$ at the sub-leading interference term proportional to $\alpha\sin^2\theta_{13}$. See Eq. 33 of Ref. \cite{Akhmedov:2004ny}.}.
The opposite behavior is seen in the case of $P_{\mu e}$ \cite{Coloma:2014kca,Minakata:2013eoa}. Therefore, we see significant degeneracies among the oscillation parameters $\Delta m^2_{31}$ and $\sin^2\theta_{23}$ in $P_{\mu\mu}$ channel. For $P_{\mu e}$ channel, the degeneracies are observed among $\delta_{\rm CP}$ and $\sin^2\theta_{23}$. 
\item 
For values of $\sin^2\theta_{23}$ in the HO that are very close to $\sin^2\theta_{23}=0.5$ $i.e.$ $\sin^2\theta_{23} \in [0.5, 0.53]$, $P_{\mu\mu}$ shows a flat behavior, $i.e.$ the slope of the $P_{\mu\mu}$ is nearly 0\footnote{Note that the minimum of $P_{\mu\mu}$ is in HO, slightly away from $\sin^2\theta_{23} = 0.5$ due to finite $\theta_{13}$ correction. See Ref. \cite{Raut:2012dm} for details.}. 
This is observed in both neutrino and antineutrino probabilities. 
\item 
For values of $\sin^2\theta_{23}$ in LO that are very close to $\sin^2\theta_{23}=0.5$, $P_{\mu\mu}$ is steep. 
\item 
For values of $\sin^2\theta_{23}$ adequately far from $\sin^2\theta_{23}=0.5$, $P_{\mu\mu}$ is very steep in both LO as well as HO. 
\item 
$P_{\mu e}$ shows a monotonic increase with respect to $\sin^2\theta_{23}$. This is true for both LO and HO, and in case of both neutrinos and antineutrinos. 
\end{itemize}

Thus, based on the observations made above, we expect the results to have the following features: 
\begin{itemize}
\item Since, we expect the combination of $P_{\mu\mu}$ and $P_{\mu e}$ channels to resolve the degeneracies that are present in each of them individually, we do not expect the sensitivity to exclude non-maximal $\theta_{23}$ be too much affected by the choice of $\Delta m^2_{31}$ and $\delta_{\rm CP}$ within the given $3\sigma$ range. 
\item For values of $\sin^2\theta_{23}$ in the HO and very close to 0.5, we expect the sensitivity to deviation from maximality to come mainly from the appearance channel. However, for $\sin^2\theta_{23}$ values farther away from 0.5, the disappearance channel will contribute significantly. In the LO, we expect the main sensitivity to come from the disappearance channel even for $\sin^2\theta_{23}$ values very close to 0.5. 
\end{itemize} 
While the above arguments have been made using the probabilities calculated with a particular choice of $E = 2.5~\rm GeV$, we will see in the results section that these features hold in general.   

\section{Discussion at the level of events}
\label{events} 

We start this section by mentioning the salient features of DUNE which are crucial for our numerical simulations. Then, we show the total appearance and disappearance event rates in neutrino and antineutrino modes as a function of $\sin^{2}\theta_{23}$, $\Delta m^{2}_{31}$, and $\delta_{\mathrm{CP}}$ to establish some physics issues which are necessary to understand our main results. We also exhibit the bi-events plot in the plane of neutrino - antineutrino disappearance events and display their event spectra.

\subsection{Salient features of DUNE}
\label{experiment}

In order to calculate the expected event rates in DUNE and to estimate its sensitivity towards various physics issues, we use the publicaly available software GLoBES (General Long Baseline Experiment Simulator)~\cite{Huber:2004ka, Huber:2007ji}. We consider the simulation details as described in Ref.~\cite{DUNE:2021cuw}. DUNE will look for $\nu_{\mu}\rightarrow \nu_{\mu}$ (disappearance) and $\nu_{\mu}\rightarrow \nu_{e}$ (appearance) oscillations in both neutrino and antineutrino modes. Neutrinos are produced at the LBNF's Main Injector in Fermilab, Illinois, Chicago where protons of energy 120 GeV and power 1.2 MW are bombarded on a graphite target. This leads to the production of charged mesons which then decay in flight producing the neutrinos. Using the desired polarity in the horn-focusing system, neutrino or antineutrino mode can be selected. The neutrino flux at DUNE is wide-band with energies ranging from few hundreds of MeV to few tens of GeV, but the flux peaks at around $~$2.5 GeV with majority of the flux lying in $1$ GeV to $5$ GeV region. 
These neutrinos first see a near detector (ND) placed 574 m downstream from the source and a far detector (FD) located roughly 5000 ft below the Earth's surface at the Sanford Underground Research Facility (SURF) in Lead, South Dakota, USA. The main purpose of ND is to precisely measure the unoscillated neutrino flux so as to reduce the systematic uncertainties related to fluxes. The distance between the source of production of neutrinos in Fermilab and the FD is 1284.9 km and the neutrinos traverse through Earth matter of roughly constant density of around $2.848~\rm g/cm ^{3}$. The FD is a 40 kt liquid argon timeprojection chamber (LArTPC) and is placed underground in order to minimize cosmogenic and atmospheric backgrounds. 
We consider a total run-time of 7 years equally divided  in neutrino and antineutrino mode with a total of 1.1$ \times$ 10$^{21}$ protons on target (P.O.T.). This corresponds  to a net 168 kt$\cdot$MW$\cdot$years of exposure each in $\nu$ and $\bar{\nu}$ mode. 
The energy resolution of the FD in $\left(0.5 - 5\right)$ GeV range is around $\left(15 - 20\right)\%$. 
Our assumptions on systematic uncertainties are based on the material provided in Ref.~\cite{DUNE:2021cuw}. The errors are bin-to-bin correlated and are same for both neutrinos and antineutrinos. In the appearance channel $i.e.$ for the electron events, the normalization error is $2\%$ while for the disappearance channel $i.e.$ for the muon events, the normalization error is $5\%$. For the background events, the error varies from $5\%$ to $20\%$. 

\subsection{Appearance and disappearance event rates as a function of $\sin^{2}\theta_{23},\ \Delta m^{2}_{31},$ and $\delta_{\rm CP}$}

In Table~\ref{table:two}, we show the total neutrino and antineutrino event rates for DUNE as a function of the oscillation parameters for both disappearance and appearance channels. The three columns correspond to LO ($\sin^2\theta_{23} = 0.455$) on the left, MM ($\sin^2\theta_{23} = 0.5$) in the center and HO ($\sin^2\theta_{23} = 0.599$) on the right. The central number in each cell, shown in boldface corresponds to the total number of events for the best-fit values of $\delta_{\rm CP} = 223^\circ$ and $\Delta m^2_{31} = 2.522\times 10^{-3}~\rm eV^2$, while considering $\sin^2\theta_{23}$ in LO, MM, and HO in second, third, and fourth columns, respectively. Rest of the oscillation parameters are kept at their respective best-fit values (see Table~\ref{table:one} for details). We determine other two numbers in the left and right (top and bottom) of central value by varying $\delta_{\rm CP}$ ($\Delta m^2_{31}$) from its central best-fit value to $3\sigma$ lower and upper bounds, respectively (as explained in the schematic diagram above Table~\ref{table:two}).

\begin{table}[htb!]
	\begin{center}
		\begin{small}
			\begin{tabular}{||c|c|c|c|c||}
			\hline \hline
			    \multicolumn{5}{||c||}{} \\
			    \multicolumn{5}{||c||}{$\mathcal{N}\left(223, 2.436\right)$} \\
			    \multicolumn{5}{||c||}{$\uparrow$} \\
			    \multicolumn{5}{||c||}{$\mathcal{N}\left(139, 2.522\right) \leftarrow \mathcal{N}\left(223, 2.522\right) \rightarrow \mathcal{N}\left(355, 2.522\right)$} \\
			    \multicolumn{5}{||c||}{$\downarrow$} \\
			    \multicolumn{5}{||c||}{$\mathcal{N}\left(223, 2.605\right)$} \\
			    \multicolumn{5}{||c||}{} \\
			    \multicolumn{5}{||c||}{$\mathcal{N}\left(x, y\right)$ where  $\mathcal{N}$: total number of events, $x$: $\delta_{\rm CP}$ in degrees, $y$: $\Delta m^2_{31}$ in $10^{-3}~\rm eV^2$ }\\
			    \multicolumn{5}{||c||}{} \\
			    \hline \hline
			    \multicolumn{5}{c}{} \\
			
				\hline\hline
				\multicolumn{2}{||c|}{Channel} & \text{LO}  & \text{MM} & \text{HO} \\
				\hline
				 &  & 1104  & 1193 & 1383 \\
				
				& & \boldsymbol{$\uparrow$} & \boldsymbol{$\uparrow$} & \boldsymbol{$\uparrow$}\\
				
			\multirow{5}{*}{\rotatebox[origin=c]{90}{Appearance}}& \text{$\nu$} &
				{820 $\leftarrow$$\boldsymbol{1121}$$\rightarrow$ 969}  & 				{908 $\leftarrow$$\boldsymbol{1211}$$\rightarrow$ 1058}  & 
								{1107 $\leftarrow$$\boldsymbol{1403}$$\rightarrow$ 1254}   \\
				
				& & \boldsymbol{$\downarrow$}& \boldsymbol{$\downarrow$}&\boldsymbol{$\downarrow$}\\
				
				& & 1135  & 1226 &
				1421 \\
				\cline{2-5}
				&  & 206  & 227 & 277 \\
				
				& & \boldsymbol{$\uparrow$}& \boldsymbol{$\uparrow$}&\boldsymbol{$\uparrow$}\\

				& \text{$\bar\nu$} & {267 $\leftarrow$$\boldsymbol{208}$$\rightarrow$ 258}  & {289 $\leftarrow$$\boldsymbol{230}$$\rightarrow$ 280}  & {338 $\leftarrow$$\boldsymbol{279}$$\rightarrow$ 329}   \\
				
				& & \boldsymbol{$\downarrow$}&\boldsymbol{$\downarrow$}&\boldsymbol{$\downarrow$}\\

				& & 210  & 232 & 281 \\
				\hline \hline
				 &  & 11018  & 10797 & 11249 \\
				
				& & \boldsymbol{$\uparrow$}& \boldsymbol{$\uparrow$}& \boldsymbol{$\uparrow$}\\
				
				\multirow{5}{*}{\rotatebox[origin=c]{90}{Disappearance}} & \text{$\nu$} & {10870 $\leftarrow$$\boldsymbol{10870}$$\rightarrow$ 10896}  & {10646 $\leftarrow$$\boldsymbol{10646}$$\rightarrow$ 10663}  & {11100 $\leftarrow$$\boldsymbol{11100}$$\rightarrow$ 11095} \\
				
				& & \boldsymbol{$\downarrow$}& \boldsymbol{$\downarrow$}&\boldsymbol{$\downarrow$}\\

				& & 10758  & 10532 & 10986 \\
				\cline{2-5}
				&  & 6397  & 6310 & 6565 \\
				
				& & \boldsymbol{$\uparrow$}&\boldsymbol{$\uparrow$}&\boldsymbol{$\uparrow$}\\

				&\text{$\bar\nu$} & {6306 $\leftarrow$$\boldsymbol{6306}$$\rightarrow$ 6280}  & {6219 $\leftarrow$$\boldsymbol{6219}$$\rightarrow$ 6193}  & {6477 $\leftarrow$$\boldsymbol{6477}$$\rightarrow$ 6452} \\
				
				& & \boldsymbol{$\downarrow$}&\boldsymbol{$\downarrow$}&\boldsymbol{$\downarrow$}\\

				& & 6234  & 6146 & 6406 \\
				\hline\hline
				
			\end{tabular}
		\end{small}
	\end{center}
	\caption{\footnotesize{Total appearance and diappearance event rates in $\nu$ and $\bar{\nu}$ mode. We assume 3.5 years of $\nu$ run and 3.5 years of $\bar{\nu}$ run and estimate the event rates for three different choices of $\sin^{2}\theta_{23}$: 0.455 (LO), 0.5 (MM), and 0.599 (HO). The central number in each cell corresponds to the current best-fit values of $\delta_{\rm CP}$ = 223$^{\circ}$ and $\Delta m^2_{31}$ = 2.522 $\times$ 10$^{-3}$ eV$^{2}$ assuming NMO. The other four numbers in each cell show the number of events corresponding to the present $3\sigma$ lower and upper bounds in $\Delta m^2_{31}$ (up and down arrows) and $\delta_{\rm CP}$ (left and right arrows). For clarity, see the schematic diagram given above this table.}}
	\label{table:two}
\end{table}

\begin{figure}[htb!]
	\centering
	\includegraphics[width=0.49\linewidth]{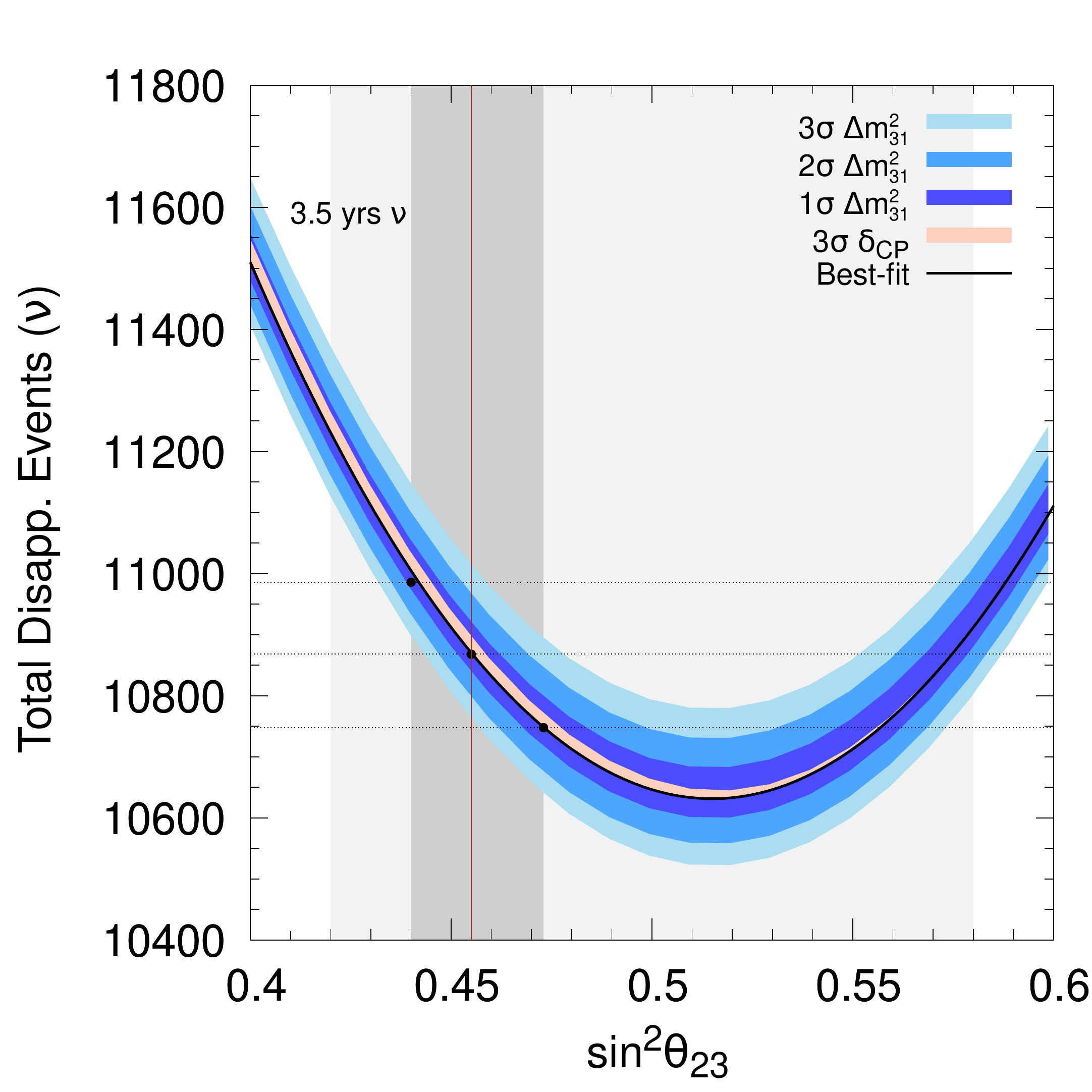}
	\includegraphics[width=0.49\linewidth]{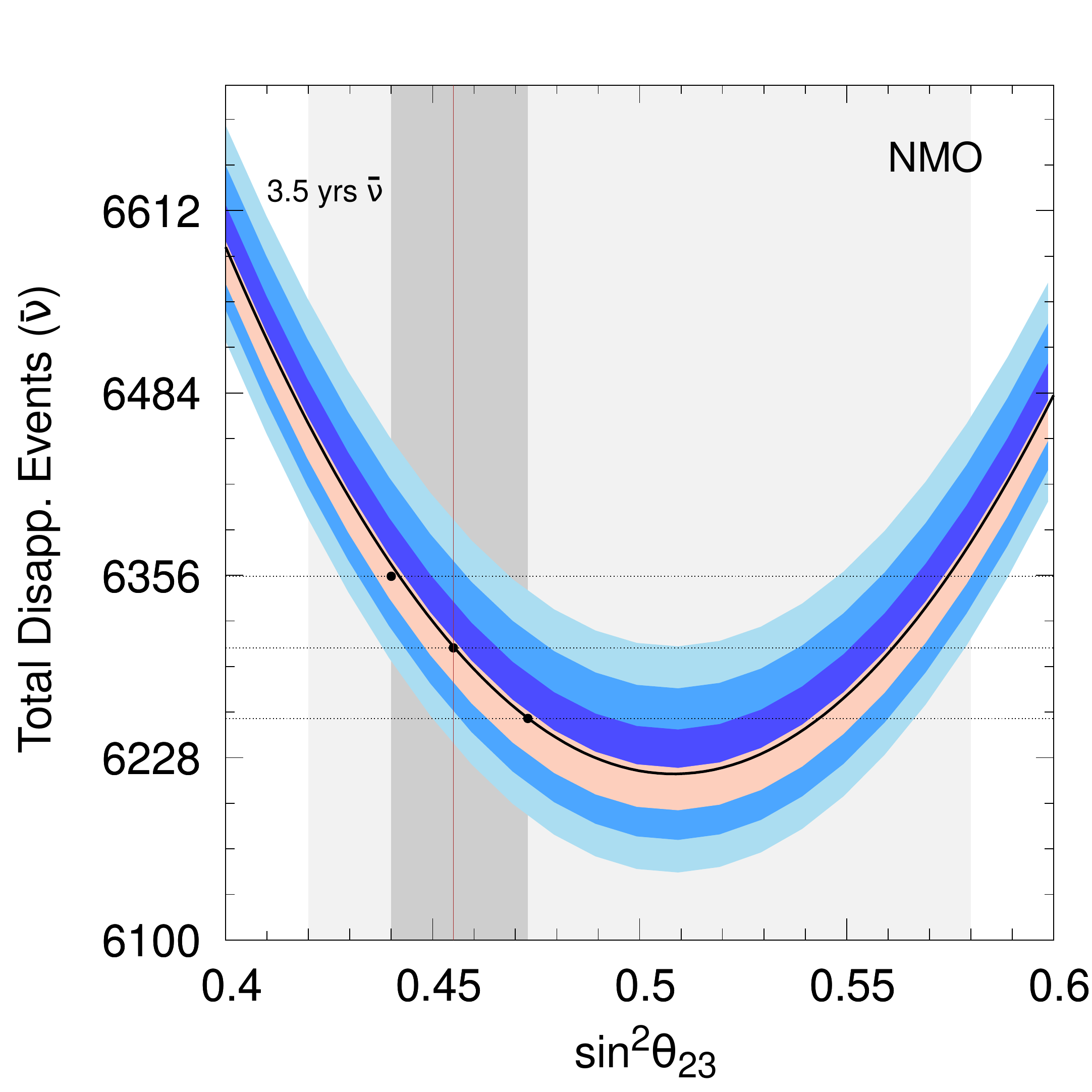}
	\includegraphics[width=0.49\linewidth]{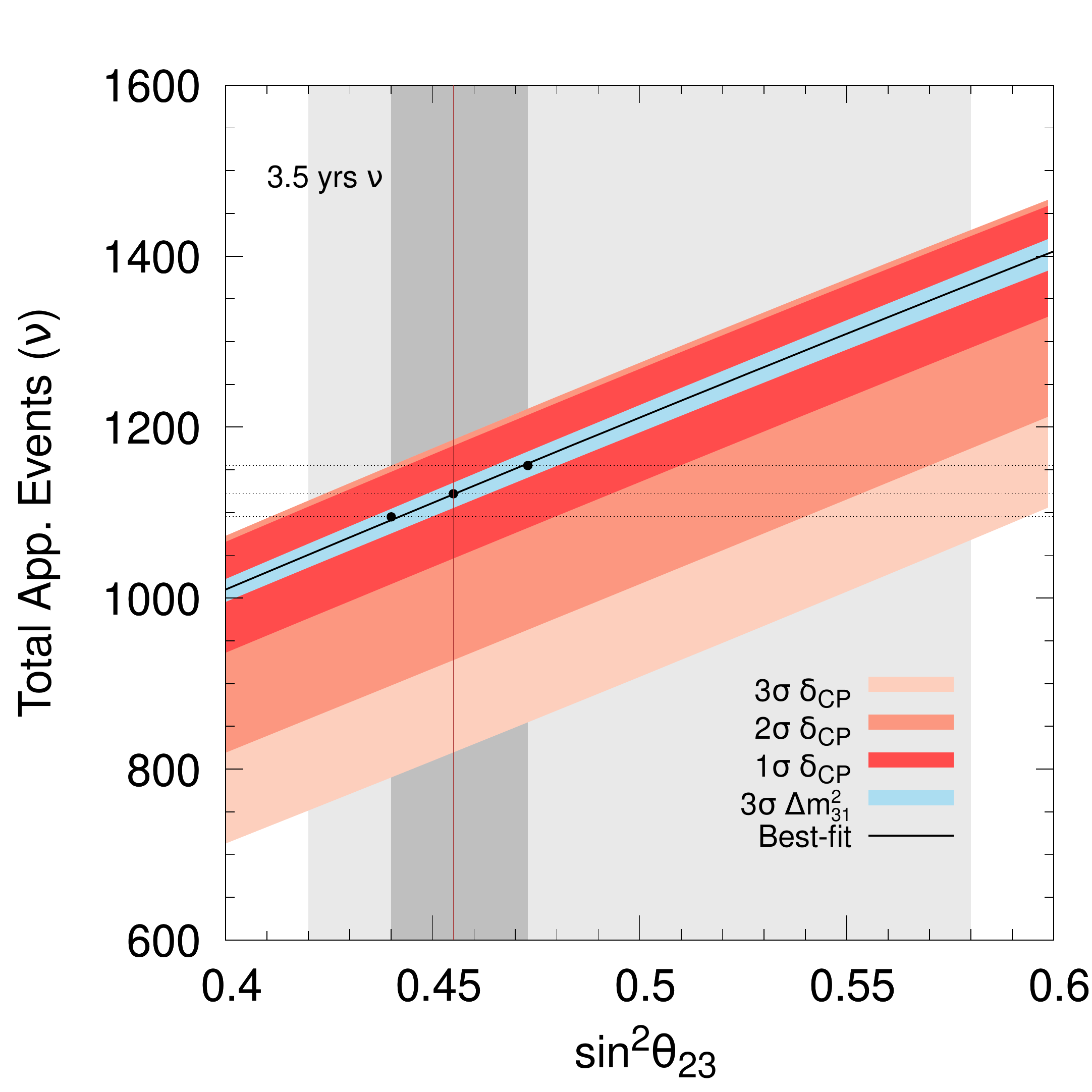}
	\includegraphics[width=0.49\linewidth]{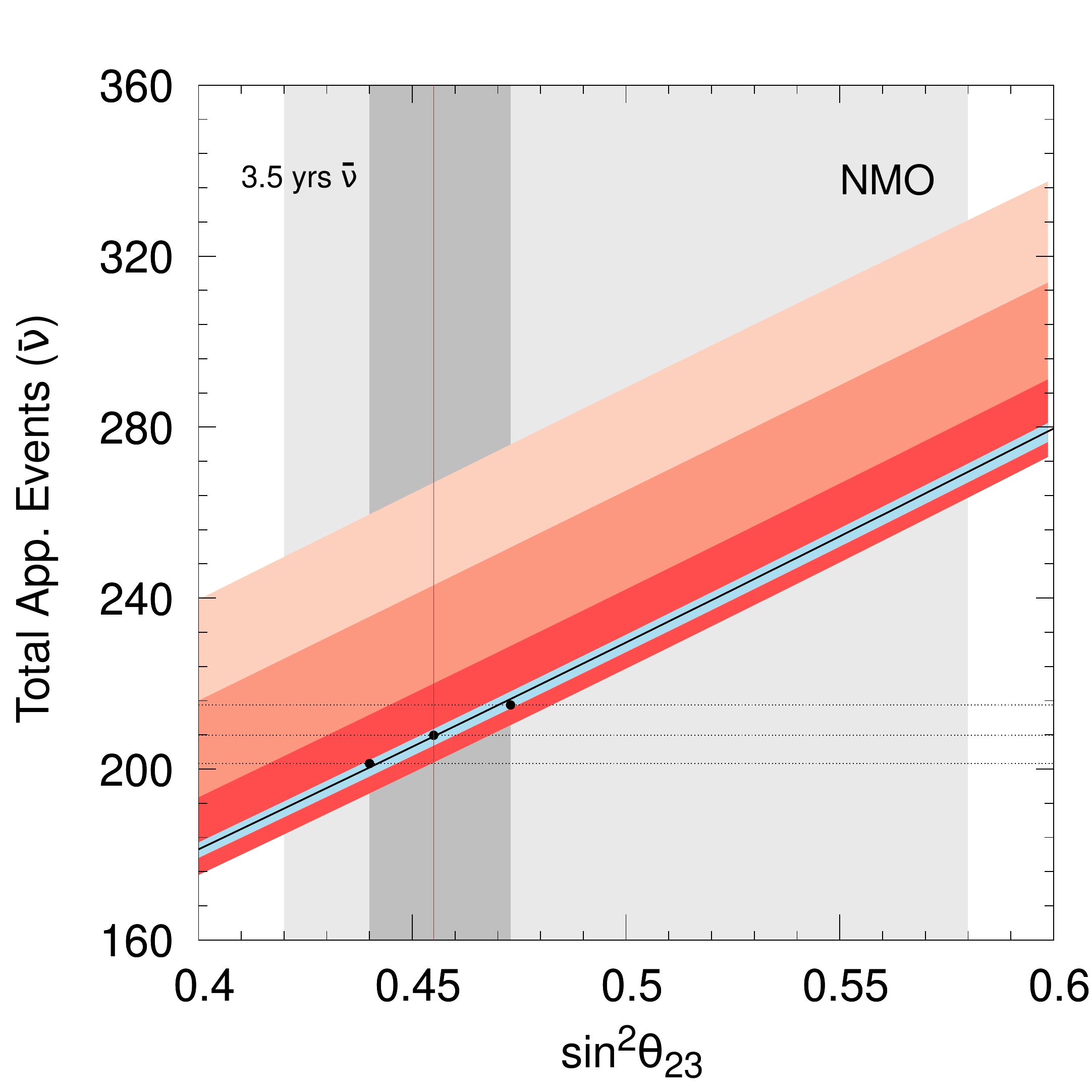}
	\caption{\footnotesize{Total event rates as a function of $\sin^2\theta_{23}$ for DUNE assuming NMO. The top (bottom) panels are for disappearance (appearance) channel. The left (right) panels are for neutrino (antineutrino) assuming 3.5 years of run. The black solid curves show the event rates considering the best-fit values of oscillation parameters as given in Table~\ref{table:one}. The three shaded blue (red) regions show the variations in events due to present $1\sigma,~2\sigma,~ \rm{and}~3\sigma$ allowed ranges in $\Delta m^2_{31}$ ($\delta_{\rm CP}$). The dark (light)-shaded grey area shows the currently allowed $1\sigma~ (2\sigma)$ region in $\sin^{2}\theta_{23}$ with the best-fit value of $\sin^{2}\theta_{23} = 0.455$ as shown by the vertical brown line. See Table~\ref{table:one} and related text for details. Note that y-ranges are different in all the four panels. }}
	\label{fig:3}
\end{figure}

From Table~\ref{table:two}, we note the following: 
\begin{itemize}
\item The appearance events change by a lot when $\sin^2\theta_{23}$ or $\delta_{\rm CP}$ is varied. This is observed in the case of both neutrino and antineutrino events. Thus, there are distinct $\theta_{23}$ - $\delta_{\rm CP}$ pairs which give same number of total events.  
\item The change in appearance events due to variation in $\Delta m^2_{31}$ is very small and the number of events in the three cases (best-fit, $3\sigma$ upper bounds, and 3$\sigma$ lower bounds) are almost degenerate. 
\item In the case of disappearance events, the central number for LO and HO are close to one another, but they are different from MM. The numbers also change significantly with respect to $\Delta m^2_{31}$. As far as $\delta_{\rm CP}$ variation is concerned, the event numbers show almost no change. Thus, in the case of disappearance events, there appears a $\sin^2\theta_{23}$ - $\Delta m^2_{31}$ degeneracy at the level of total rates.
\end{itemize} 

The observations made above are in line with the physics discussion done before in Sec.~\ref{probability} based on probabilities. As an example, we note that for 
$\nu$ appearance events, the numbers can vary between 820 to 969 for LO and 908 to 1058 for MM as $\delta_{\rm CP}$ is varied in the current $3\sigma$ range. Thus, there is a significant overlap in $\nu$ appearance events for LO and MM due to unknown $\delta_{\mathrm{CP}}$. However, for $\nu$ disappearance channel, the same oscillation parameter sets give number of events in the range 10870 to 10896 and 10646 to 10663, respectively, which may help to reduce the degenearcy as observed in appearance channel. The reverse argument can also be made where, for $\nu$ disappearance events, the numbers vary between 10758 to 11018 for LO and 10532 to 10797 for MM as $\Delta m^2_{31}$ is varied in its current $3\sigma$ range. But in the case of neutrino appearance channel, the corresponding events are lie in the range of 1104 to 1135 for LO and 1193 to 1226 for MM. Thus, the degeneracy that exists in the disappearance channel is partially resolved through the measurements made using appearance channel.

In Fig.~\ref{fig:3}, we show the disappearance (top panels) and appearance (bottom panels) event rates for various choices of $\sin^2\theta_{23}$ in the range $[0.4, 0.6]$. To generate this figure, we use the benchmark values of the oscillation parameters and their corresponding ranges as given in Table~\ref{table:one}. We see that the total event rates follow the same behavior as previously seen in Fig.~\ref{fig:2}, where we show $P_{\mu\mu}$ and $P_{\mu e}$ as a function of $\sin^2\theta_{23}$ assuming $E= 2.5~\rm GeV$. Note that though in Sec.~\ref{probability}, we discuss the main physics issues assuming a particular value of $E=2.5$ GeV, similar features are retained at the total event rates level as well in Fig.~\ref{fig:3}.

\begin{figure}[htb!]
\centering
\includegraphics[width=0.63\linewidth]{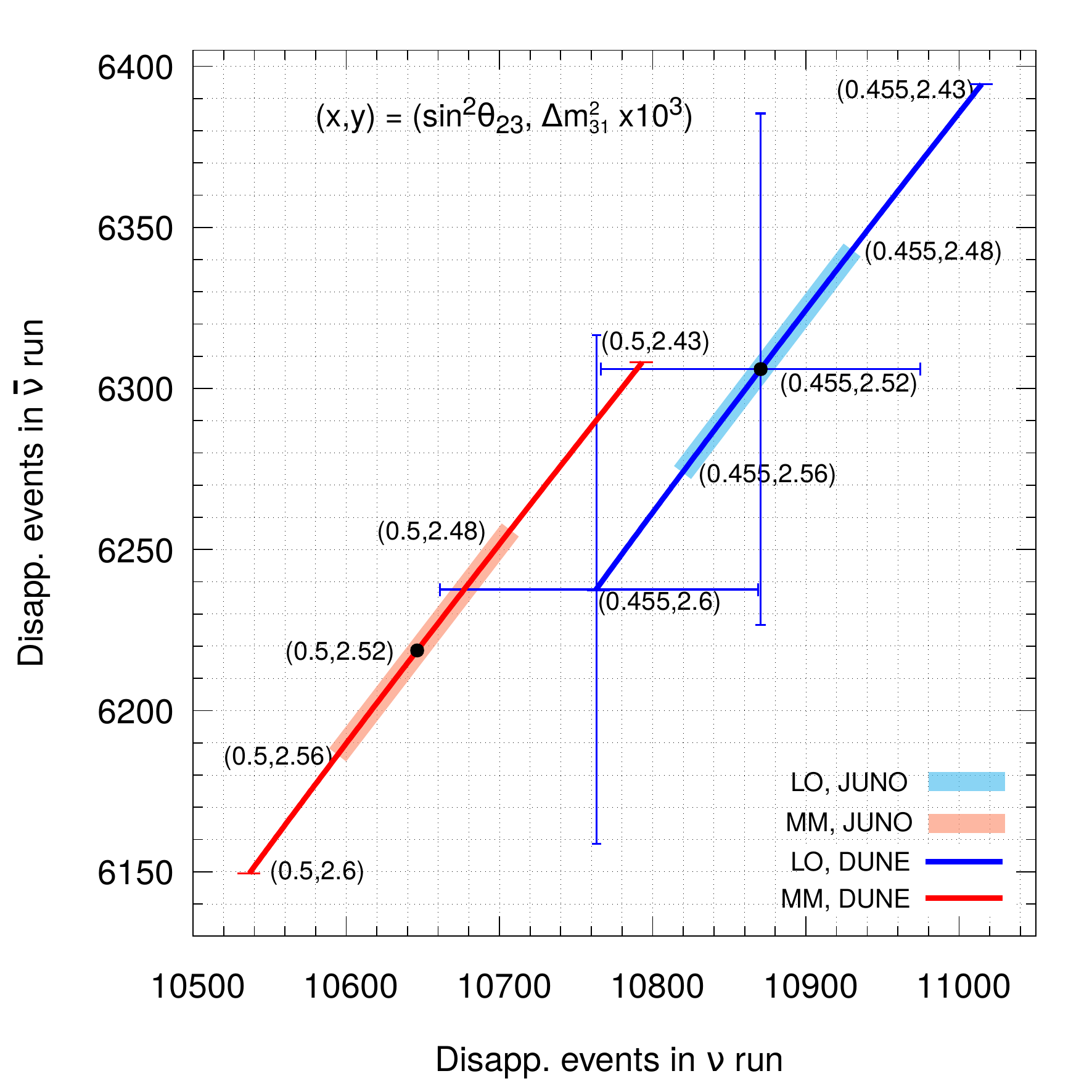}
\caption{\footnotesize{Bi-events plot for DUNE in the plane of neutrino - antineutrino disappearance events assuming 336 kt$\cdot$MW$\cdot$years of exposure equally divided in neutrino and antineutrino modes. The blue line is obtained by varying $\Delta m^{2}_{31}$ in its 3$\sigma$ range of $[2.43 : 2.6] \times 10^{-3}~\rm{eV}^{2}$ with $\sin^2\theta_{23}=0.455$ (LO). The red line depicts the same with $\sin^2\theta_{23}=0.5$ (MM). The black dot on each line shows the disappearance events corresponding to the best-fit value of $\Delta m^2_{31} = 2.52 \times 10^{-3}$ eV$^{2}$. The values of other oscillation parameters are taken from Table~\ref{table:one} assuming NMO. The blue (red) rectangular region on blue (red) line portrays the variation in event rates due to 3$\sigma$ range in $\Delta m^{2}_{31}$ expected from JUNO~\cite{EPS-HEP-Conference2021, JUNO:2015zny}. The horizontal (vertical) error bars for the points [($\sin^{2}\theta_{23} = 0.455, \Delta m^{2}_{31} = 2.52 \times 10^{-3} \rm{eV}^{2}$) and ($\sin^{2}\theta_{23} = 0.455, \Delta m^{2}_{31} = 2.6 \times 10^{-3} \rm{eV}^{2}$)] show the 1$\sigma$ statistical uncertainties which are obtained by taking the square root of the $\nu \,(\bar{\nu})$ disappearance events.}} 
\label{fig:4}
\end{figure}

In Fig.~\ref{fig:4}, we show the dependence of disappearance events on the oscillation parameters $\Delta m^2_{31}$ and $\sin^2\theta_{23}$ through the bi-events plot where we display the total neutrino (antineutrino) disappearance events on the x-axis (y-axis) assuming 3.5 years of run. We obtain blue curve by varying $\Delta m^{2}_{31}$ in its current 3$\sigma$ range of $[2.43 : 2.6] \times 10^{-3}~\rm{eV}^{2}$ assuming the current best-fit of $\sin^2\theta_{23}=0.455$ (see LO in legends). The red curve portrays the same with $\sin^2\theta_{23}=0.5$ (see MM in legends). The black dot on each line shows the disappearance events corresponding to the best-fit value of $\Delta m^2_{31} = 2.52 \times 10^{-3}$ eV$^{2}$. The values of other oscillation parameters are taken from Table~\ref{table:one} assuming NMO and $\delta_{\rm CP}$ = 223$^{\circ}$. The blue (red) rectangular region on blue (red) curve shows the variation in event rates due to allowed 3$\sigma$ range in $\Delta m^{2}_{31}$ as expected from JUNO~\cite{EPS-HEP-Conference2021, JUNO:2015zny}. The horizontal (vertical) error bars for the points [($\sin^{2}\theta_{23} = 0.455, \, \Delta m^{2}_{31} = 2.52 \times 10^{-3} \rm{eV}^{2}$) and ($\sin^{2}\theta_{23} = 0.455, \, \Delta m^{2}_{31} = 2.6 \times 10^{-3} \rm{eV}^{2}$)] show the 1$\sigma$ statistical uncertainties which are obtained by taking the square root of the neutrino (antineutrino) disappearance events.

The 1$\sigma$ statistical uncertainty in neutrino disappearance events corresponding to the benchmark oscillation parameters $\sin^2\theta_{23} = 0.455$ and $\Delta m^2_{31} = 2.522 \times 10^{-3}~\rm eV^2$ (see horizontal error bar around the black dot on blue line) has some overlap with the neutrino events on the red line. The same is true for antineutrino disappearance events (see vertical error bar around the black dot on blue line). These overlapping regions due to 1$\sigma$ statistical fluctuations get reduced when we consider the variation in event rates for the allowed 3$\sigma$ range in $\Delta m^{2}_{31}$  (2.48 $\times$ 10$^{-3}$ eV$^2$ to 2.56$ \times$ 10$^{-3}$ eV$^2$, see rectangular red region) as expected from JUNO~\cite{EPS-HEP-Conference2021, JUNO:2015zny}. Therefore, we can conclude that based on only total event rates, a definitive exclusion of MM is not possible at high confidence level for the current best-fit values of oscillation parameters. We demonstrate later that the $\sin^2\theta_{23}$ - $\Delta m^2_{31}$ degeneracy that is present at the total event rates level can be resolved by including the spectral shape information along with total event rates and we can establish the deviation from maximal $\theta_{23}$ at high confidence level in DUNE. If we consider the event rates and their 1$\sigma$ statistical uncertainties corresponding to the oscillation parameters $\sin^2\theta_{23} = 0.455$ (current best-fit) and $\Delta m^2_{31} = 2.6 \times 10^{-3}~\rm eV^2$ (current 1$\sigma$ upper bound) on the blue line then we see more overlap with the event rates on the red line corresponding to MM solution. The overlap is less for the benchmark oscillation parameters $\sin^2\theta_{23} = 0.455$ (current best-fit) and $\Delta m^2_{31} = 2.43 \times 10^{-3}~\rm eV^2$ (current 1$\sigma$ lower bound) on the blue line.

\subsection{Disappearance event spectra to resolve $\sin^{2} \theta_{23}-\Delta m^{2}_{31}$ degeneracy}

\begin{figure}[htb!]
	\centering
	\includegraphics[width=0.49\linewidth]{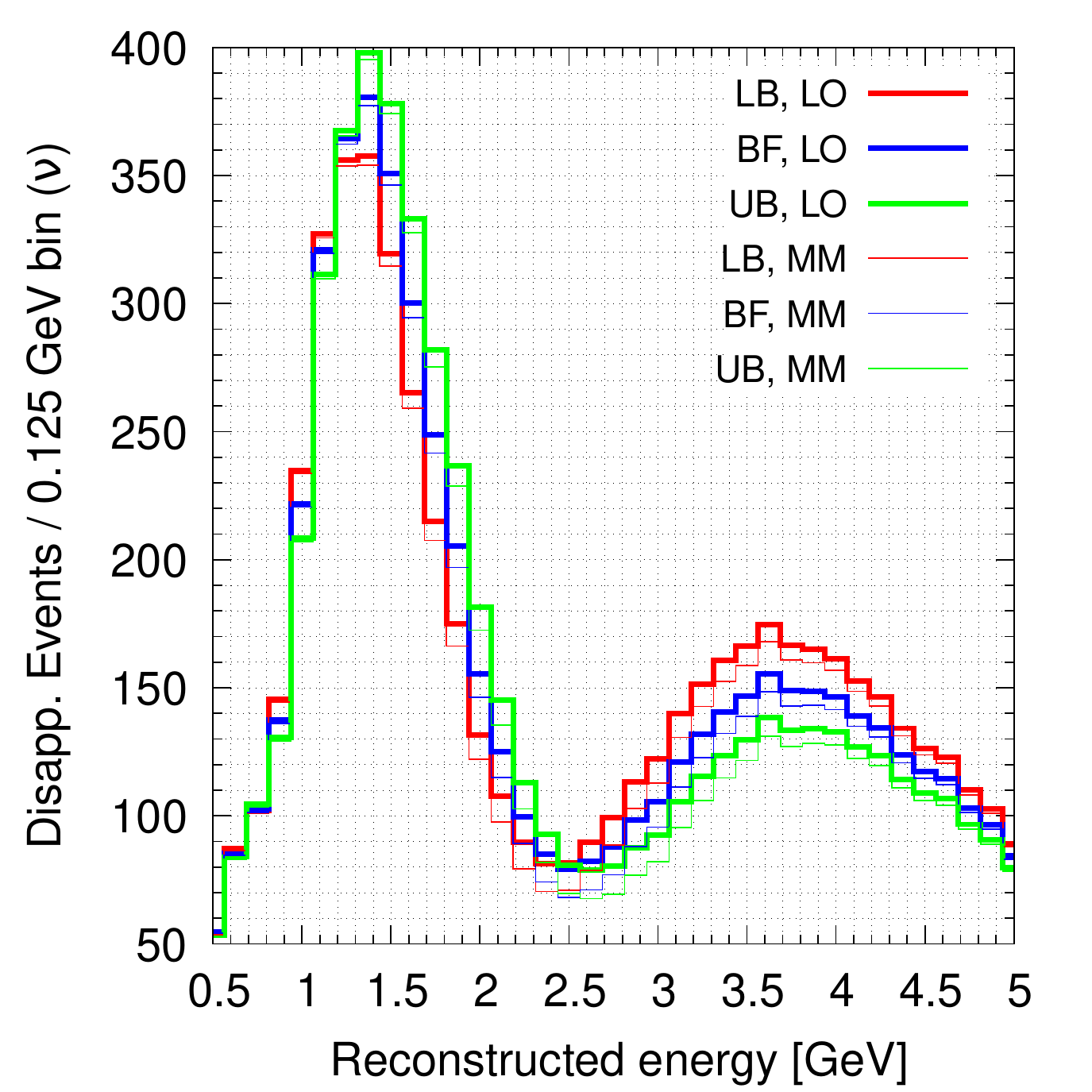}
	\includegraphics[width=0.49\linewidth]{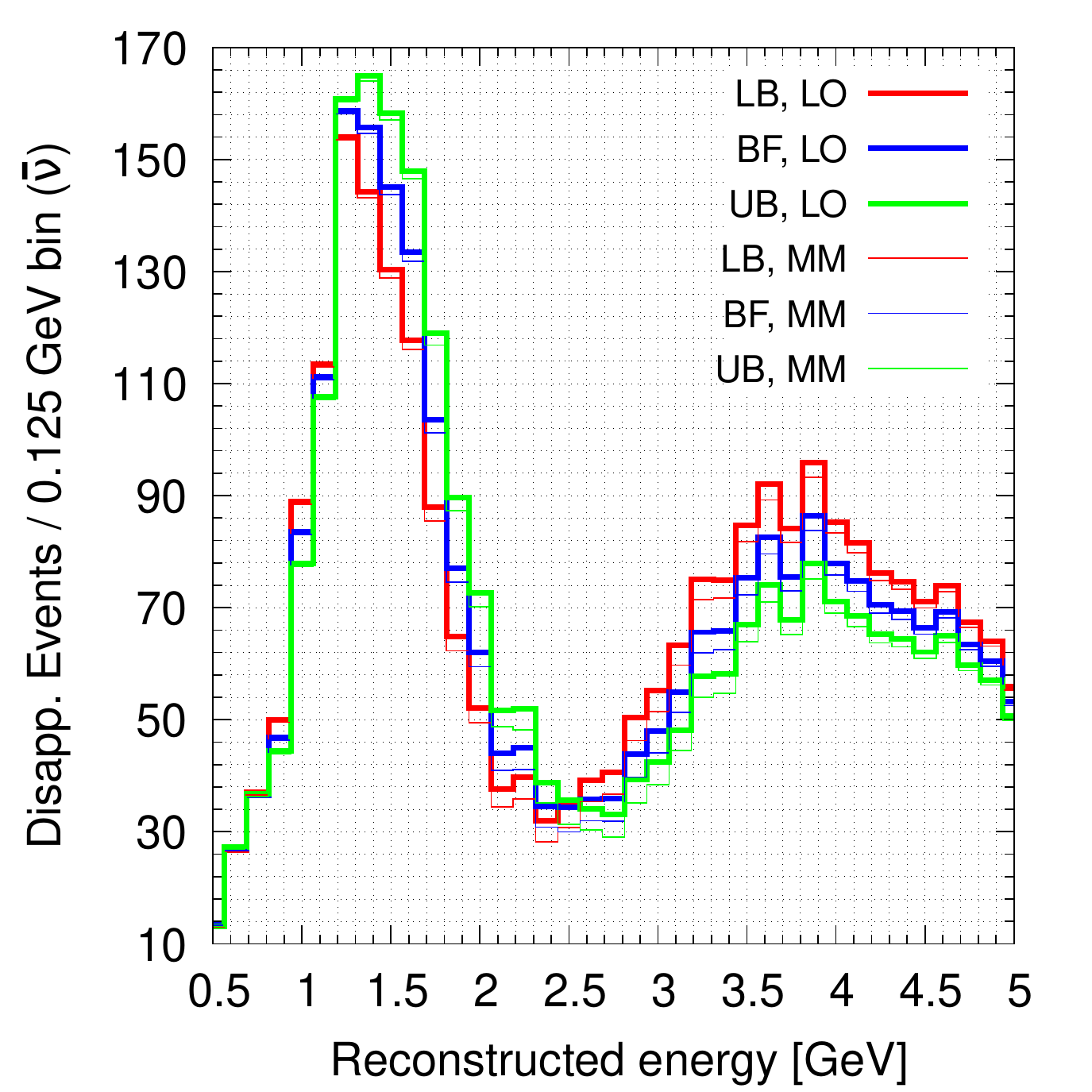}
	\includegraphics[width=0.49\linewidth]{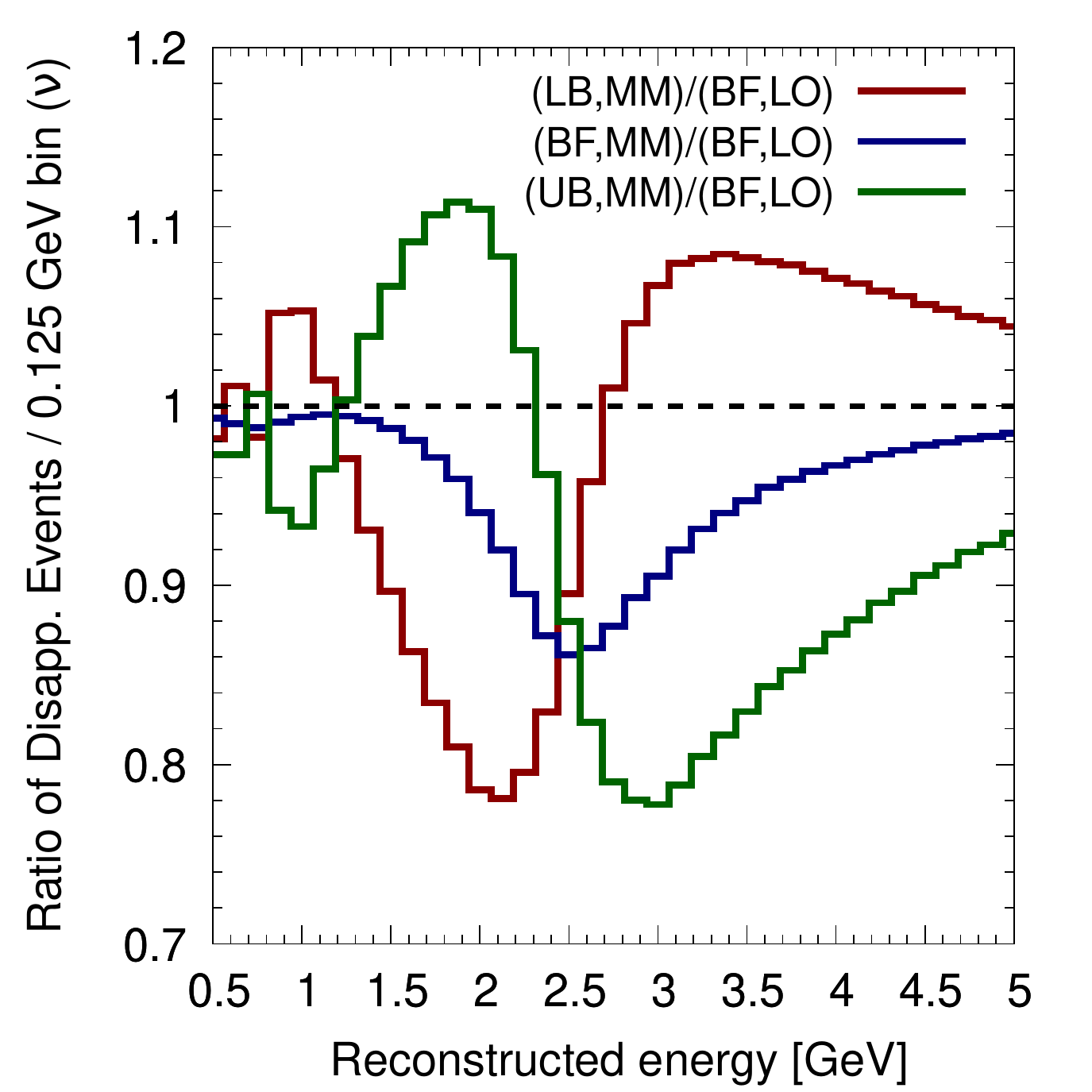}
	\includegraphics[width=0.49\linewidth]{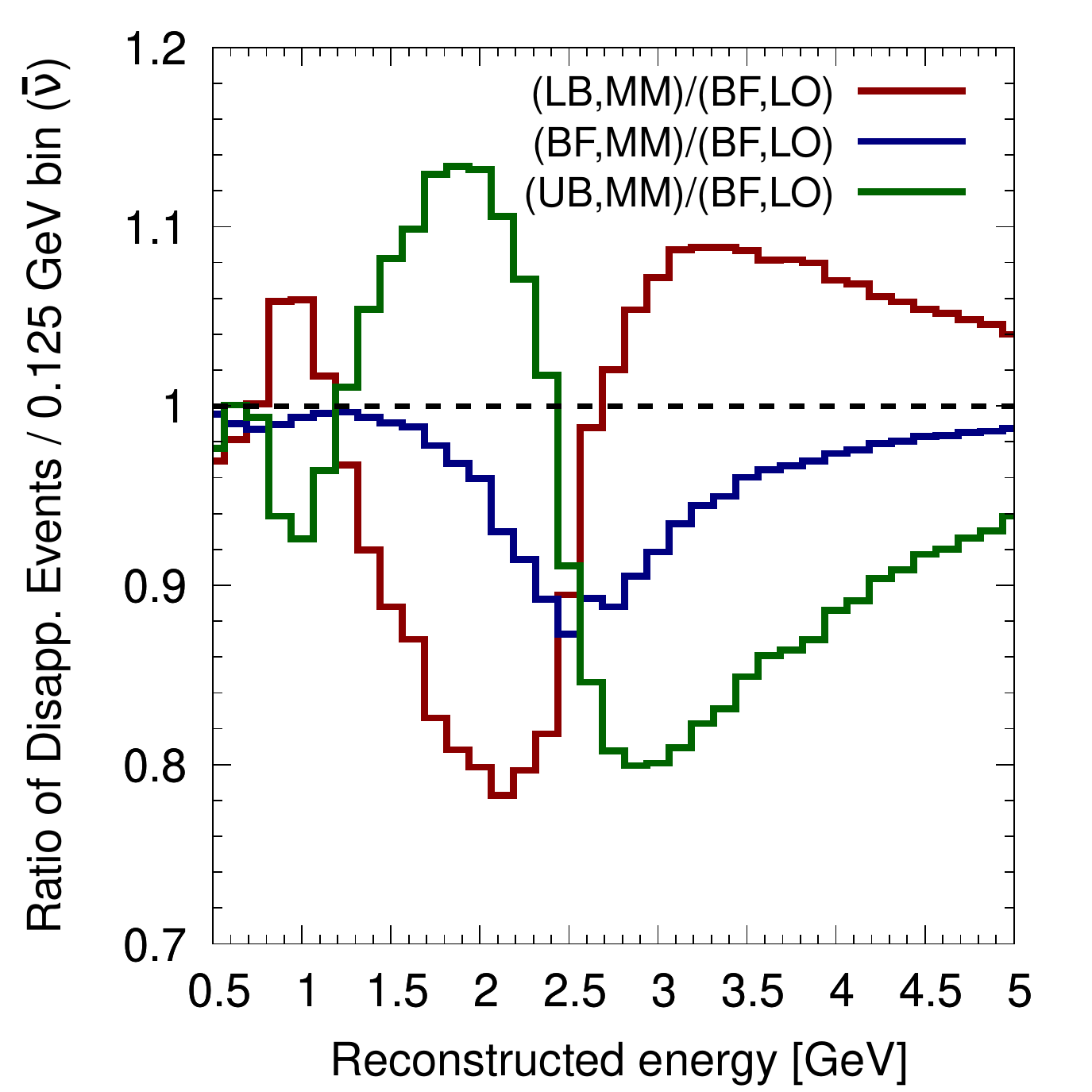}
	\caption{\footnotesize{In the top left (right) panel, we show the expected neutrino (antineutrino) disappearance event spectra as a function of reconstructed neutrino (antineutrino) energy assuming 3.5 years of $\nu$ ($\bar{\nu}$) run. The thick (thin) colored histograms correspond to a LO (MM) value of $\sin^{2}\theta_{23}$ = 0.455 (0.5). For a given value of $\sin^{2}\theta_{23}$, we depict the event spectra for three different choices of $\Delta m^2_{31}$: 2.522$\times 10^{-3}$ eV$^2$ [current best-fit (BF), blue lines], 2.436$\times 10^{-3}$ eV$^2$ [current 3$\sigma$ lower bound (LB), red lines], and 2.605$\times 10^{-3}$ eV$^2$ [current 3$\sigma$ upper bound (UB), green lines]. In the bottom left panel, we show the ratio of neutrino disappearance events in each energy bin as a function of reconstructed neutrino energy assuming 3.5 years of $\nu$ run. We present the same for antineutrino in the bottom right panel. The brown, blue, and green curves show the ratio of $N_{1}/{N}$, $N_{2}/{N}$, and $N_{3}/{N}$, respectively, where $N$: events in a given energy bin for $\left(\Delta m^2_{31} = 2.522\times 10^{-3}~\rm eV^2, \sin^2\theta_{23}=0.455\right)$, $N_{1}$: events in a given energy bin for $\left(\Delta m^2_{31} = 2.436\times 10^{-3}~\rm eV^2, \sin^2\theta_{23}=0.5\right)$, $N_{2}$: events in a given energy bin for $\left(\Delta m^2_{31} = 2.522\times 10^{-3}~\rm eV^2, \sin^2\theta_{23}=0.5\right)$, and $N_{3}$: events in a given energy bin for $\left(\Delta m^2_{31} = 2.605\times 10^{-3}~\rm eV^2, \sin^2\theta_{23}=0.5\right)$.
}}
	\label{fig:5}
\end{figure}

In Fig.~\ref{fig:5}, we show the event spectra for $\nu_{\mu}\rightarrow\nu_{\mu}$ disappearance channel as a function of the reconstructed neutrino energy. The top left (top right) panel is for neutrino (antineutrino) disappearance events. In the top panel, the thick and thin colored histograms correspond to a LO and MM value of $\sin^{2}\theta_{23}$, respectively. For a given value of $\sin^{2}\theta_{23}$, we exhibit the event spectra for three different choices of $\Delta m^2_{31}$: 2.522 $\times 10^{-3}$ eV$^2$ (BF, see blue lines), 2.436 $\times 10^{-3}$ eV$^2$ (LB, see red lines), and 2.605 $\times 10^{-3}$ eV$^2$ (UB, see green lines). Looking at Fig.~\ref{fig:5}, it seems that there is significant degeneracy between LO and MM when a full $3\sigma$ variation in $\Delta m^2_{31}$ is considered. However, on observing closely, it can be seen that the events for energy bins on either side of the oscillation minimum (maximum in the case of $P_{\mu e}$) at $ E = 2.5~GeV$ behave oppositely when $\Delta m^2_{31}$ is varied. This is made more evident in the lower panel of Fig.~\ref{fig:5} where the left (right) figure correspond to neutrino (antineutrino). The three set of curves correspond to ratio of events in each reconstructed energy bin - $N_{i}/N$ (for $i =1,2,3$) where N is the number of events when $\sin^2\theta_{23} = 0.455$ and $\Delta m^2_{31} = 2.522 \, \times  10^{-3}~\rm eV^2$. In the ratio, $N_{1}, N_{2}, N_{3}$ are number of events when $\sin^2\theta_{23} = 0.5$ and $\Delta m^2_{31} = 2.438 \, \times 10^{-3}~\rm eV^2$, $2.522 \times 10^{-3}~\rm eV^2$, and $2.602 \times 10^{-3}~\rm eV^2$, respectively. It can be seen from the lower panels in Fig.~\ref{fig:5}, that the ratio of events approach 1 (reduction in sensitivity towards exclusion of maximality) on one side of the oscillation minimum while moving farther away from 1 compared to the blue line on the other side of the oscillation minimum. This explains that, while some of the energy bins decreases the sensitivity in deviation from maximal choice of $\theta_{23}$, the other energy bins help to  increase.  Therefore, we conclude that doing a spectral analysis further breaks the $\sin^2\theta_{23}$ - $\Delta m^2_{31}$ degeneracy seen in total event rates and therefore current 3$\sigma$ uncertainty in $\Delta m^2_{31}$ will not affect the sensitivity of DUNE in establishing non-maximal mixing of $\theta_{23}$. 

\section{Our findings}
\label{results}

In this section, we demonstrate the capability of DUNE to address three important issues related to atmospheric oscilation parameter: (i) possible deviation of $\theta_{23}$ from maximal mixing (45$^{\circ}$), (ii) the correct octant of $\theta_{23}$ if it turns out to be non-maximal in Nature, and (iii) the achievable precision on the atmospheric oscillation parameters $\sin^2\theta_{23}$ and $\Delta m^2_{31}$ in light of current neutrino oscillation data. To estimate the sensitivities, we use the following definition of Poissonian $\chi^2$

\begin{equation}
\chi^2 (\vec{\omega}, \, \kappa_{s}, \, \kappa_{b})= \underset{( \vec{\lambda}, \, \kappa_{s}, \,\kappa_{b})}{\mathrm{min}}\left\{  2\sum_{i=1}^{n}(\tilde{y_i}-x_i-x_i\mathrm{ln}\frac{\tilde{y_i}}{x_i})+\kappa^2_{s}+ \kappa^2_{b}\right\}\, , 
\label{chi2} 
\end{equation}

where, n is the total number of reconstructed energy bins and 

\begin{equation}
\tilde{y_i}\,(\vec{\omega},\{\kappa_{s},\kappa_{b}\}) = N^{th}_i(\vec{\omega})[1+\pi^s\kappa_{s}]+N^b_i(\vec{\omega})[1+\pi^b\kappa_{b}]\, .
\label{chi}
\end{equation}
In the above equation, $N^{th}_i\,(\vec{\omega})$ denotes the predicted number of signal events in the $i$-th energy bin for a set of oscillation parameters $\vec{\omega} = \left\lbrace \theta_{23},\theta_{13},\theta_{12},\Delta \mathrm{m}^{2}_{21}, \Delta \mathrm{m}^{2}_{31},\delta_{\mathrm{CP}}\right\rbrace$ and $\vec{\lambda}$ is the set of oscillation parameters which we have marginalized in the fit. For an instance, when we address the issue of deviation from maximality, $\vec{\lambda} = \{\Delta m_{31}^2,~\delta_{\mathrm{CP}}\}$. $N^{b}_i\,(\vec{\omega})$ represents the number of background events in the $i$-th energy bin where the neutral (charged) current backgrounds are independent (dependent) on $\vec{\omega}$. The quantity $\pi^s$ ($\pi^b$) is the normalization uncertainty on signal (background). The quantities $\kappa_{s}$ and $\kappa_{b}$ are the systematic pulls~\cite{Huber:2002mx,Fogli:2002pt,Gonzalez-Garcia:2004pka} on signal and background, respectively. We incorporate the prospective data in Eq.~\ref{chi2} using the variable $x_{i}= N^{ex}_i\, +\, N^b_i$, where $N^{ex}_i$ denotes the observed charged current signal events in the $i$-th energy bin and $N^b_i$  represents the background as mentioned earlier.

\subsection{Deviation from maximal $\theta_{23}$}

As discussed previously in Sec.~\ref{introduction}, the three global analyses of the oscillation data do not agree on the octant in which the best-fit value of $\theta_{23}$ lies. Further, they all find $\sin^2\theta_{23}=0.5$ to be allowed at $3\sigma$ confidence level.
Therefore, before we address the issue of resolving the octant of $\theta_{23}$, it is imperative to question at what confidece level maximal 2-3 mixing can be ruled out. We define $\Delta\chi^2$ for deviation from maximal $\theta_{23}$ as follows: 
\begin{equation}
\Delta \chi^2_{\rm DM}= \underset{(\vec{\lambda},~ \kappa_{s},\kappa_{b})}{\mathrm{min}}\left\{ \chi^2\left(\sin^2\theta_{23}^{\mathrm{true}} \in [0.4,0.6]\right) - \chi^2\left(\sin^2\theta_{23}^{\mathrm{test}} = 0.5\right)\right\},  
\end{equation} 
Here, $\vec{\lambda} = \{\delta_{\mathrm{CP}}, \, \Delta m^{2}_{31}\}$ are the oscillation parameters over which the $\Delta\chi^2$ has been marginalized, while $\kappa_{s}~\text{and}~ \kappa_{b}$ are the systematic pulls on signal and background, respectively.

\begin{figure}[htb!]
	\centering
	\includegraphics[width=0.49\linewidth]{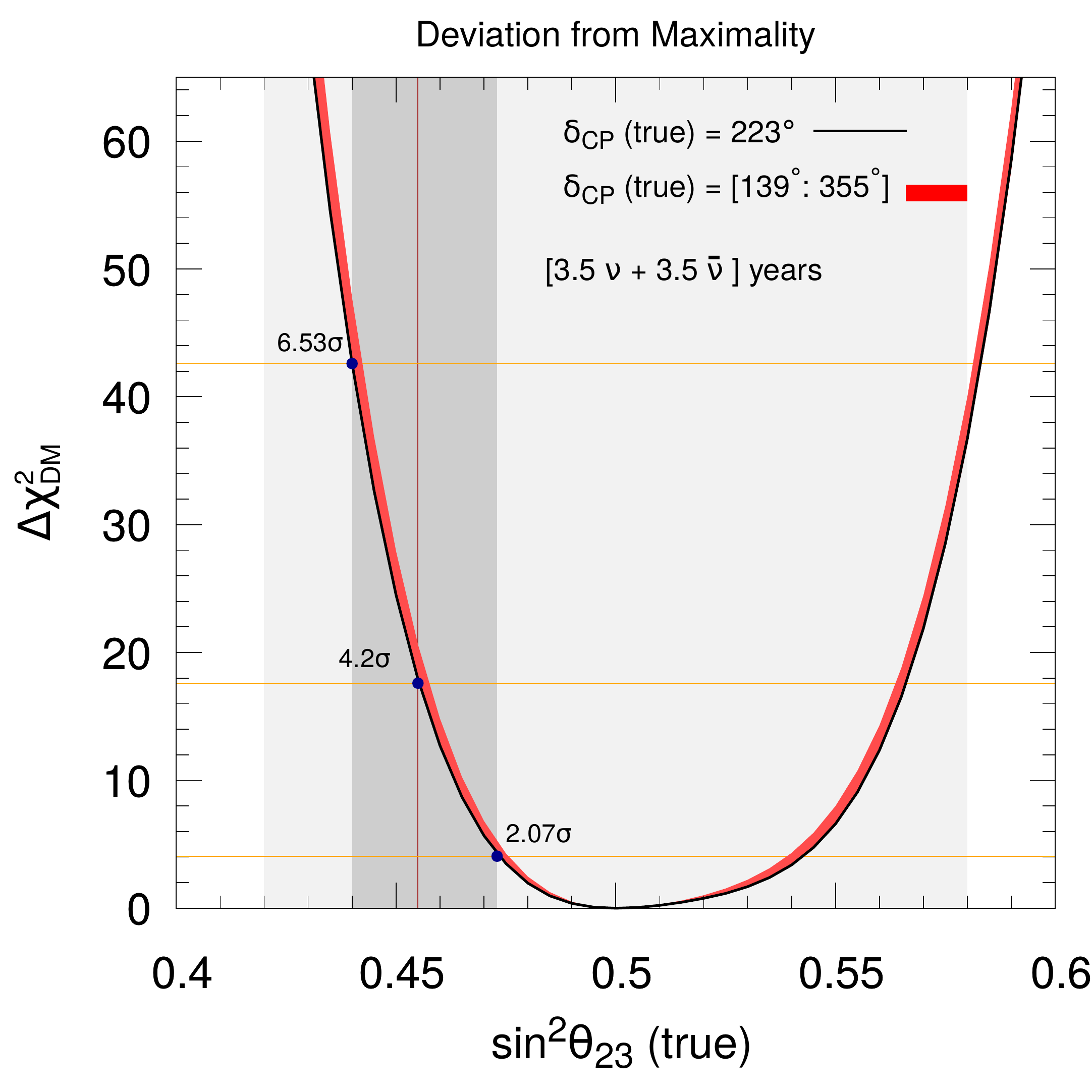}
	\includegraphics[width=0.49\linewidth]{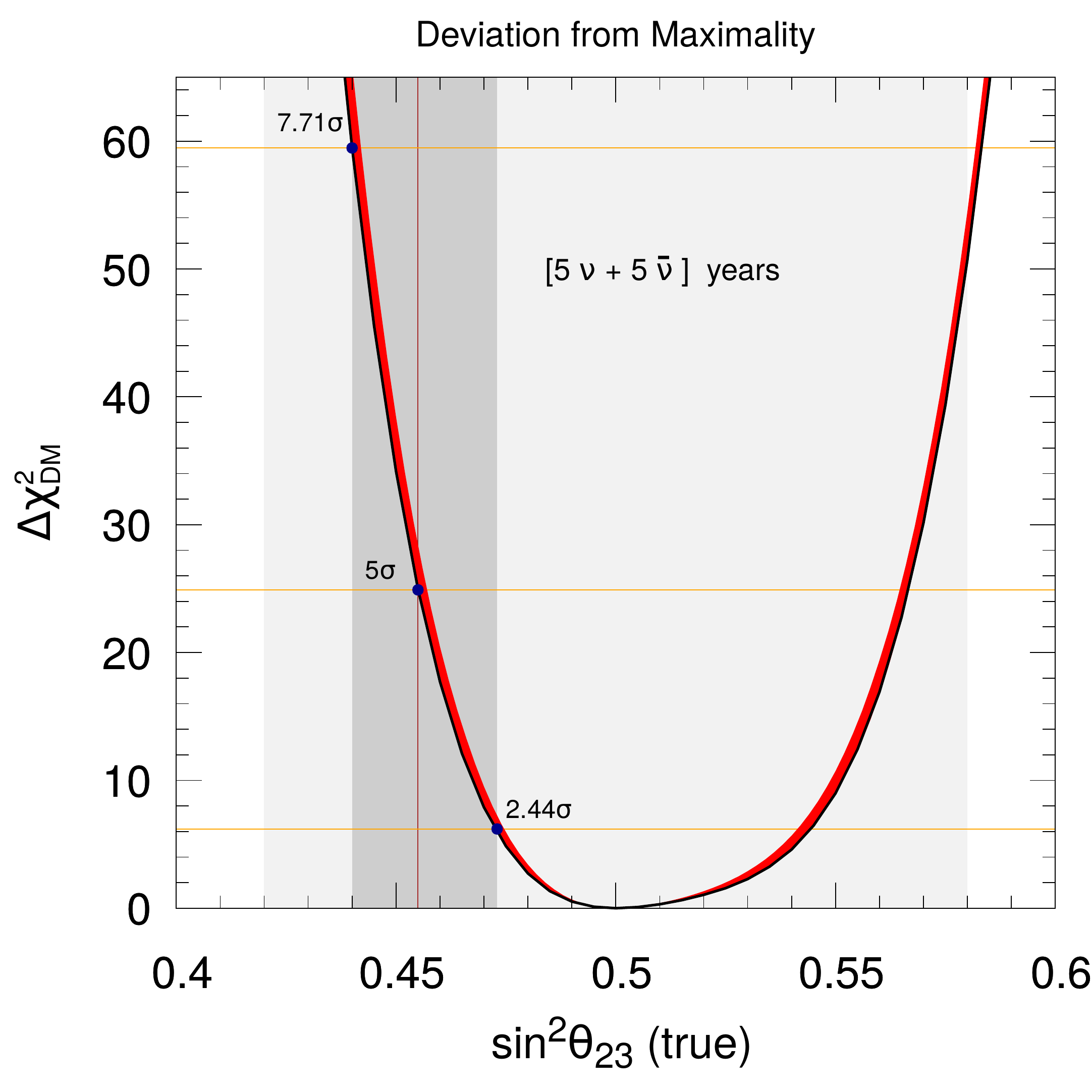}
	\caption{\footnotesize{The black curve in the left (right) panel shows the potential of DUNE to establish the deviation from maximal $\theta_{23}$ as a function of true $\sin^2\theta_{23}$ assuming true NMO and $\delta_{\mathrm{CP}}$ (true) = 223$^{\circ}$ with 7 (10) years of total exposure equally divided in $\nu$ and $\bar{\nu}$ modes. The red bands portray the same for true $\delta_{\mathrm{CP}}$ in the range of 139$^{\circ}$ to 355$^{\circ}$. In the fit, we marginalize over the current 3$\sigma$ range of $\Delta m^{2}_{31} $ and $\delta_{\rm CP}$, while keeping rest of the oscillation parameters fixed at their present best-fit values as shown in Table~\ref{table:one}. The dark (light)-shaded grey area shows the currently allowed $1\sigma$ $(2\sigma)$ region in $\sin^{2}\theta_{23}$ as obtained in the global fit study~\cite{Capozzi:2021fjo} with the best-fit value of $\sin^{2}\theta_{23} = 0.455$ as shown by vertical brown line. The horizontal orange lines show the sensitivity (experessed in $\sigma = \sqrt{\Delta\chi^2_{\rm DM}}$) for the current best-fit and 1$\sigma$ upper and lower bounds of $\sin^2\theta_{23}$.}}
	\label{fig:6}
\end{figure}

In Fig.~\ref{fig:6}, we show the potential of DUNE to establish deviation from maximal $\theta_{23}$ as a function of the true $\sin^2\theta_{23}$ in the range of 0.4 to 0.6. The black lines in both left and right panels display the ability of DUNE in establishing deviation from maximal $\theta_{23}$ assuming true NMO and $\delta_{\rm CP}$ (true) = 223$^{\circ}$. In the left panel, we show the results with nominal neutrino and antineutrino runs of 3.5 years each, while in the right panel we show results with 5 years of running in each mode. The red bands in Fig.~\ref{fig:6} portray the variation in $\Delta \chi^2_{\rm DM}$ for true $\delta_{\rm CP}$  in its current 3$\sigma$ allowed range of 139$^{\circ}$ to 355$^{\circ}$ (see Table~\ref{table:one}). Left panel reveals that for $\sin^2\theta_{23}$ (true) = 0.47 (current 1$\sigma$ upper bound), 0.455 (current best-fit), and 0.44 (current 1$\sigma$ lower bound), DUNE can exclude maximal mixing solution at 2.07$\sigma$, 4.2$\sigma$, and 6.5$\sigma$, respectively assuming true NMO and with a total 7 years of run equally divided in neutrino and antineutrino modes. For a total 10 years of run, the above sensitivities get enhanced to 2.44$\sigma$, 5$\sigma$, and 7.71$\sigma$, respectively (see right panel). We observe that a 3$\sigma$ (5$\sigma$) determination of non-maximal $\theta_{23}$ is possible in DUNE with a total exposure of 7 years if the true value of $\sin^2\theta_{23} \lesssim 0.465~(0.450)$ or $\sin^2\theta_{23} \gtrsim 0.554~(0.572)$ for any value of true $\delta_{\mathrm{CP}}$ in its present 3$\sigma$ range and true NMO (see left panel). When we increase the total exposure from 7 years to 10 years, we see a marginal enhancement in the sensitivity (see right panel).

\subsubsection{Contributions from appearance and disappearance channels and role of systematics}
\label{appdisapp}

\begin{figure}[htb!]
	\centering
	\includegraphics[width=0.65\linewidth]{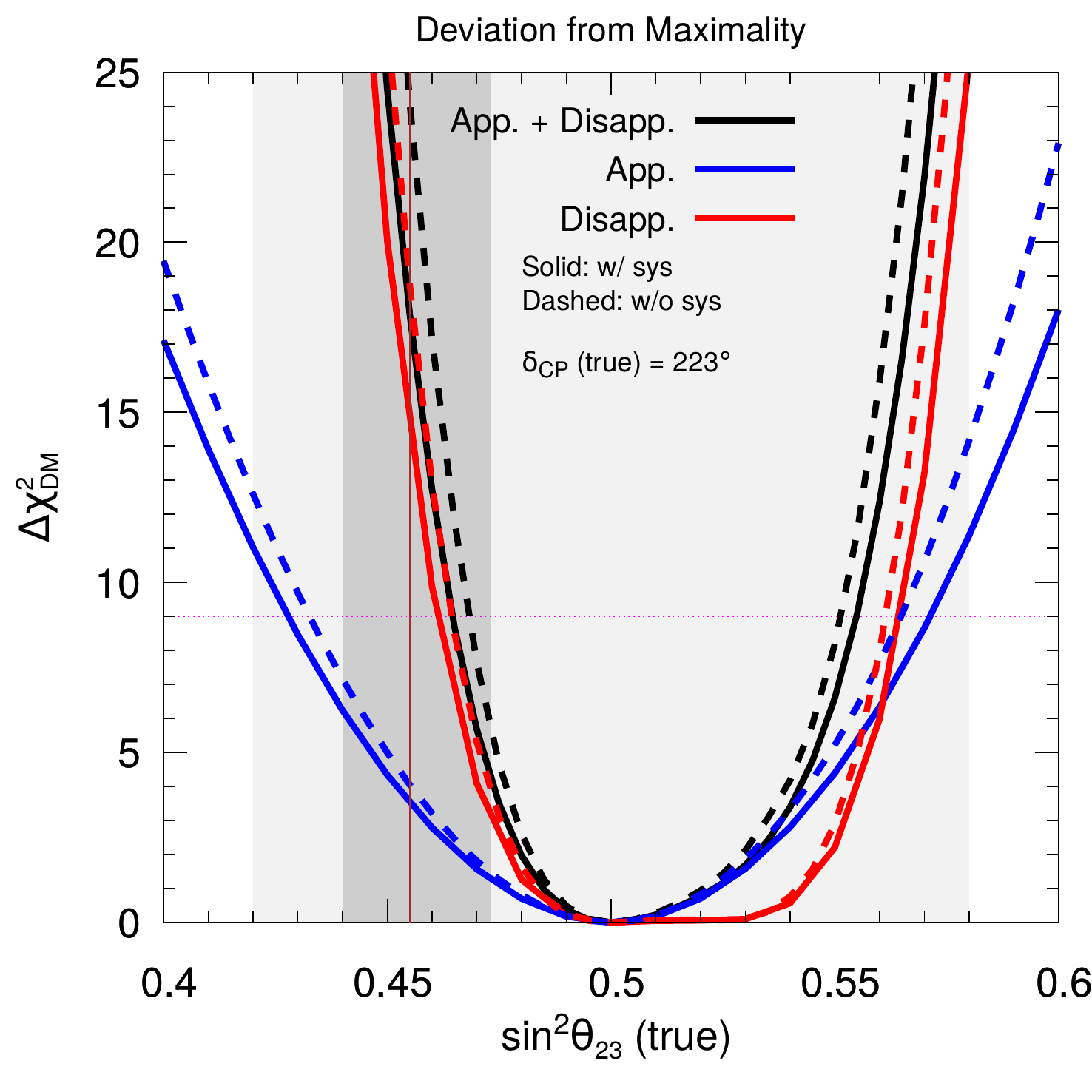}
	\caption{\footnotesize{Performance of DUNE to establish deviation from maximality as a function of true $\sin^2\theta_{23}$ assuming true NMO and $\delta_{\rm CP}$ (true) = 223$^\circ$ with 7 years of exposure equally divided in $\nu$ and $\bar{\nu}$ modes. Red, blue, and black curves show the sensitivity obtained using disappearance channel, appearance channel and their combinations, respectively. The solid (dashed) lines depict the results with (without) systematic uncertainties. In the fit, we marginalize over the current 3$\sigma$ range of $\Delta m^{2}_{31} $ and $\delta_{\rm CP}$, while keeping rest of the oscillation parameters fixed at their present best-fit values as shown in Table~\ref{table:one}. The dark (light)-shaded grey area shows the currently allowed $1\sigma$ $(2\sigma)$ region in $\sin^{2}\theta_{23}$ as obtained in the global fit study~\cite{Capozzi:2021fjo} assuming NMO with the best-fit value of $\sin^{2}\theta_{23} = 0.455$ as shown by vertical brown line. The capability of DUNE to establish non-maximal $\theta_{23}$ at 3$\sigma$ ($\Delta \chi^{2}_{\mathrm{DM}} = 9$) confidence level is shown by horizontal pink dotted line.}}
	\label{fig:7}
\end{figure}

We now explore how the appearance and disappearance channels individually contribute towards the exclusion of MM. In Fig.~\ref{fig:7}, we show the $\Delta \chi^2_{\rm DM}$ as a function of true $\sin^2\theta_{23}$ for appearance (in solid blue), disappearance (in solid red) and combined (in solid black). It is interesting to note that for true  values of $\sin^2\theta_{23}$ in HO that are very close to MM, the appearance channel provides better sensitivity towards the exclusion of MM. However, for $\sin^2\theta_{23} \gtrsim 0.56$, the $\Delta \chi^2$ increases very rapidly. In the case of LO, we see that it is mainly the disappearance channel that contributes to the exclusion of MM.  
In order to understand such a behavior, we refer to Section~\ref{probability}, where we showed that for $\sin^2\theta_{23} \gtrsim 0.5$, $P_{\mu\mu}$ shows a flat behavior while $P_{\mu e}$ increases linearly. It is only when $\sin^2\theta_{23}$ is a little away from $0.5$ that $P_{\mu\mu}$ increases steeply. On the other hand, in LO, $P_{\mu\mu}$ is very steep even for values which are close to $0.5$. 
In this figure, we also discuss the role that systematic uncertainties play in deteriorating the sensitivity of DUNE towards exclusion of MM.
We consider two scenarios here which are shown in Fig.~\ref{fig:7}. In the first case, we consider an ideal experimental setup with no systematic uncertainties (shown with dashed curves in Fig.~\ref{fig:7}). In the second case, we consider the DUNE's nominal systematic uncertainties (shown by solid curves in Fig.~\ref{fig:7}) described in Ref. \cite{DUNE:2021cuw}. Looking at Fig.~\ref{fig:7}, it appears that both appearance and disappearance channels are affected by the systematics especially when going from a no systematics ideal experimental setup to the realistic situation. For example, at true $\sin^2\theta_{23} = 0.455$, systematic uncertainties deteriorate the MM-exclusion from $\Delta\chi^2_{\rm DM} = 25$ to $\Delta\chi^2 = 20$. In order to explore this point further, we generate results for three more choices of systematic uncertainties. The results are shown in Table~\ref{table:three}.

\begin{table}[htb!]
	\begin{center}
		\begin{small}
			\begin{tabular}{|c|c|c|c|c|c|c|}
				\hline\hline
                 {True $\sin^2\theta_{23}$} & Channels & 2\%, 5\% & 0\%, 0\% & 5\%, 5\% & 5\%, 10\% & 10\%, 10\%\\
 				\hline
				 & App.+Disapp.&
				$\boldsymbol{17.64}$ &
				$\boldsymbol{24.13}$  & 
				$\boldsymbol{16.88}$ &
				$\boldsymbol{16.74}$ &
				$\boldsymbol{15.42}$\\
				
				$\boldsymbol{0.455}$ & App.&	 $\boldsymbol{3.52}$ & $\boldsymbol{4.05}$ & $\boldsymbol{2.33}$  & $\boldsymbol{2.33}$ & $\boldsymbol{1.05}$ \\
				
			(Best-fit)	& 	Disapp.& $\boldsymbol{14.31}$ & $\boldsymbol{18.79}$ & $\boldsymbol{14.31}$  & $\boldsymbol{14.16}$ & $\boldsymbol{14.16}$ \\

				\hline \hline
				 & App.+Disapp.&
				$\boldsymbol{4.28}$ &
				$\boldsymbol{5.72}$ & 
				$\boldsymbol{3.88}$ &
				$\boldsymbol{3.84}$ &
				$\boldsymbol{3.42}$\\
				
				$\boldsymbol{0.473}$ & 	App.& $\boldsymbol{1.27}$ & $\boldsymbol{1.47}$ & $\boldsymbol{0.84}$  & $\boldsymbol{0.84}$ & $\boldsymbol{0.38}$ \\
				
				($1\sigma$ upper bound) & 	Disapp.& $\boldsymbol{2.99}$ & $\boldsymbol{3.88}$ &
				$\boldsymbol{2.99}$  & 
				$\boldsymbol{2.97}$ & 
				$\boldsymbol{2.97}$ \\
				
				\hline\hline
				
			\end{tabular}
		\end{small}
	\end{center}
	\caption{\footnotesize{Impact of systematics uncertainties on the determination of non-maximal $\sin^{2}\theta_{23}$. We show results for $\sin^2\theta_{23}$ (true) = 0.455 (current best-fit) and  $\sin^2\theta_{23}$ (true) = 0.473 (current $1\sigma$ upper bound) assuming true NMO and $\delta_{\rm CP}$ (true) = 223$^\circ$ with 7 years of exposure equally shared in $\nu$ and $\bar{\nu}$ modes. We estimate the sensitivity for different choices of normalization uncertainties on appearance and disappearance events [(2\%, 5\%), (0\%, 0\%), (5\%, 5\%), (5\%, 10\%), and (10\%, 10\%)], where (2\%, 5\%) is the benchmark choice~\cite{DUNE:2021cuw}. We keep the normalization uncertainties on various backgrounds fixed at their default values as given in Ref.~\cite{DUNE:2021cuw}. Results are given for appearance channel, disappearance channel, and their combination. In the fit, we marginalize over the current 3$\sigma$ range of $\Delta m^{2}_{31}$ and $\delta_{\rm CP}$, keeping rest of the oscillation parameters fixed at their present best-fit values as shown in Table~\ref{table:one}.}}
	\label{table:three}
\end{table}

We show the results for two values of $\sin^2\theta_{23}$ corresponding to the current best-fit of 0.455 and the present $1\sigma$ upper bound of 0.473 (see Table~\ref{table:one}). The three rows in Table~\ref{table:three} correspond to combined data from appearance and disappearance, only appearance data and only disappearance data. The different columns correspond to various choice of the systematic errors denoted as $(x\%,y\%)$ where $x\%$ denotes the normalization error in the measurement of electron-like events (due to appearance) while $y\%$ denotes the normalization error in the measurement of muon-like events (due to disappearance). We do not change the systematic uncertainties for background events and consider the same systematic uncertainty values for both neutrino and antineutrino channels. Our results show that while the sensitivity certainly deteriorates as we go from the ideal case of $(0\%, 0\%)$ to the nominal values of $(2\%, 5\%)$, there is negligible decrease in sensitivity due to only disappearance channel as the errors are increased further. The appearance channels are affected more because of systematic uncertainties, but since their contribution to the overall sensitivity to establish deviation from maximal $\theta_{23}$ is marginal, it does not affect much. Therefore, we conclude that DUNE's sensitivity to MM exclusion will not be systematics dominated and similar performance can be expected even with somewhat worse systematics. 

\subsubsection{Advantage due to spectral analysis 
and impact of marginalization over oscillation parameters}  

\begin{figure}[htb!]
	\centering
	\includegraphics[width=0.7\textwidth]{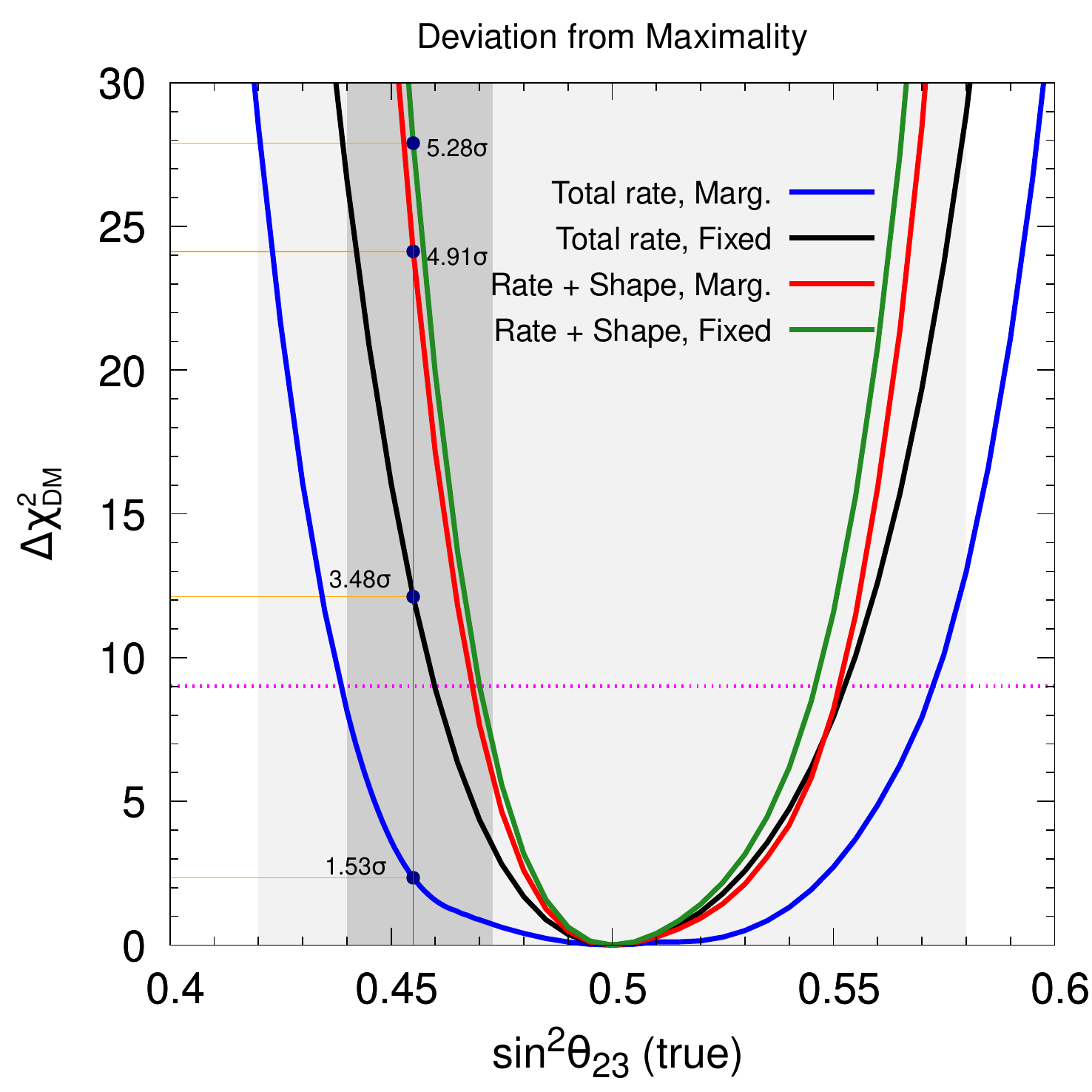}
	\caption{\footnotesize{Potential of DUNE to establish deviation from maximality as a function of true $\sin^2\theta_{23}$ assuming true NMO and $\delta_{\rm CP}$ (true) = 223$^\circ$ with the combined 3.5 years $\nu$ + 3.5 years $\bar{\nu}$ run. Blue (Black) curve shows the performance based on total event rates when $\Delta m^{2}_{31}$ and $\delta_{\mathrm{CP}}$ are marginalized (fixed) in the fit. Red (green) curve depicts the sensitivity based on rate + shape analysis when $\Delta m^{2}_{31}$ and $\delta_{\mathrm{CP}}$ are marginalized (fixed) in the fit. See text for details. The dark (light)-shaded grey area shows the currently allowed $1\sigma$ $(2\sigma)$ region in $\sin^{2}\theta_{23}$ as obtained in the global fit study~\cite{Capozzi:2021fjo} with the best-fit value of $\sin^{2}\theta_{23} = 0.455$ as shown by vertical brown line. The horizontal orange lines show the sensitivity (experessed in $\sigma = \sqrt{\Delta\chi^2_{\rm DM}}$) due to individual runs for the current best-fit value of $\sin^2\theta_{23}$. The capability of DUNE to establish non-maximal $\theta_{23}$ at 3$\sigma$ ($\Delta \chi^{2}_{\mathrm{DM}} = 9$) confidence level is shown by horizontal pink dotted line. For simplicity, we do not consider systematic uncertainties in this figure.}}
	\label{fig:8}
\end{figure}

We now explore the benefit of spectral shape information in DUNE on top of total event rates in establishing deviation from maximality. In Fig.~\ref{fig:8}, we show $\Delta\chi^2_{\rm DM}$ as a function of true $\sin^2\theta_{23}$ for four cases. The blue and black curves are obtained based on total event rates, while the red and green curves show the sensitivity when we include the spectral shape information along with total event rates. For both total rate and rate + shape analyses, we estimate the sensitivities in the fixed-parameter and marginalized scenarios. In the fixed-parameter case, we keep all the oscillation parameters fixed at their best-fit values (see second column in Table~\ref{table:one}) in both data and fit, while in the marginalized case, we minimize over $\Delta m^2_{31}$ and $\delta_{\rm CP}$ in their current 3$\sigma$ ranges. This comparison between fixed-parameter (see black and green lines) and marginalized (see blue and red lines) scenarios enable us to see how much the sensitivity gets deteriorated due to the uncertainties on $\Delta m^2_{31}$ and $\delta_{\rm CP}$. While establishing non-maximal $\theta_{23}$, the bulk of the sensitivity stems from the disappearance channel (see Fig.~\ref{fig:7}) and the uncertainty on $\Delta m^{2}_{31}$ affects this channel more than $\delta_{\rm CP}$. This can be seen from the top panels of Figs.~\ref{fig:2} and~\ref{fig:3}, and Fig.~\ref{fig:6} also confirms that the impact of $\delta_{\rm CP}$ is minimal in establishing the deviation from maximality. At the same time, we expect that the upcoming medium-baseline reactor experiment JUNO will measure $\Delta m^{2}_{31}$ with utmost precision~\cite{EPS-HEP-Conference2021,JUNO:2015zny} before DUNE will start taking data. Therefore, it makes complete sense to analyze the potential of DUNE to establish non-maximal $\theta_{23}$ in the fixed-parameter scenario (see black and green lines). However, Fig.~\ref{fig:8} also reveals that the impact of uncertainty on $\Delta m^{2}_{31}$ in the marginalized case is substantialy reduced when we exploit the spectral shape information (see red line) $-$ thanks to the intense wide-band beam resulting into high-statitstics in disappearance mode and excellent energy resolution of LArTPC detector in DUNE~\cite{Friedland:2018vry,DeRomeri:2016qwo}. 

We see from Fig.~\ref{fig:8} that the ability of DUNE to exclude maximal mixing solution in the fit for $\sin^{2}\theta_{23}$ (true) = 0.455, gets significantly enhanced from 1.53$\sigma$ (see blue line) to 4.91$\sigma$ (see red line) when we include spectral shape information in the analysis. We see this improvement in the sensitivity because the impact of $\Delta m^2_{31} - \sin^2\theta_{23}$ degeneracy gets reduced substantially when we perform rate + shape analysis instead of using only total rates. As we already demonstrate before using Fig.~\ref{fig:5} in Sec.~\ref{events} that we can reduce the impact of this degeneracy because of the fact that energy bins on either side of the oscillation minimum in disappearance events show different behavior with respect to a change in the value of $\Delta m^2_{31}$. For this reason, in the fit, the test value of $\Delta m^2_{31}$ does not get deviate much from its central best-fit value. Nevertheless, we observe that even in the case of rate + shape analysis, the uncertainty in $\Delta m^{2}_{31}$ reduces the potential of DUNE to establish deviation from maximality for $\sin^{2}\theta_{23}$ (true) = 0.455 from 5.28$\sigma$ to 4.91$\sigma$ while going from fixed-parameter case to marginalized scenario. So, an ultra-precise measurement of $\Delta m^{2}_{31}$ in future will undoubtedly enhance DUNE's capability to establish deviation from maximal $\theta_{23}$. For the sake of simplicity, while addressing the advantage due to spectral analysis and the impact of marginalization over oscillation parameters in Fig.~\ref{fig:8},  we do not take into account the systematic uncertainties in the analysis.

\subsubsection{Individual contributions from  neutrino and antineutrino runs}

\begin{figure}[htb!]
	\centering
	\includegraphics[width = 0.49\linewidth,height=0.51\linewidth]{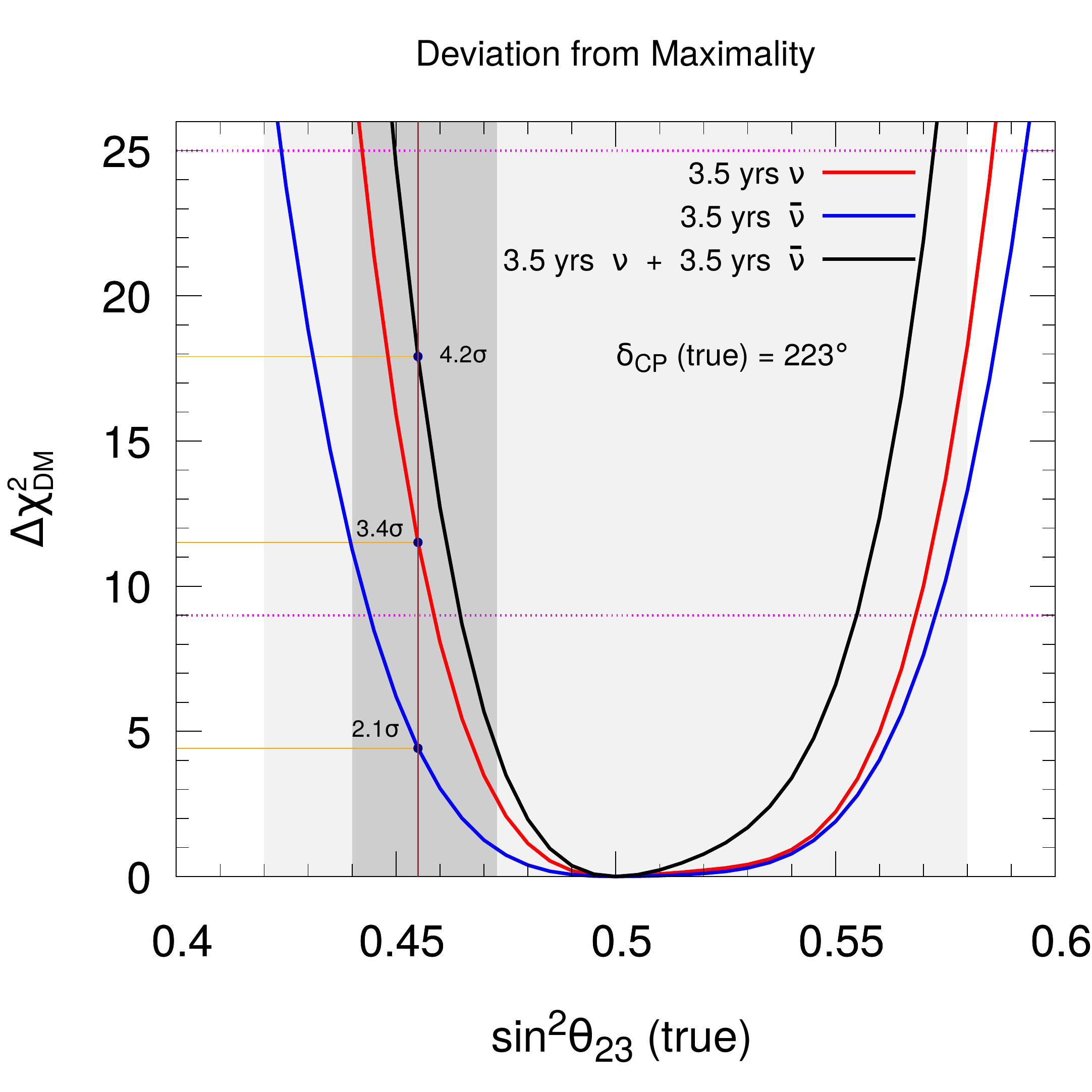}
	\includegraphics[width = 0.49\linewidth,height=0.51\linewidth]{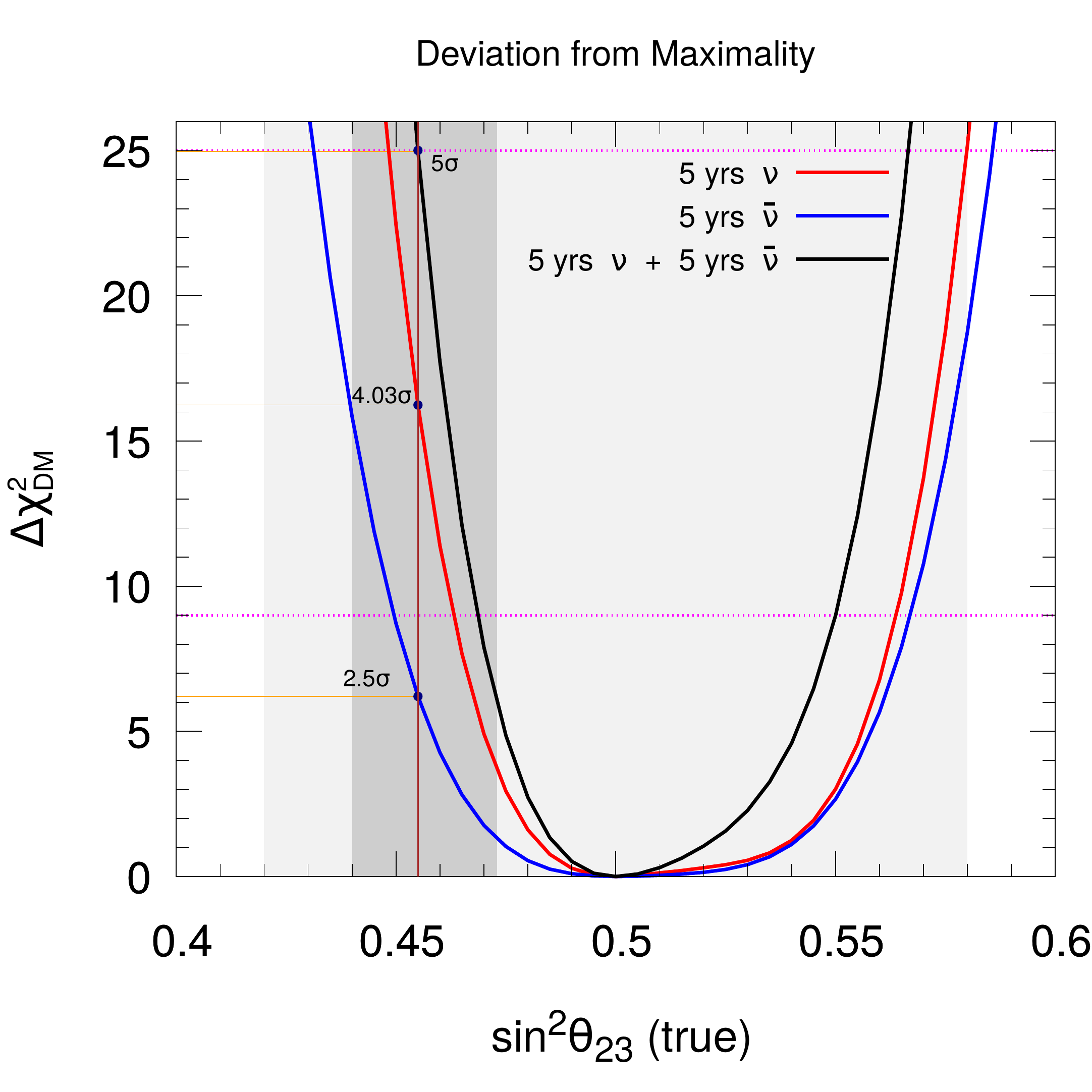}
	\caption{\footnotesize{Potential of DUNE to establish the deviation from maximal $\theta_{23}$ as a function of true $\sin^2\theta_{23}$ assuming true NMO and $\delta_{\rm CP}\, \mathrm{(true)} = 223^\circ$. The red, blue, and black curves in the left (right) panel are drawn assuming 3.5 (5) years of neutrino run, 3.5 (5) years of antineutrino run, and the combined 7 (10) years of $\nu + \bar{\nu}$ run, respectively. In the fit, we marginalize over the current 3$\sigma$ range of $\Delta m^{2}_{31}$ and $\delta_{\rm CP}$, while keeping rest of the oscillation parameters fixed at their present best-fit values as shown in Table~\ref{table:one}. The dark (light)-shaded grey area shows the currently allowed $1\sigma$ $(2\sigma)$ region in $\sin^{2}\theta_{23}$ as obtained in the global fit study~\cite{Capozzi:2021fjo} assuming NMO with the best-fit value of $\sin^{2}\theta_{23} = 0.455$ as shown by vertical brown line. The horizontal orange lines show the sensitivity (experessed in $\sigma = \sqrt{\Delta\chi^2_{\rm DM}}$) due to individual runs for the current best-fit value of $\sin^2\theta_{23}$. The capability of DUNE to establish non-maximal $\theta_{23}$ at 3$\sigma$ ($\Delta \chi^{2}_{\mathrm{DM}} = 9$) and 5$\sigma$ ($\Delta \chi^{2}_{\mathrm{DM}} = 25$) confidence levels are shown by horizontal pink dotted lines.
	}}
	\label{fig:9}
\end{figure} 

In Fig.~\ref{fig:9}, we demonstrate the capability of DUNE to establish non-maximal $\theta_{23}$ assuming true NMO and $\delta_{\rm CP}\, \mathrm{(true)} = 223^\circ$. The red, blue, and black curves in the left (right) panel are drawn assuming 3.5 (5) years of neutrino run, 3.5 (5) years of antineutrino run, and the combined 7 (10) years of $\nu + \bar{\nu}$ run, respectively. We observe that the sensitivity of DUNE to exclude maximal $\theta_{23}$ gets improved significantly when we combine the data from both neutrino and antineutrino modes (see black curves) as compared to the stand-alone neutrino (see red curves) or antineutrino (see blue curves) run. Mostly, the data from neutrino run contributes in the combined analysis due to their superior statistics. We notice from the left panel that a 2.1$\sigma$, 3.4$\sigma$, and 4.2$\sigma$ determination of non-maximal $\theta_{23}$ is possible in DUNE considering 3.5 years of $\bar{\nu}$ run, 3.5 years of $\nu$ run, and the combined 3.5 years $\nu$ + 3.5 years $\bar{\nu}$ run, respectively assuming the present best-fit values of $\sin^{2}\theta_{23}$ (0.455) and $\delta_{\rm CP}$ (223$^\circ$) as their true choices and with true NMO. In the right panel, the sensitivities get improved to 2.5$\sigma$, 4$\sigma$, and 5$\sigma$ with 5 years of $\bar{\nu}$ run, 5 years of $\nu$ run, and the combined 5 years $\nu$ + 5 years $\bar{\nu}$ run, respectively.

\subsubsection{Performance as a function of exposure}

\begin{figure}[htb!]
	\centering
	\includegraphics[width = 0.7\linewidth]{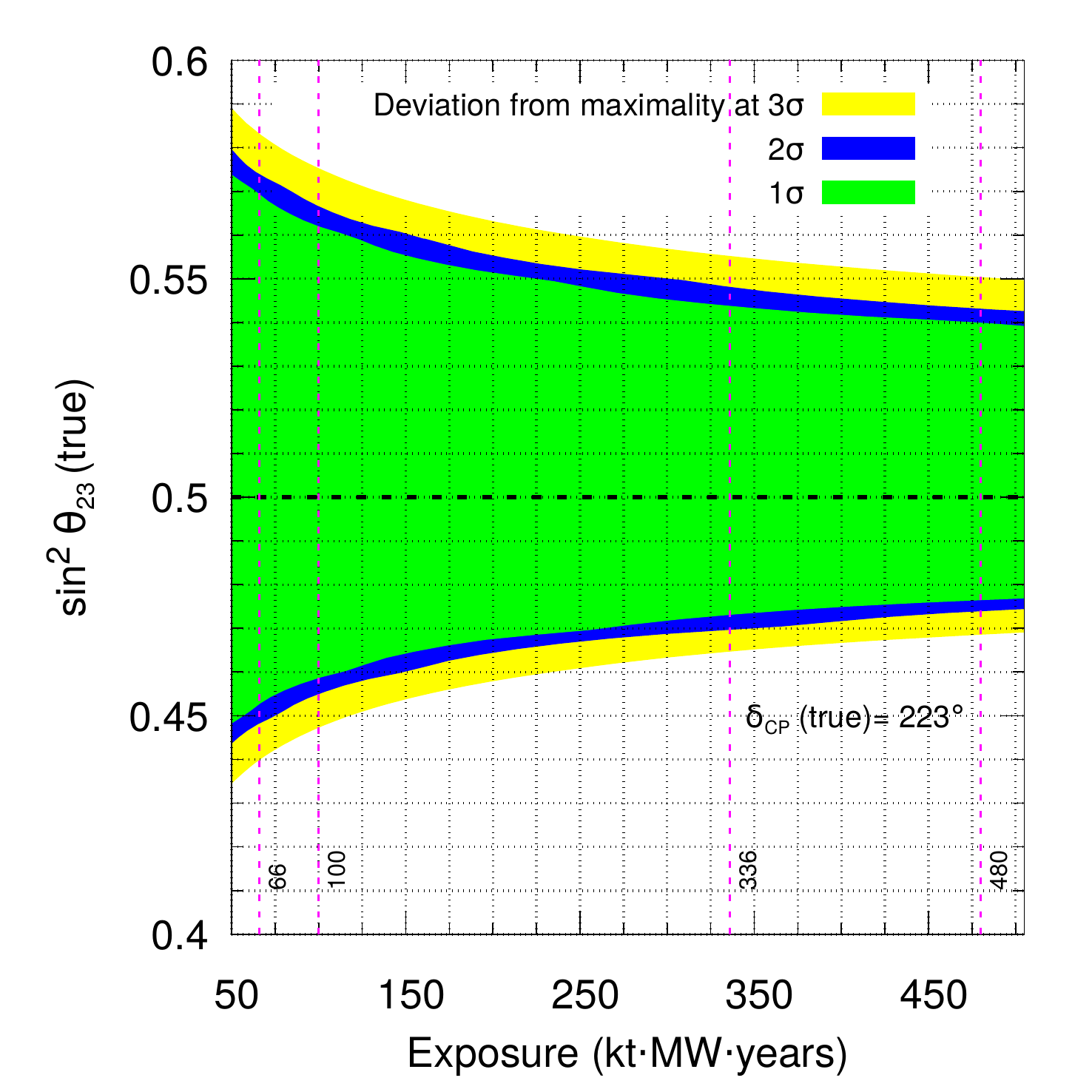}
	\caption{\footnotesize{The discovery of true values of non-maximal $\sin^{2}\theta_{23}$ as a function of exposure (kt$\cdot$MW$\cdot$years) at $3\sigma$ (yellow curves), $2\sigma$ (blue curves), and $1\sigma$ (green curves) confidence levels. For a given exposure, we assume equal run-time in both $\nu$ and $\bar{\nu}$ modes. We consider true NMO and $\delta_{\mathrm{CP}}$ (true) = 223$^{\circ}$. We marginalize over $\delta_{\mathrm{CP}}$ and $\Delta m^{2}_{31}$ in the fit in their allowed 3$\sigma$ ranges as given in Table~\ref{table:one}.}
	}
	\label{fig:10}
\end{figure}

 The DUNE collaboration is planning to adopt an incremental approach where they will gradually increase the exposure by adding the new detector modules to their setup and will also upgrade the beam power from 1.2 MW to 2.4 MW after 6 years \cite{DUNE:2020jqi,DUNE:2021mtg} 
. This staging approach is well justified in light of the challenges that appear while operating a high-power superbeam and in constructing a massive underground 40 kt liquid argon detector. A nominal deployment plan is discussed in Ref. \cite{DUNE:2020jqi}, where the collaboration plans to start the experiment with two far detector (FD) modules having a total fiducial mass of 20 kt and with a beam power of 1.2 MW. After one year, they plan to add one more FD module of 10 kt fiducial mass and after two more years, they will add another 10 kt FD module to have the total fiducial mass of 40 kt. After operating the experiment for six years with a beam power of 1.2 MW, there is also a plan to upgrade the beam power to 2.4 MW \cite{DUNE:2020jqi}.

In Fig.~\ref{fig:10}, we exhibit the performance of DUNE to establish the possible deviation of true values of $\sin^{2}\theta_{23}$ from MM choice ($\sin^{2}\theta_{23}^{\mathrm{test}}$ = 0.5) in the fit as a function of exposure expressed in the units of kt$\cdot$MW$\cdot$years. We show the results at 3$\sigma$ (see yellow curves), 2$\sigma$(see blue curves), and 1$\sigma$ (see green curves) confidence levels assuming $\delta_{\mathrm{CP}}$ (true) = 223$^{\circ}$, and true NMO. While obtaining the results, we marginalize over $\delta_{\mathrm{CP}}$ and $\Delta m^{2}_{31}$ in their presently allowed 3$\sigma$ ranges as given in Table~\ref{table:one}. We see a significant improvement in the discovery of a non-maximal $\theta_{23}$ while increasing the exposure from 50 kt$\cdot$MW$\cdot$years to 100 kt$\cdot$MW$\cdot$years. In Fig.~\ref{fig:10} we show for the first time the true values of $\sin^{2}\theta_{23}$ that DUNE can distinguish from $\sin^{2}\theta_{23}^{\mathrm{test}}= 0.5$ using these exposures at various confidence levels (see dashed vertical lines). While further increasing the exposure from 100 kt$\cdot$MW$\cdot$years to 336kt$\cdot$MW$\cdot$years  which is our benchmark choice, we see a marginal increment in the performance. Note that we hardly see any improvement in the sensitivity if we increase the exposure further which suggest that the statistics is not a limiting factor anymore and any possible reduction in the systematic uncertainties may enhance the results further.

\subsection{Octant of $\theta_{23}$} 

\begin{figure}[htb!]
	\centering
	\includegraphics[width=0.49\linewidth]{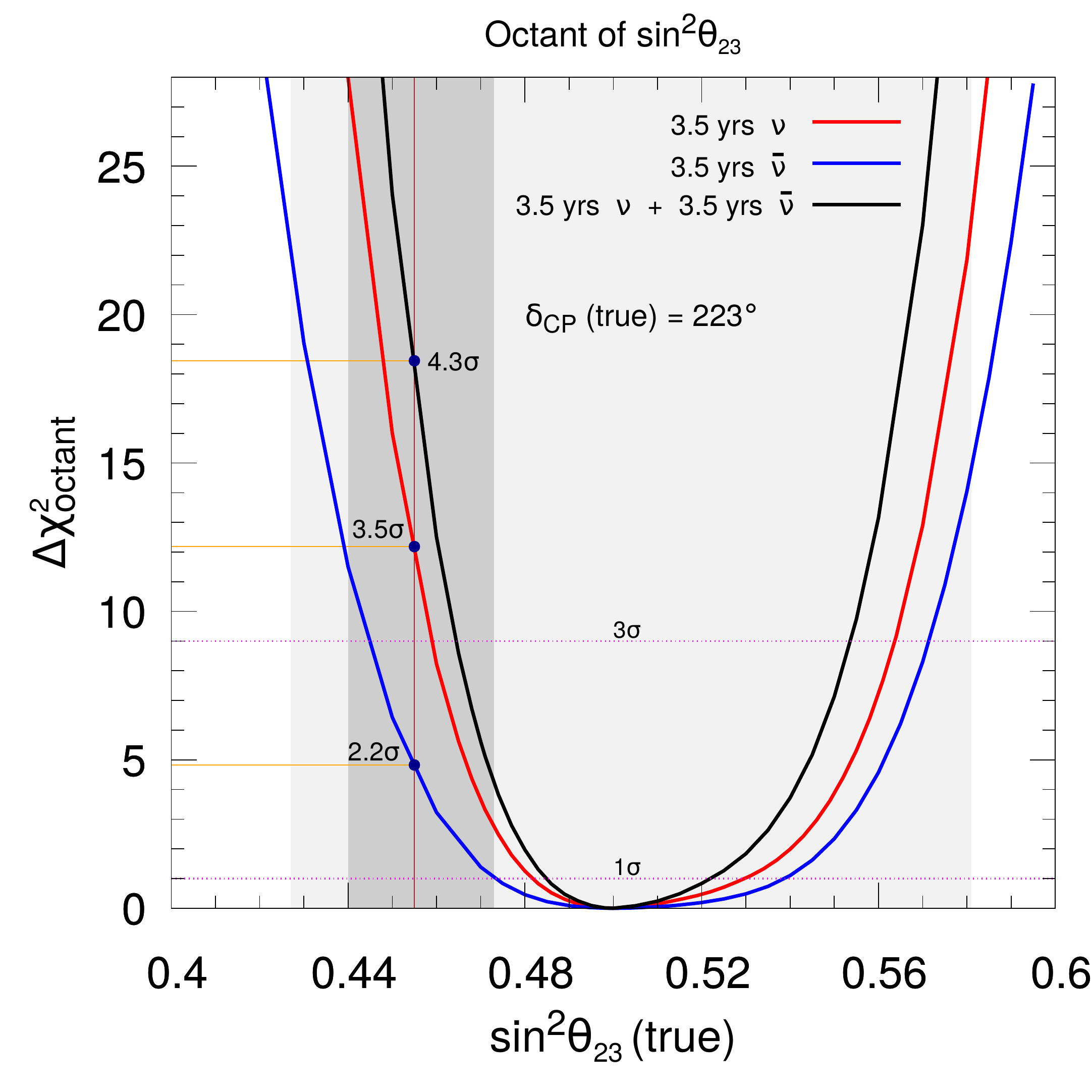}
	\includegraphics[width=0.49\linewidth]{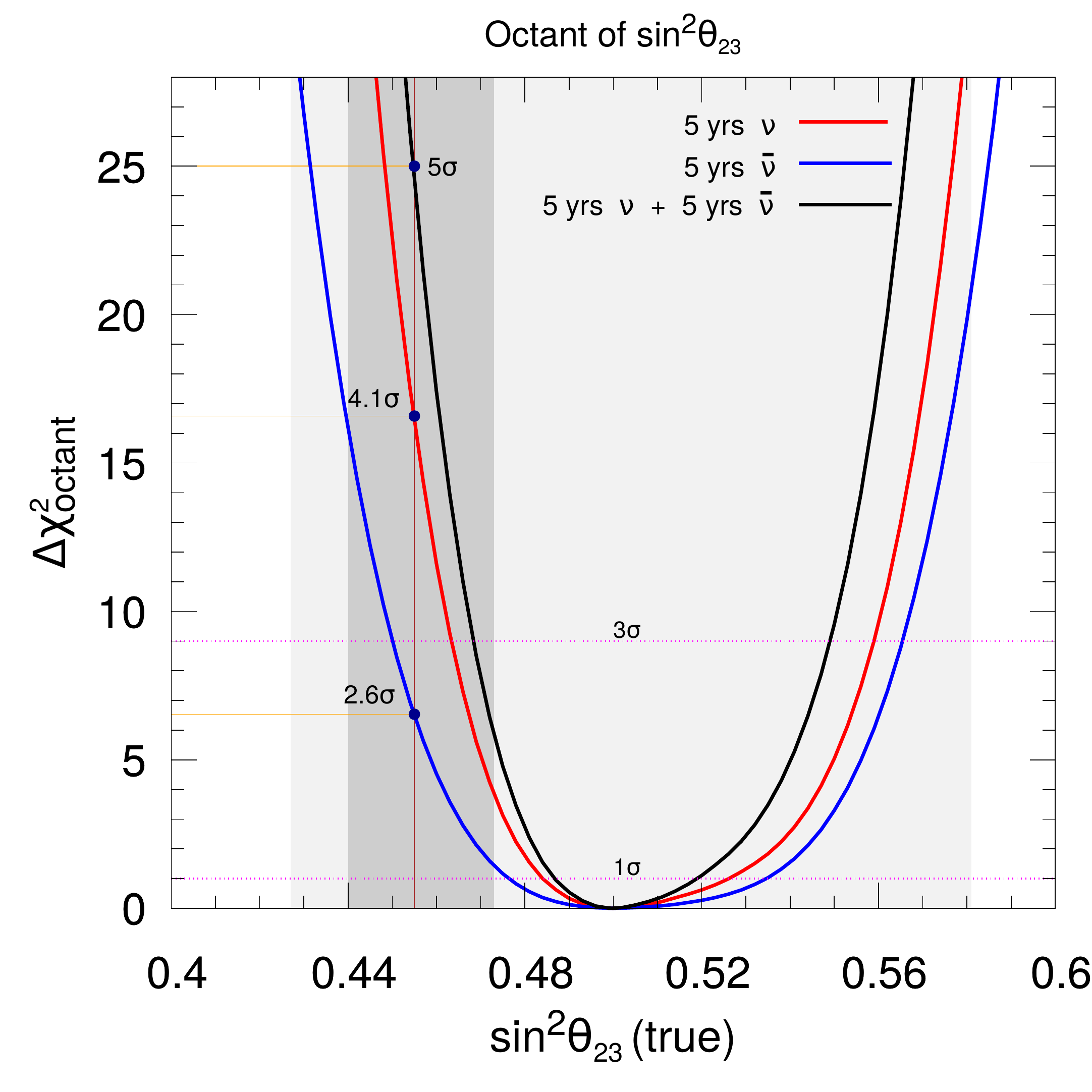}
	\caption{\footnotesize{Octant discovery potential of DUNE as a function of true $\sin^2\theta_{23}$ assuming true NMO and $\delta_{\rm CP}\, \mathrm{(true)} = 223^\circ$. The red, blue, and black curves in the left (right) panel are drawn assuming 3.5 (5) years of neutrino run, 3.5 (5) years of antineutrino run, and the combined 7 (10) years of $\nu + \bar{\nu}$ run, respectively. In the fit, we marginalize over the current 3$\sigma$ range of $\Delta m^{2}_{31}$ and $\delta_{\rm CP}$, while keeping rest of the oscillation parameters fixed at their present best-fit values as shown in Table~\ref{table:one}. The dark (light)-shaded grey area shows the currently allowed $1\sigma$ $(2\sigma)$ region in $\sin^{2}\theta_{23}$ as obtained in the global fit study~\cite{Capozzi:2021fjo} assuming NMO with the best-fit value of $\sin^{2}\theta_{23} = 0.455$ as shown by vertical brown line. The horizontal orange lines show the sensitivity (experessed in $\sigma = \sqrt{\Delta\chi^2_{\rm octant}}$) due to individual runs for the current best-fit value of $\sin^2\theta_{23}$. The octant discovery potential at 1$\sigma$ ($\Delta \chi^{2}_{\mathrm{octant}} = 1$) and 3$\sigma$ ($\Delta \chi^{2}_{\mathrm{octant}} = 9$) confidence levels are shown by horizontal pink dotted lines.}}
	\label{fig:11}
\end{figure}

In this subsection, we study the potential of DUNE to resolve the octant of 2-3 mixing angle. We define $\Delta\chi^2_{\rm octant}$ in the following fashion
\begin{equation}
\Delta \chi^2_{\text{\rm octant}}(\zeta) = \underset{(\delta_{\mathrm{CP}},~\Delta m^{2}_{31},~ \kappa_{s},~\kappa_{b})}{\mathrm{min}}\left\{ \chi^2\left(\zeta^{\mathrm{true}}\right) - \chi^2\left(\zeta^{\mathrm{test}} \right)\right\}.  
\end{equation}
Here, $\zeta^{\text{true}}$ is the true value of $\sin^2\theta_{23}$ in lower or upper octant  and $\zeta^{\text{test}}$ is the test value of $\sin^2\theta_{23}$ in opposite octant including the test value of $\sin^2\theta_{23}= 0.5$. $\delta_{\mathrm{CP}}$ and $\Delta m^{2}_{31}$ are the oscillation parameters over which $\Delta\chi^2_{\mathrm{octant}}$ has been marginalized in the fit, while $\kappa_{s},~\text{and}~ \kappa_{b}$ are the systematic pulls~\cite{Huber:2002mx,Fogli:2002pt,Gonzalez-Garcia:2004pka} on signal and background, respectively.

In Fig.~\ref{fig:11}, we show $\Delta\chi^2_{\rm octant}$ as a function of true $\sin^2\theta_{23}$, where, for each true value of $\sin^{2}\theta_{23}$, we consider test values of $\sin^2\theta_{23}$ in its present 3$\sigma$ range in the opposite octant including $\sin^2\theta_{23}\, \mathrm{(test)}\,= 0.5$ in the fit and pick up the minimum value of $\Delta \chi^{2}_{\mathrm{octant}}$. Here, we consider true NMO and $\delta_{\rm CP}\, \mathrm{(true)} = 223^\circ$. The red, blue, and black curves in the left (right) panel are obtained assuming 3.5 (5) years of neutrino run, 3.5 (5) years of antineutrino run, and the combined 7 (10) years of $\nu + \bar{\nu}$ run, respectively. In the fit, we marginalize over the present 3$\sigma$ range of $\Delta m^{2}_{31}$ and $\delta_{\rm CP}$, while keeping rest of the oscillation parameters fixed at their present best-fit values as shown in Table~\ref{table:one}. The dark (light)-shaded grey area shows the currently allowed $1\sigma$ $(2\sigma)$ region in $\sin^{2}\theta_{23}$ as obtained in the global fit study~\cite{Capozzi:2021fjo} assuming NMO with the best-fit value of $\sin^{2}\theta_{23} = 0.455$ as shown by vertical brown line. The horizontal orange lines show the sensitivity (experessed in $\sigma = \sqrt{\Delta\chi^2_{\rm octant}}$) due to individual runs for the current best-fit value of $\sin^2\theta_{23}$. The octant discovery potential at 1$\sigma$ ($\Delta \chi^{2}_{\mathrm{octant}} = 1$) and 3$\sigma$ ($\Delta \chi^{2}_{\mathrm{octant}} = 9$) confidence levels are shown by horizontal pink dotted lines.

It is evident from Fig.~\ref{fig:11} that the combined data from neutrino and antineutrino modes (see black curves) significantly improve the result by breaking the octant - $\delta_{\rm CP}$ degeneracy as discussed before in~\cite{Agarwalla:2013ju}. Assuming the current best-fit values of $\sin^{2}\theta_{23}$ = 0.455 and $\delta_{\mathrm{CP}}$ = 223$^{\circ}$ as their true choices and with true NMO, the octant of $\theta_{23}$ can be settled in DUNE at 4.3$\sigma$ (5$\sigma$) using 336 (480) kt$\cdot$MW$\cdot$years of exposure which corresponds to 7 years (10 years) of data taking with equal sharing in neutrino and antineutrino modes. A 3$\sigma$ (5$\sigma$) resolution of $\theta_{23}$ octant is possible in DUNE with an exposure of 336 kt$\cdot$MW$\cdot$years if the true value of $\sin^{2}\theta_{23}$ $\lesssim~0.462~(0.450)$ or $\sin^{2}\theta_{23}$ $\gtrsim~0.553~(0.569)$ assuming true NMO and $\delta_{\rm CP}~(\rm true)~ = 223^{\circ}$. The same is possible with 480 kt$\cdot$MW$\cdot$years of exposure if the true value of $\sin^{2}\theta_{23}$ $\lesssim~0.466~(0.454)$ or $\sin^{2}\theta_{23}$ $\gtrsim~0.548~(0.565)$.

\subsection{Precision measurements of $\sin^2\theta_{23}$ and $\Delta m^2_{31}$}
\label{precisionapp}

\begin{figure}[htb!]
\includegraphics[width=0.49\linewidth]{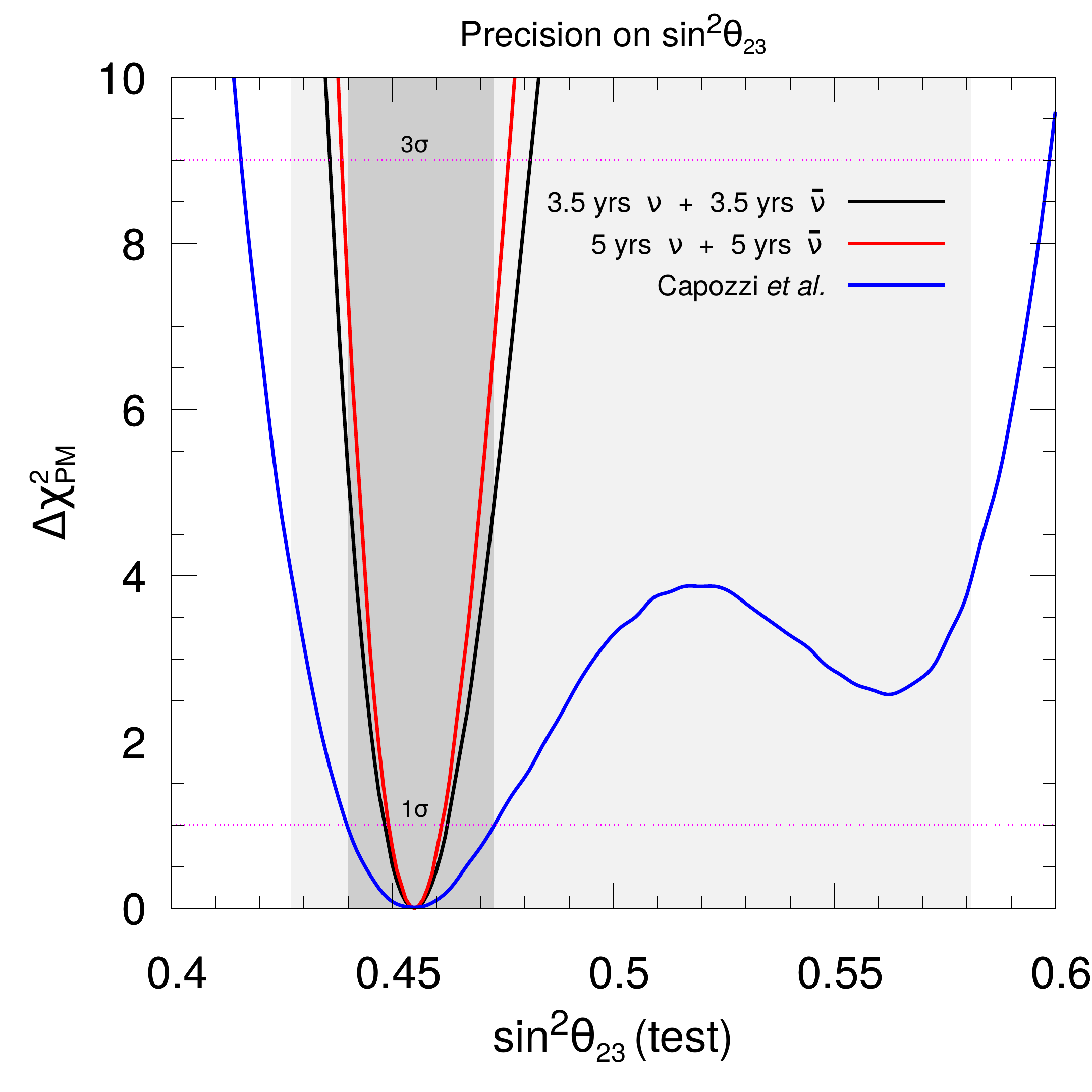}
\includegraphics[width=0.49\linewidth]{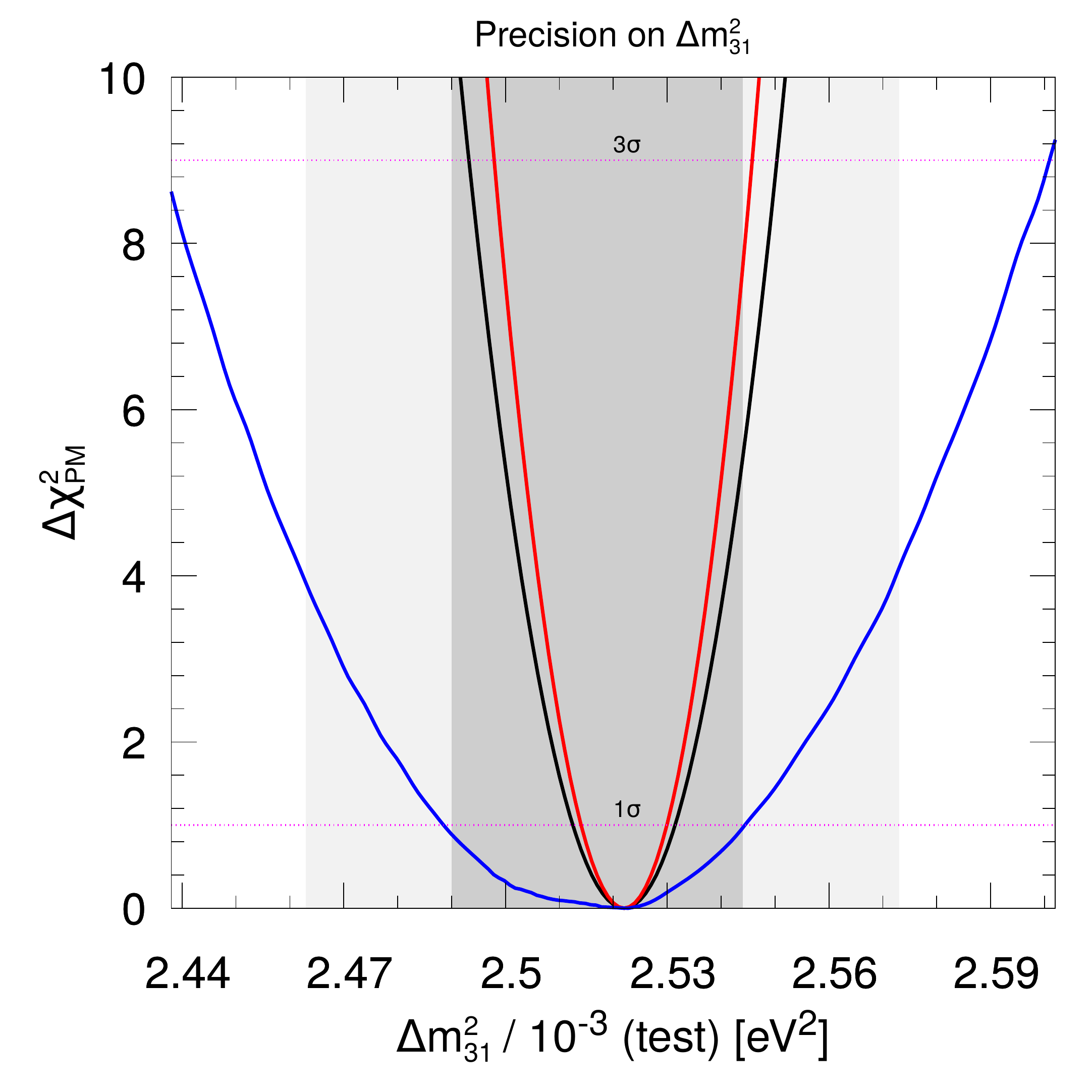}
\caption{\footnotesize{The left panel shows $\Delta\chi^2_{\rm PM}$ around $\sin^2\theta_{23}\, \mathrm{(true)}\, = 0.455$ and the right panel depicts $\Delta \chi^{2}_{\mathrm{PM}}$ for $\Delta m^2_{31}\, \mathrm{(true)} = 2.522\times 10^{-3}~\rm eV^2$ assuming true NMO and $\delta_{\rm CP}\, \mathrm{(true)} = 223^\circ$. The black (red) curves show the precision with 7 (10) years of exposure, equally divided in neutrino and antineutrino modes. Blue lines portray the present precision from the global fit study~\cite{Capozzi:2021fjo}. In the fit, we marginalize over the current 3$\sigma$ range of $\Delta m^{2}_{31}$ $(\sin^{2}\theta_{23})$ in the left (right) panel. In both the panels, we also marginalize over the current 3$\sigma$ range of $\delta_{\rm CP}$, while keeping rest of the oscillation parameters fixed at their present best-fit values as shown in Table~\ref{table:one}. The dark (light)-shaded grey area shows the currently allowed $1\sigma$ $(2\sigma)$ region in $\sin^{2}\theta_{23}$ ($\Delta m^2_{31}$) in the left (right) panel as obtained in the global fit study~\cite{Capozzi:2021fjo} assuming NMO. The precision at 1$\sigma$ ($\Delta \chi^{2}_{\mathrm{PM}} = 1$) and 3$\sigma$ ($\Delta \chi^{2}_{\mathrm{PM}} = 9$) confidence levels are shown by horizontal pink dotted lines.}}
\label{fig:12}
\end{figure}

In this subsection, we estimate the sensitivity of the DUNE experiment to constrain the oscillation parameters $\Delta m^2_{31}$ and $\sin^2\theta_{23}$. We calculate the relative $1\sigma$-precision with which DUNE can measure these oscillation parameters and compare them with the existing constraints. 
In Fig.~\ref{fig:12}, we show $\Delta\chi^2_{\rm PM}$ as a function of the test oscillation parameters $\sin^2\theta_{23}$ (left panel) and $\Delta m^2_{31}$ (right panel). $\Delta\chi^2_{\rm PM}$ is computed as follows:

\begin{equation}
	\Delta \chi^2_{\text{PM}}\,(\zeta^{\rm test}) = \underset{(\vec{\lambda},~ \kappa_{s},\kappa_{b})}{\mathrm{min}}\left\{ \chi^2\left(\zeta^{\mathrm{test}}\right) - \chi^2\left(\zeta^{\mathrm{true}} \right)\right\}.
	\label{precision-chisq}
	\end{equation}
	
Here, $\zeta^{\text{true}}$ is the best-fit value of the oscillation parameter under consideration and $\zeta^{\text{test}}$ represents a test value of the same oscillation parameter in its currently allowed 3$\sigma$ range~\cite{Capozzi:2021fjo}. $\vec{\lambda}$ denotes the set of oscillation parameters over which we perform marginalization in the fit for a given analysis. $\kappa_{s}$ and $\kappa_{b}$ are the systematic pulls on signal and background, respectively. In the fit, we minimize $\chi^2\left(\zeta^{\rm test}\right)$ over the systematic uncertainties to obtain $\Delta \chi^2_{\text{PM}}(\zeta^{\rm test})$. We show the results in Table~\ref{table:four}. The relative 1$\sigma$ precision in the measurement of oscillation parameters $\zeta$ is estimated as follows:

\begin{equation}
p(\zeta)\, =\, \frac{\zeta^{\rm max} - \zeta^{\rm min}}{6.0 \,\times \, \zeta^{\rm true}}\, \times \, 100\%\, .
\label{precision1}
\end{equation} 

Here, $\zeta^{\rm max}$ and $\zeta^{\rm min}$ represent the allowed $3\sigma$ upper and lower bounds, respectively. In the fourth column of Table~\ref{table:four}, we mention the current relative 1$\sigma$ precision on $\sin^2\theta_{23}$ and $\Delta m^2_{31}$ from the recent global fit study~\cite{Capozzi:2021fjo}. The achievable relative 1$\sigma$ precision\footnote{The achievable precision on atmospheric oscillation parameters using the full exposure of T2K and NO$\nu$A is discussed in Ref.~\cite{Agarwalla:2013qfa}.} on $\Delta m^2_{31}$ from the upcoming medium-baseline reactor experiment JUNO~\cite{JUNO:2015zny} is mentioned in the fifth column. We observe that DUNE can improve the current relative 1$\sigma$ precision on $\sin^2\theta_{23}$ ($\Delta m^2_{31}$) by a factor of 4.4 (2.8) using 3.5 years of neutrino and 3.5 years of antineutrino runs. The total exposure of 10 years equally shared in neutrino and antineutrino modes further improves the precision on these parameters. 

\begin{table}[htb!]
\centering
\resizebox{\columnwidth}{!}{%
\begin{tabular}{|c||c|c|c|c|}
\hline
\hline
\multirow{3}{*}{Parameter} & \multicolumn{4}{c|}{Relative 1$\sigma$ precision (\%)}\\
\cline{2-5}
& DUNE & DUNE &  \multirow{2}{*}{Capozzi $et~al.$ \cite{Capozzi:2021fjo}} & \multirow{2}{*}{JUNO \cite{EPS-HEP-Conference2021,JUNO:2015zny}} \\
& ($3.5~\nu + 3.5~\bar{\nu}$) yrs  & ($5~\nu + 5~\bar{\nu}$) yrs & &  \\
\hline
$\sin^{2}\theta_{23}$ & 1.53 & 1.31 & 6.72 & ---\\
\hline
$\Delta m^{2}_{31}$ & 0.39 & 0.31 & 1.09 & 0.50\\
\hline
\hline
\end{tabular}
}
\caption{\footnotesize{Relative 1$\sigma$ precision on $\sin^2\theta_{23}$ and $\Delta m^2_{31}$ around the true choices $\sin^{2}\theta_{23} = 0.455$ and $\Delta m^2_{31} = 2.522\times 10^{-3}~\rm eV^2$. The second and third columns show the performance of DUNE with 7 and 10 years of exposures, respectively, equally divided in neutrino and antineutrino modes. The fourth coulumn depicts the current relative 1$\sigma$ precision on these parameters from the global fit study~\cite{Capozzi:2021fjo}. The achievable precision on $\Delta m^{2}_{31}$ from the upcoming JUNO experiment~\cite{EPS-HEP-Conference2021,JUNO:2015zny} is mentioned in the fifth column. Note that JUNO is insensitive to $\theta_{23}$. }}
\label{table:four}
\end{table}

\begin{figure}[htb!]
\centering
\includegraphics[width=0.7\linewidth]{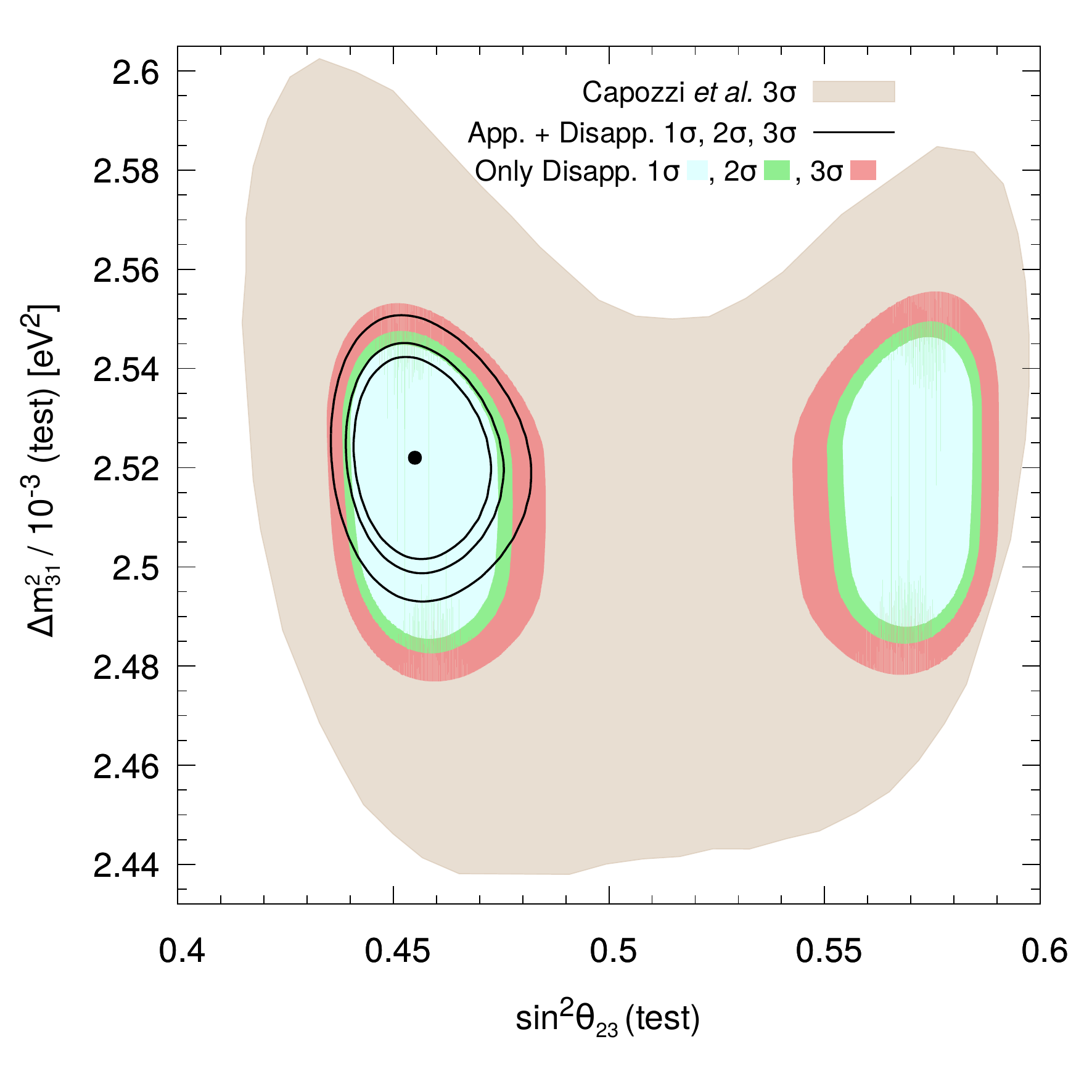}
\caption{\footnotesize{Allowed regions in the test ($\sin^2\theta_{23}$ - $\Delta m^2_{31}$) plane. The shaded light-grey region shows the current $3\sigma$ allowed ranges from Ref. \cite{Capozzi:2021fjo}. Blue, green, and red contours portray the $1\sigma$, $2\sigma$, and $3\sigma$ allowed regions, respectively using 3.5 years of neutrino and 3.5 years of antineutrino disappearance data in DUNE. Solid black lines exhibit the performance of DUNE at $1\sigma$, $2\sigma$, and $3\sigma$ combining both appearance and disappearance data in 3.5 years of neutrino and 3.5 years of antineutrino runs. The black dot depicts the true choices of $\sin^2\theta_{23} = 0.455$ and $\Delta m^2_{31} = 2.522\times 10^{-3}~\rm eV^2$ assuming true NMO and $\delta_{\rm CP}$ = 223$^\circ$. In the fit, we marginalize over the current 3$\sigma$ range of $\delta_{\rm CP} = [139^{\circ} : 355^{\circ}]$ and the rest of the oscillation parameters are kept fixed at their present best-fit values as shown in Table~\ref{table:one}.}}
\label{fig:13}
\end{figure}

In Fig.~\ref{fig:13}, we show the allowed regions in the test $\sin^2\theta_{23}$ - $\Delta m^2_{31}$ at 1$\sigma$, 2$\sigma$, and 3$\sigma$ for 1 degree of freedom. The shaded light-grey region shows the currently allowed $3\sigma$ values due to the existing oscillation data~\cite{Capozzi:2021fjo}. As can be observed, the allowed region is quite large especially in the parameter $\sin^2\theta_{23}$. Further, both maximal mixing and wrong octant solutions are allowed at the $3\sigma$ C.L. The current best-fit to the global data is the lower octant solution of $\sin^2\theta_{23}=0.455$ shown by the black dot. We give the results for DUNE with 7 years of equally shared exposure in neutrino and antineutrino modes, with only disappearance data (blue, green, and red contours) and combined appearance and disappearance data (solid black contours) considering the true value of $\sin^2\theta_{23}$ to be $0.455$. It can be seen that the disappearance data significantly constrains the allowed range of $\sin^2\theta_{23}$. However, it is still unable to rule out the wrong octant solution even at $1\sigma$. On the other hand, though the appearance data only marginally improves the $\sin^2\theta_{23}$ precision in the right octant, it plays the main role in completely ruling out the wrong octant solution. We also find that the combined appearance and disappearance data improves the precision in measurement of both $\sin^{2}\theta_{23}$ in correct octant and $\Delta m^2_{31}$ when compared to the precision when obtained with only disappearance data.

\begin{figure}[htb!]
	\centering
	\includegraphics[width=\linewidth]{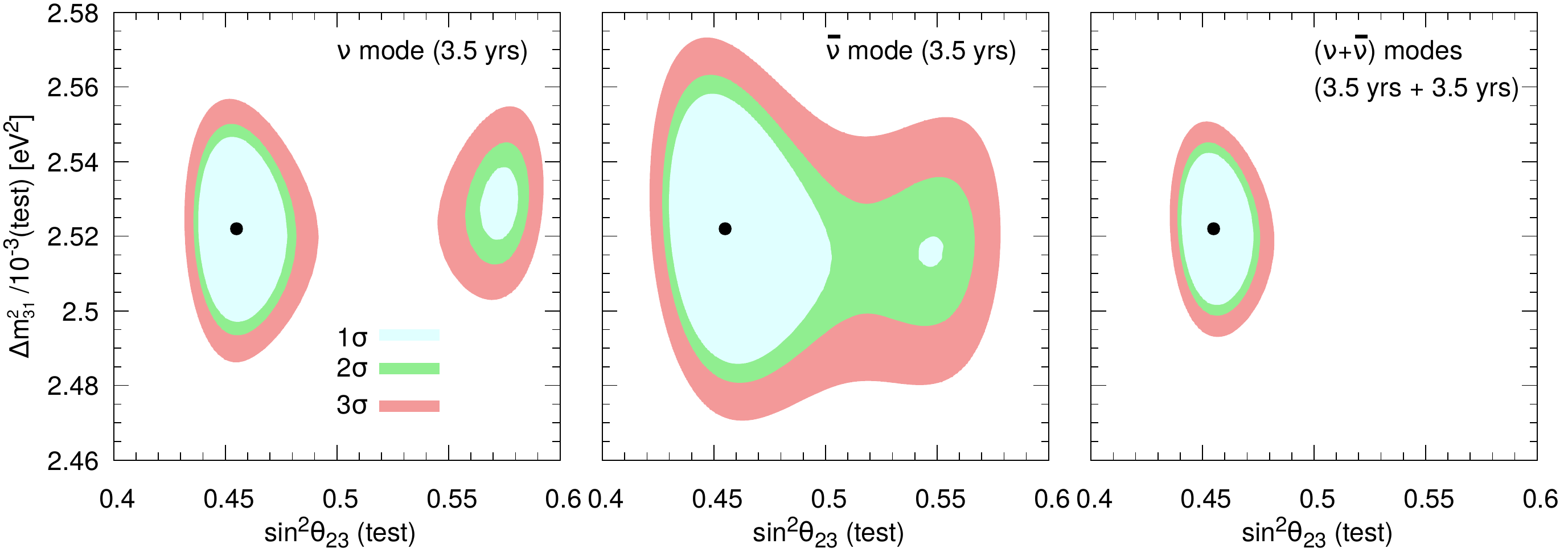}
	\caption{\footnotesize{ Allowed regions in the test ($\sin^2\theta_{23}$ - $\Delta m^2_{31}$) plane at 1$\sigma$ (blue), 2$\sigma$ (green), and 3$\sigma$ (red) C.L. combining appearance and disappearance data in DUNE.  Left (Middle) panel is for 3.5 years of neutrino (antineutrino) run. The right panel shows the performance of combined neutrino (3.5 years) and antineutrino (3.5 years) runs. The black dot depicts the true choices of $\sin^2\theta_{23} = 0.455$ and $\Delta m^2_{31} = 2.522\times 10^{-3}~\rm eV^2$ assuming true NMO and $\delta_{\rm CP}$ = 223$^\circ$. In the fit, we marginalize over the current 3$\sigma$ range of $\delta_{\rm CP} = [139^{\circ} : 355^{\circ}]$ and the rest of the oscillation parameters are kept fixed at their present best-fit values as shown in Table~\ref{table:one}.}}
	\label{fig:14}
\end{figure}

In Fig.~\ref{fig:14}, we show the benefit of having data from both neutrino and antineutrino modes, the merit of which was discussed elaborately in the context of T2K and NO$\nu$A in Ref.~\cite{Agarwalla:2013ju}. The left panel explores the allowed region in the test ($\sin^2\theta_{23}-\Delta m^2_{31}$) plane considering 3.5 years of neutrino run and having contributions from both disappearance and appearance channels. The middle panel depicts the same for 3.5 years of antineutrino run. In the right panel, we demonstrate how the allowed region in the test ($\sin^2\theta_{23}-\Delta m^2_{31}$) plane gets shrinked when we combine the data from 3.5 years of neutrino and 3.5 years of antineutrino runs. From the left and middle panels, we observe that the prospective data from only neutrino or only antineutrino run cannot rule out the wrong octant solution even at $1\sigma$ confidence level, while with only antineutrino run, even maximal mixing solution of $\theta_{23}$ is allowed at $2\sigma$. However, from the right panel of Fig.~\ref{fig:14}, it is evident that the data from both neutrino and antineutrino runs are quite effective in ruling out the wrong octant solution as well as the maximal mixing at $3\sigma$ confidence level. This happens because the combined neutrino and antineutrino data can resolve the octant\,-\,$\delta_{\mathrm{CP}}$ degeneracy~\cite{Agarwalla:2013ju} that exists in the stand-alone neutrino or antineutrino data. We also notice that the allowed regions for $\sin^2\theta_{23}\, \mathrm{and}\,\Delta m^2_{31}$ around the correct octant get reduced when we combine the data from both neutrino and antineutrino modes (see right panel). The increase in statistics due to both neutrino and antineutrino runs and the possible complementarity between them lead to these improvements in the sensitivity.

\section{Summary and conclusions}
\label{conclusion}

We have achieved remarkable precision on the solar ($\theta_{12}, \, \Delta m^{2}_{21}$) and atmospheric ($\theta_{23}, \, \Delta m^{2}_{31}$) oscillation parameters over the last few years. According to Ref.~\cite{Capozzi:2021fjo}, the current relative 1$\sigma$ errors on $\sin^{2}\theta_{12}, \, \Delta m^{2}_{21}, \, \sin^{2}\theta_{23}$, and $\Delta m^{2}_{31}$ are 4.5\%, 2.3\%, 6.7\%, and 1.1\%, respectively. The recent hints for normal mass ordering (at $\sim$ 2.5$\sigma$), as well as for lower octant $\theta_{23}$ ($\theta_{23} <$ 45$^{\circ}$) and for $\delta_{\mathrm{CP}}$ in the lower half-plane ($\sin \delta_{\mathrm{CP}} < 0$) signify major developments in the three-flavor neutrino oscillation paradigm. The high-precision measurement of $\theta_{13}$ from the Daya Bay reactor experiment and the possible complementarities between the recent Super-K Phase I-IV atmospheric data and the appearance and disappearance data from the ongoing long-baseline oscillation experiments - NO$\nu$A and T2K, play an important role in providing these crucial hints. An accurate measurement of $\theta_{23}$ and resolution of its octant (if $\theta_{23}$ turns out to be non-maximal) are crucial to transform these preliminary hints into 5$\sigma$ discoveries. A discovery of non-maximal $\theta_{23}$ at high confidence level will serve as a crucial input to the theories of neutrino masses and mixings and it will certainly be a major breakthrough in addressing the the age-old flavor problem. In this paper, we explore in detail the sensitivities of the upcoming high-precision long-baseline experiment DUNE to establish the possible deviation from maximal $\theta_{23}$ and to resolve its octant at high confidence level in light of the recent neutrino oscillation data. 

We start the paper by showing the possible correlations and degeneracies among the oscillation parameters $\sin^{2}\theta_{23}, \, \Delta
m^{2}_{31}$, and $\delta_{\mathrm{CP}}$ in the context of $\nu_{\mu} \rightarrow \nu_{\mu}$ disappearance channel and $\nu_{\mu} \rightarrow \nu_{e}$ appearance channel at the probability and event levels. We introduce for the first time, a bi-events plot in the plane of total neutrino and antineutrino disappearance events to demonstrate the impact of $\sin^{2}\theta_{23} - \Delta m^{2}_{31}$ degeneracy in establishing deviation from maximality. Next, we show how the spectral shape information in neutrino and antineutrino disappearance events can play an important role to resolve this degeneracy.

Using the latest simulation details of DUNE~\cite{DUNE:2021cuw}, we observe that a 3$\sigma$ (5$\sigma$) determination of non-maximal $\theta_{23}$ is possible in DUNE with an exposure of 336 kt$\cdot$MW$\cdot$years if the true value of $\sin^2\theta_{23} \lesssim 0.465~(0.450)$ or $\sin^2\theta_{23} \gtrsim 0.554~(0.572)$ for any value of true $\delta_{\mathrm{CP}}$ in the present 3$\sigma$ range and true NMO. DUNE can exclude the maximal mixing solution of $\theta_{23}$ at 4.2$\sigma$ (5$\sigma$) with a total 7 (10) years of run (equally divided in neutrino and antineutrino modes) assuming the present best-fit values of $\sin^{2}\theta_{23}$ (0.455) and $\delta_{\mathrm{CP}}$ (223$^{\circ}$) as their true choices with true NMO. The same can be enhanced to 6.5$\sigma$ (7.7$\sigma$) if we assume $\sin^{2}\theta_{23}$ (true) = 0.44, which is the current 1$\sigma$ lower bound. On the other hand, the sensitivity can be reduced to 2.07$\sigma$ (2.44$\sigma$) if $\sin^{2}\theta_{23}$ (true) turns out to be 0.473, which is the current 1$\sigma$ upper bound.

We study the role that systematic uncertainties play in establishing  deviation from maximality by varying the normalization errors in both appearance and disappearance channels. We explore the contribution that each oscillation channel has on the sensitivity and show how performing a spectral analysis alleviates the possible reduction in sensitivity due to the marginalization over $\Delta m^2_{31}$ that is present when only total event rates are considered. We also explore the effect of exposure and the individual contributions from neutrino and antineutrino modes. 

We notice that both neutrino and antineutrino data are needed to reduce the impact of octant - $\delta_{\mathrm{CP}}$ degeneracy, which in turn allows us to resolve the $\theta_{23}$ octant at a high confidence level. DUNE can settle the issue of $\theta_{23}$ octant at 4.3$\sigma$ (5$\sigma)$ using 336 (480) kt$\cdot$MW$\cdot$years of exposure assuming $\sin^{2}\theta_{23}$ (true) = 0.455, $\delta_{\mathrm{CP}}$ (true) = 223$^{\circ}$, and true NMO. On the other hand, the octant ambiguity of $\theta_{23}$ can be resolved at 3$\sigma$ (5$\sigma$) in DUNE with an exposure of 336 kt$\cdot$MW$\cdot$years if the true value of $\sin^{2}\theta_{23}$ $\lesssim 0.462\, (0.450)$ or $\sin^{2}\theta_{23}$ $\gtrsim 0.553\, (0.569)$ assuming $\delta_{\mathrm{CP}}$ (true) = 223$^{\circ}$ and true NMO. If we increase the exposure to 480 kt$\cdot$MW$\cdot$years (corresponding to 10 years of run), the wrong octant solution can be excluded if $\sin^{2}\theta_{23}$ (true) $\lesssim 0.466 \,(0.454)$ or $\sin^{2}\theta_{23}$ (true) $\gtrsim 0.548 \,(0.565)$ keeping the assumptions on other oscillation parameters same. 

Finally, we quote how accurately DUNE can measure the atmospheric oscillation parameters. We observe that DUNE can improve the current relative 1$\sigma$ precision on $\sin^2\theta_{23}$ ($\Delta m^2_{31}$) by a factor of 4.4 (2.8) using 336 kt$\cdot$MW$\cdot$years of exposure. We analyze how much contribution we obtain from individual appearance and disappearance oscillation channels and also study the importance of having data from both neutrino and antineutrino modes while measuring these parameters.

We hope that this study serves as an important addition to several fundamental physics issues that can be explored by the high-precision long-baseline experiment DUNE and provides a boost to the physics reach of DUNE.


\appendix 

\section{Comparison of global neutrino data analyses and current bounds on the neutrino oscillation parameters}
\label{appendix_a}

	\begin{table}[H]
 \centering
 \begin{tabular}{|c||c|c|c|}
 \hline
 Reference & Esteban $et~al.$ \cite{Esteban:2020cvm, NuFIT} & de Salas $et~al.$ \cite{deSalas:2020pgw} & Capozzi $et~al.$ \cite{Capozzi:2021fjo} \\
 \hline
 \hline
 $\sin^{2}\theta_{12}$ & 0.304$\,^{+0.012}_{-0.012}$ & 0.318$\,^{+0.016}_{-0.016}$ & 0.303$\,^{+0.013}_{-0.013}$ \\
 3$\sigma$ range & 0.269 $\rightarrow$ 0.343 & 0.271 $\rightarrow$ 0.369 & 0.263 $\rightarrow$ 0.345 \\
 \hline \hline
 $\sin^{2}\theta_{13}$ (NMO) & 0.02246$\,^{+0.00062}_{\,-0.00062}$ & 0.02200$^{\,+0.00069}_{\,-0.00062}$ & 0.02230$\,^{+0.00070}_{-0.00060}$  \\
 3$\sigma$ range & 0.02060 $\rightarrow$ 0.02435 & 0.02000 $\rightarrow$ 0.02405 & 0.02040 $\rightarrow$ 0.02440 \\
 \hline
 $\sin^{2}\theta_{13}$ (IMO) & 0.02241$\,^{+0.00074}_{-0.00062}$ & 0.02250$\,^{+0.00064}_{-0.00070}$ & 0.02230$\,^{+0.00060}_{-0.00060}$ \\
 3$\sigma$ range & 0.02055 $\rightarrow$ 0.02457 & 0.02018 $\rightarrow$ 0.02424 & 0.02030 $\rightarrow$ 0.02450 \\
 \hline \hline
 $\sin^{2}\theta_{23}$ (NMO) & 0.450$\,^{+0.019}_{-0.016}$ & 0.574$\,^{+0.014}_{-0.014}$ & 0.455$\,^{+0.018}_{-0.015}$ \\
 3$\sigma$ range & 0.408 $\rightarrow$ 0.603 & 0.434 $\rightarrow$ 0.610 & 0.416 $\rightarrow$ 0.599 \\
 \hline
 $\sin^{2}\theta_{23}$ (IMO) & 0.570$\,^{+0.016}_{-0.022}$ & 0.578$\,^{+0.010}_{-0.017}$ & 0.569$\,^{+0.013}_{-0.021}$  \\
 3$\sigma$ range & 0.410 $\rightarrow$ 0.613 & 0.433 $\rightarrow$ 0.608 & 0.417 $\rightarrow$ 0.606 \\
 \hline
 \hline
 $\dfrac{\Delta m^{2}_{\text{sol}}}{10^{-5}\,\text{eV}^{2}}$ & 7.42$\,^{+0.21}_{-0.20}$ & 7.50$\,^{+0.22}_{-0.20}$ & 7.36$\,^{+0.16}_{-0.15}$  \\
 3$\sigma$ range & 6.82 $\rightarrow$ 8.04 & 6.94 $\rightarrow$ 8.14 & 6.93 $\rightarrow$ 7.93 \\
 \hline
 \hline
 $\dfrac{\Delta m^{2}_{\text{atm}}}{10^{-3}\,\text{eV}^{2}}$ (NMO) & 2.55$\,^{+0.02}_{-0.03}$ & 2.56$\,^{+0.03}_{-0.04}$ & 2.522$\,^{+0.023}_{-0.03}$ \\
 3$\sigma$ range & 2.430 $\rightarrow$ 2.593 & 2.47 $\rightarrow$ 2.63 & 2.436 $\rightarrow$ 2.605  \\
 \hline
 $\dfrac{|\Delta m^{2}_{\text{atm}}|}{10^{-3}\,\text{eV}^{2}}$ (IMO) & 2.45$\,^{+0.02}_{-0.03}$ & 2.46$\,^{+0.03}_{-0.03}$ & 2.418$\,^{+0.0304}_{-0.024}$ \\
 3$\sigma$ range & 2.410 $\rightarrow$ 2.574 & 2.37 $\rightarrow$ 2.53  & 2.341 $\rightarrow$ 2.501  \\
 \hline \hline
 $\delta_{\mathrm{CP}}/^{\circ}$ (NMO) & 230$\,^{+36}_{-25}$ & 194$\,^{+24}_{-22}$ & 223$\,^{+33}_{-23}$ \\
 3$\sigma$ range & 144 $\rightarrow$ 350 & 128 $\rightarrow$ 359 & 139 $\rightarrow$ 355 \\
 \hline
 $\delta_{\mathrm{CP}}/^{\circ}$ (IMO) & 278$\,^{+22}_{-30}$ & 284$\,^{+26}_{-28}$ & 274$\,^{+25}_{-27}$   \\
 3$\sigma$ range & 194 $\rightarrow$ 345 & 200 $\rightarrow$ 353 & 193 $\rightarrow$ 342 \\
 \hline
 \end{tabular}
 \caption{\footnotesize{Best-fit values, $\pm$ $1\sigma$ uncertainties, and $3\sigma$ allowed ranges of the three-flavor neutrino oscillation parameters from the three global fit analyses of world neutrino data~\cite{deSalas:2020pgw,Esteban:2020cvm,NuFIT,Capozzi:2021fjo}. For Esteban $et~al.$~\cite{Esteban:2020cvm, NuFIT}, $\Delta m^{2}_{\text{atm}}$ stands for $\Delta m^{2}_{31}\, (\Delta m^{2}_{32})$ for NMO (IMO). For de Salas $et~al.$ \cite{deSalas:2020pgw} and Capozzi $et~al.$ \cite{Capozzi:2021fjo} $\Delta m^{2}_{\text{atm}}$ signifies $\Delta m^{2}_{31}$ for both NMO and IMO. Note that for Capozzi $et~al.$, we estimate the values of $\Delta m^{2}_{31}$ for both NMO and IMO using the relation $\Delta m^{2}_{31} = \Delta m^{2} + \Delta m^{2}_{21}/2$ where $\Delta m^{2} = m_{3}^{2} - (m^{2}_{1} + m^{2}_{1})/2$\,. 
}}
 \label{table:five}
 \end{table}

Table~\ref{table:five} shows the best-fit values and current $1\sigma$ and $3\sigma$ allowed ranges of various oscillation parameters as obtained in the three golbal fit studies~\cite{deSalas:2020pgw,Esteban:2020cvm,NuFIT,Capozzi:2021fjo} of world neutrino data. Note that for Esteban $et~al.$~\cite{Esteban:2020cvm, NuFIT}, $\Delta m^{2}_{\text{atm}}$ stands for $\Delta m^{2}_{31}\, (\Delta m^{2}_{32})$ for NMO (IMO). $\Delta m^{2}_{\text{atm}}$ represents $\Delta m^{2}_{31}$ for both NMO and IMO in the studies performed by de Salas $et~al.$ \cite{deSalas:2020pgw} and Capozzi $et~al.$ \cite{Capozzi:2021fjo}. For Capozzi $et~al.$, we estimate the values of $\Delta m^{2}_{31}$ for both NMO and IMO using the relation $\Delta m^{2}_{31} = \Delta m^{2} + \Delta m^{2}_{21}/2$ where $\Delta m^{2} = m_{3}^{2} - (m^{2}_{1} + m^{2}_{1})/2$\,. As far as the measurement of $\theta_{23}$ is concerned, there is a slight disagreement between the three global fit studies. de Salas $et~ al.$ in Ref. \cite{deSalas:2020pgw}, obtain a best-fit value of $\sin^2\theta_{23}$ in HO around $\sim 0.57$ assuming NMO, while Capozzi $et~ al.$ in Ref.~\cite{Capozzi:2021fjo} and Esteban $et~ al.$ in Ref.~\cite{NuFIT,Esteban:2020cvm} find the best-fit value of $\sin^2\theta_{23}$ in LO around $\sim 0.45$. This discrepency maybe due to the recent Super-K Phase I-IV 364.8 kt$\cdot$yrs of atmospheric data~\cite{sk} that only Capozzi $et~ al.$ and Esteban $et~ al.$ include in their latest analyses. It is indeed impressive to see that all the three global fit analyses indicate towards leptonic CP violation ($\sin \delta_{\mathrm{CP}} < 0$). The recent analysis by Capozzi $et~ al.$~\cite{Capozzi:2021fjo} finds a preference for $\delta_{\mathrm{CP}} \simeq \, 223^{\circ}$ with respect to the CP-conserving value of $\delta_{\mathrm{CP}}$ = 180$^{\circ}$ at 1.6$\sigma$ C.L. under NMO scheme and disfavors the values of $\delta_{\mathrm{CP}}$ in the range of 0$^{\circ}$ to 139$^{\circ}$ at more than 3$\sigma$ C.L. assuming NMO.

\acknowledgments 
We thank the organizers of the IRCHEP 1400 conference where the preliminary results from this work were presented. We acknowledge the support of the Department of Atomic Energy (DAE), Govt. of India. S.K.A. is supported by the Young Scientist Project [INSA/SP/YSP/144/2017/1578] from the Indian National Science Academy (INSA). S.K.A. acknowledges the financial support from the Swarnajayanti Fellowship Research Grant (No. DST/SJF/PSA-05/2019-20) provided by the Department of Science and Technology (DST), Govt. of India and the Research Grant (File no. SB/SJF/2020-21/21) from the Science and Engineering Research Board (SERB) under the Swarnajayanti Fellowship by the DST, Govt. of India. M.S. acknowledges financial support from the DST, Govt. of India (DST/INSPIRE Fellowship/2018/IF180059). The numerical simulations are carried out using SAMKHYA: High-Performance Computing Facility at Institute of Physics, Bhubaneswar.


\bibliographystyle{JHEP}
\bibliography{matter}

\end{document}